\documentclass[aoas]{imsart}
\RequirePackage[OT1]{fontenc}
\RequirePackage{amscd,amsfonts,amsmath,amssymb,amsthm,bbm,bm,latexsym,mathrsfs}
\RequirePackage{natbib}
\RequirePackage{xr}
\RequirePackage[colorlinks,citecolor=blue,urlcolor=blue]{hyperref}
\RequirePackage{epsfig,graphics,graphicx,subfigure}
\RequirePackage[english]{babel}
\RequirePackage{rotating}
\RequirePackage{epstopdf}
\RequirePackage{longtable}
\RequirePackage{tabularx}

\RequirePackage[usenames]{color}
\RequirePackage{bookmark}
\RequirePackage{sgame}
\allowdisplaybreaks

\RequirePackage[export]{adjustbox}
\RequirePackage{gensymb}
\RequirePackage{epstopdf}
\RequirePackage{tikz}
\hypersetup{hidelinks}
\usetikzlibrary{shapes,arrows}	% flow charts
\RequirePackage{tkz-berge}
\usetikzlibrary{bayesnet}
\RequirePackage[toc,page]{appendix}
\RequirePackage{chapterbib}
\RequirePackage[normalem]{ulem}% to highlight cancelled text
\RequirePackage{cancel}

\makeatother

\theoremstyle{remark}

\theoremstyle{plain}
\arxiv{arXiv:0000.0000}

\startlocaldefs
\theoremstyle{plain}

\endlocaldefs

\begin{document}

\begin{frontmatter}
\title{Bayesian nonparametric panel Markov-switching GARCH models\thanksref{T1}}
\runtitle{BNP panel MS-GARCH}
\thankstext{T1}{Corresponding author: Roberto Casarin E-mail: \texttt{r.casarin@unive.it}. The code is available upon request.}

\begin{aug}
\author{\fnms{Roberto} \snm{Casarin}\thanksref{m1}\ead[label=e1]{r.casarin@unive.it}},
\author{\fnms{Mauro} \snm{Costantini}\thanksref{m2}\ead[label=e2]{mauro.costantini@univaq.it}}
\and
\author{\fnms{Anthony} \snm{Osuntuyi}\thanksref{m1}\ead[label=e3]{ayokunle.osuntuyi@unive.it}}
%\ead[label=u1,url]{http://www.foo.com}

\runauthor{R. Casarin et al.}

\affiliation{University Ca' Foscari of Venice\thanksmark{m1} and University of L'Aquila \thanksmark{m2}}

%\address{Roberto Casarin\\
%Department of Economics\\
%University Ca' Foscari of Venice\\
%San Giobbe 873/b\\
%30121 Venezia, Italy\\
%E-mail: \printead*{e1}}
%
%\address{Mauro Costantini\\
%Department of Industrial and Information Engineering and Economics\\
%University of L'Aquila\\
%Via Giuseppe Mezzanotte\\
%67100, L'Aquila, Italy\\
%E-mail: \printead*{e2}}
%
%\address{Anthony Osuntuyi\\
%Department of Economics\\
%University Ca' Foscari of Venice\\
%San Giobbe 873/b\\
%30121 Venezia, Italy\\
%E-mail: \printead*{e3}}

\end{aug}

\begin{abstract}  \enspace This paper introduces a new model for panel data with Markov-switching GARCH effects. The model incorporates a series-specific hidden Markov chain process that drives the GARCH parameters. To cope with the high-dimensionality of the parameter space, the paper exploits the cross-sectional clustering of the series by first assuming a soft parameter pooling through a hierarchical prior distribution with two-step procedure, and then introducing clustering effects in the parameter space through a nonparametric prior distribution. The model and the proposed inference are evaluated through a simulation experiment. The results suggest that the inference is able to recover the true value of the parameters and the number of groups in each regime. An empirical application to 78 assets of the SP\&100 index from $6^{th}$ January 2000 to $3^{rd}$ October 2020 is also carried out by using a two-regime Markov switching GARCH model. The findings shows the presence of 2 and 3 clusters among the constituents in the first and second regime, respectively.
\end{abstract}



\end{frontmatter}

\section{Introduction}

Over the last ten years, there has been an increasing interest in the study of volatility of large panels of asset returns, with a special focus on dynamic dependence and heterogeneity across assets \citep[][]{Pakel2011,Barigozzi2014,Ardia2018,Bollerslev2020}. The empirical evidence has also shown the presence of regimes in the volatility of financial returns \citep[see][among others]{Ardia08,Ang2012, Bauwens2016,Haas2018} and Markov switching (MS) GARCH models have been used to cope with regime changes and temporal clustering of the conditional volatility.

Several GARCH models have been proposed to account for dependence  \citep[for a review, see][]{Conchi15,Bauwens2016,Bauwens2020}, but the estimation of a large number of parameters with the available data dimension remains an open issue. In this respect, evidence of cluster-wise dependence in the distribution of financial asset returns \citep[see][]{Bauwens2007} has prompted researcher to adopt cross-sectional clustering of the time series as a building block for a dimensionality reduction step in large dimensional problems of the parameter space \citep[see, for example,][]{Hir02,Bil19}.

%,Caporin2014,Hansen2016,,Clements2019,
%\textcolor{red}{We propose a new Bayesian model for panel data which account for temporal and cross-sectional clustering effects in the volatility dynamics. Modelling and forecasting time variations in the volatility is crucial in many macroeconomic [REFERENCES] and financial applications such as asset pricing and asset allocation [REFERENCES] BECAUSE MOTIVARE BENE... Cross-sectional clustering of the time series is usually employed as a building block for a dimensionality reduction step in large dimensional problems. In forecasting with multiple time series clustering is used to reduce the dimension of the parameter space \citep[e.g., see][]{Bil19} or the latent space \cite[e.g., see][]{CasRavGraVan20}. In asset allocation clustering can be used to reduce the portfolio size [REFERENCES]... OTHERS MAURO WILL PROVIDE ATHER REFERENCES AND MOTIVATIONS}.
%\textcolor{red}{Among volatility models, the the class of Generalized Autoregressive Conditional Heteroskedastic (GARCH) model introduced by \cite{Boll86} become very popular. \cite{HamSum94} and \cite{Cai94} propose a Markov Switching-Autoregressive Conditional Heteroskedastic (MS-ARCH) model to account for the structural changes in the variance process. \cite{Gray96} instead considers a MS-GARCH model which is more parsimonious than the MS-ARCH model.

In this paper, we propose to model the cross-sectional clustering effects with a Bayesian nonparametric technique \citep{Ferguson73,Lo1984} where a hierarchical Pitman-Yor process prior \citep{PitmanYor1997} for the MS-GARCH parameters is considered. Non-parametric Bayesian techniques have been largely and successfully used in different fields such as biostatistics \citep[][]{Do2005},  biology \cite[][]{Arbel2016}, medicine \citep[][]{Xu2016}, and neuroimaging \citep[][]{Zhang2016}. For an introduction to Bayesian non-parametrics see \cite{BNP2010} and for a review of models and applications in different fields see \cite{muller2013}.

In our panel model, the first stage of the hierarchical prior allows for cross-unit heterogeneity, while shrinking all unit-specific parameters towards a common mean. The second stage of the hierarchy allows for mixed effects in the common mean. There are many advantages in using this hierarchical nonparametric prior. First, our approach allows for making inference on the number of mixture components in the cross-sectional clustering. Second, it adds flexibility to the model allowing for different shapes of the prior and posterior predictive distributions. Third, the predictive distribution incorporates uncertainty in the parameters and in the number of mixture components. Lastly, the Bayesian nonparametric combined with a data-augmentation strategy makes the inference more tractable for our high dimensional model.

The model and inference proposed in this paper are novel in some respects. As such, the paper contributes to the literature on Bayesian semiparametrics and nonparametrics for time series analysis \citep[e.g., see][]{TaddyKottas2009,JensenMaheu2010,GriffinSteel2011,DiLucca2013,Bas13,casarin19,Bil19,Nieto2016,Griffin18}. The paper also innovates the Bayesian nonparametric dynamic panel model in \cite{Hir02} by introducing Markov-switching and GARCH dynamics.

The paper also extends the nonparametric switching regression in \cite{TaddyKottas2009} to a panel model with GARCH dynamics. Our approach differs from those in \cite{Hir02} and \cite{TaddyKottas2009}, and is in line with the strategies for large dimensional and over-parametrized models \cite[e.g., see][]{MacDun10,wang2010sparse,Bil19}, where a multiple-stage hierarchical prior is used to combine partial pooling and clustering effects in the parameter space. Further, differently from \cite{Hir02} and \cite{TaddyKottas2009}, the paper uses a MCMC algorithm for posterior approximation that relies on the efficient sampling method developed in \cite{walker2007,walker2011,HatjispyrosaNicolerisaWalker}. Lastly, the paper makes a contribution to the literature on Bayesian Markov-switching panel models \citep[e.g., see][]{Kau10, Kau11, Bil13, CFMR2019} by introducing GARCH effects and allowing for a flexible nonparametric specification.

The estimation of MS-GARCH models is also a difficult task given the path dependence problem \citep{Gray96} and approximation methods have been considered \citep[e.g., see][]{Bau10, Hen11, Ardia08,HMP04,HeMah10,Bau11,Elliot12,Du12,Wee2020}. In this paper, we extend the univariate Gibbs sampler by \cite{Antho14} to a multiple time series set-up and provide an efficient MCMC procedure for the hidden states of a panel MS-GARCH model. The proposed method relies on a combination of Gibbs and Metropolis samplers. The model and the proposed inference are evaluated through simulation experiments. The results show that the inference is able to recover the true value of the parameters and the number of groups in each regime.

Our model is applied to 78 assets of the SP\&100 index from $6^{th}$ January 2000 to $3^{rd}$ October 2020. The analysis can be useful for portfolio making and style investing decisions. 
In particular, the analysis aims to identify the under- and over-performance regimes in expected returns. Then, clusters of assets within each regime are identified. Lastly, we use the sector classification and some fundamental financial ratios to study the composition of the clusters. The main empirical results are as follows. We find evidence of different clustering structure across regimes. Regime 1 (over-performance phase) and regime 2 (under-performance phase) comprise 2 and 3 clusters, respectively. While the composition of clusters varies across regimes, some common features are observed. In both regimes, medium size companies represent the largest majority and assets in cluster 2 seem to be overvalued by their Price-to-Earning ratio. 
%The pattern of the market capitalization is similar across the two regimes as Cluster 2 shows lower(larger) values compared to Cluster 1 before (after) the 2008/09 Global financial crisis. 

\indent The paper is organized as follows. Section \ref{pan} introduces our MS-GARCH panel model and the Bayesian nonparametric prior distribution. Section \ref{CompDet} presents the data augmentation strategy and the posterior approximation method. In Section \ref{EmpApp}, we present a simulation study and an empirical application to the financial returns data. Section \ref{Concl} concludes.

\section{A Bayesian nonparametric MS-GARCH model}\label{pan}
We assume that the observable variable $y_{it}$ for the $i$-th unit of the panel at time $t$ satisfies
\begin{eqnarray}
y_{it}&=&\mu_{i}(s_{it})+\sigma_{it}\varepsilon_{it},\quad \varepsilon_{it}\overset{iid}{\sim} \mathcal{N}(0,1)\label{MeaEqn} 
\end{eqnarray}
for $t=1,\ldots,T$ and $i=1,\ldots,N$, where $\mathcal{N}(\mu,\sigma^2)$ denotes the Gaussian distribution with location $\mu$ and scale $\sigma$. The conditional variance is as follows:
\begin{eqnarray}
\sigma_{it}^{2}&=&\gamma_{i}(s_{it})+ \alpha_{i}(s_{it})\varepsilon_{it-1}^2+\beta_{i}(s_{it})\sigma_{it-1}^{2}\label{GARCHEqn} 
\end{eqnarray}
which is the MS-GARCH model, and $s_{it}$, $t=1\ldots,T$ is a hidden Markov chain process with transition probability
\begin{equation}
P(s_{it}=k|s_{it-1}=l)=p_{i,kl}
\end{equation}
where $k,l=1,\ldots,K$ with $K$ the number of states.
The following functional form for the switching parameters is specified as follows:
\begin{eqnarray}
&&\mu_{i}(s_{it})=\sum_{k=1}^{K}\mu_{ik}\mathbb{I}(s_{it}=k),\quad\alpha_{i}(s_{it})=\sum_{k=1}^{K}\alpha_{ik}\mathbb{I}(s_{it}=k)\label{prior11}\\
&&\beta_{i}(s_{it})=\sum_{k=1}^{K}\beta_{ik}\mathbb{I}(s_{it}=k),\quad\gamma_{i}(s_{it})=\sum_{k=1}^{K}\gamma_{ik}\mathbb{I}(s_{it}=k)\label{prior12}
\end{eqnarray}
We cope with the high-dimensionality of the parameter space due to the large cross-section dimension $N$, and related overfitting issues of the model, by exploiting cross-sectional clustering of the series. More specifically we propose to combine two modelling strategies. First, we assume soft parameter pooling through a hierarchical prior distribution with two stages, and second we introduce clustering effects in the parameter space through a nonparametric prior. The resulting joint prior distribution for the MS-GARCH parameters is given by the following.

In the first stage, the rows of the transition matrix are assumed to follow a Dirichlet distribution: 
\begin{equation}
(p_{i,k1},\ldots,p_{i,kK})\overset{iid}{\sim} \mathcal{D}(\phi r_{k1},\ldots,\phi r_{kK})
\end{equation}
for all units $i=1,\ldots,N$ and regimes $k=1,\ldots,K$, where the precision parameter $\phi$ shrinks the unit-specific probabilities toward a common value $(r_{k1},\ldots,r_{kK})$. For the second stage we assume
\begin{equation}
(r_{k1},\ldots,r_{kK})\overset{iid}{\sim}\mathcal{D}(d,\ldots,d)
\end{equation}
with $d=1/K$. 

To cope with the high dimension of the parameter space, in the first stage of the hierarchical prior, we shrink the switching parameters toward some common values, and in the second stage we introduce a regime-specific process which is clustering the units in $M_k$ groups $C_{1,k},\ldots,C_{M_k,k}$ such that $C_{h,k}\cap C_{l,k}=\emptyset$ for $h\neq l$ and $\cup_{h=1}^{M_k}C_{h,k}=\{1,\ldots,N\}$.
In the first stage, we assume the following
\begin{eqnarray}
%&&\mu_{ik}\sim \mathcal{N}(\tilde{\mu}_{ik}^{*},s),\quad \gamma_{ik}\sim \mathcal{G}a(v \tilde{\gamma}_{ik}^{*},v),\\
&&\mu_{ik}\sim \mathcal{N}(\tilde{\mu}_{ik}^{*},s),\quad \gamma_{ik}/a\sim \mathcal{B}e(r \tilde{\gamma}_{ik}^{*}/a,r (1-\tilde{\gamma}_{ik}^{*}/a)),\label{prior11}\\
&&\alpha_{ik}\sim \mathcal{B}e(r\tilde{\alpha}_{ik}^{*},r(1-\tilde{\alpha}_{ik}^{*})),\quad \beta_{ik}\sim \mathcal{B}e(r\tilde{\beta}_{ik}^{*},r(1-\tilde{\beta}_{ik}^{*}))\label{prior12}
\end{eqnarray}
for $k=1,\ldots,K$ where $\mathcal{B}e(\alpha,\beta)$ denotes the beta distribution with mean $\alpha/(\alpha+\beta)$ and $a$ is a real positive constant.
%and $\mathcal{G}a(\alpha,\beta)$ the gamma distribution with mean $\alpha/\beta$. 
The scale hyper-parameters $s$ and $r$ are shrinking $\boldsymbol{\theta}_{ik}=(\mu_{ik},\gamma_{ik},\alpha_{ik},\beta_{ik})\in \mathbb{R}\times [0,a]\times[0,1]^2$ toward the parameter $\tilde{\boldsymbol{\theta}}_{ik}^{*}=(\tilde{\mu}_{ik}^{*},\tilde{\gamma}_{ik}^{*},\tilde{\alpha}_{ik}^{*},\tilde{\beta}_{ik}^{*})\in \mathbb{R}\times [0,a]\times[0,1]^2$ which is assumed to be constant for all units in the same cluster, that is for all $i\in C_{hk}$ where $h=1,\ldots, M_k$ (for further details see Section \ref{CompDet} and Eq. \ref{ClustPar}). Since the parameters are non-identified due to the label switching problem, we follow a commonly used approach and impose a prior restriction on the intercepts $\mu_{i1}>\mu_{i2} >\ldots >\mu_{iK}$  \citep[e.g., see][]{Celeux1998,Fruhwirth2001,Fruhwirth2006}.

The second stage of the hierarchy is generating the clusters of parameters. For each regime $k$ we assume a Pitman-Yor process (PYP) prior
\begin{eqnarray}
\tilde{\boldsymbol{\theta}}_{ik}^{*}|G_k \overset{iid}{\sim} G_k,\quad G_k&\sim& \hbox{PYP}(\nu,\psi, H_0)\label{prior21}
\end{eqnarray}
with base measure $H_0$ and concentration and dispersion parameters $\nu\in[0,1]$ and $\psi>-\nu$, respectively. We assume $H_0(\boldsymbol{\theta})$ is the product measure of the following independent normal and uniform distributions
\begin{eqnarray}
&&\mathcal{N}(\mu;m^{*},s^{*}),\quad \mathcal{U}(\gamma;0,a),\quad \mathcal{U}(\alpha;0,1),\quad\mathcal{U}(\beta;0,1)\label{prior22}
\end{eqnarray} 
which are usually chosen as prior distributions in parametric Bayesian inference for MS-GARCH \cite[e.g., see][]{Antho14}. The PYP introduced in \cite{PitmanYor1997} is  a generalization of the Dirichlet process (DP) defined in  \cite{Ferguson73} which can be obtained for $\nu=0$.

%The clustering structure of the PYP process is defined by a Polya-Urn sampling scheme and can be illustrated through the Chinese Restaurant metaphor. The parameter $\boldsymbol{\theta}_{i}^{*}$ of the $i$-th unit is either equal to one of the other units or a new one from the base distribution $H_0$, i.e.:
Through the illustration of the Chinese Restaurant metaphor, the clustering structure of the PYP is defined by a Polya-Urn sampling scheme. The parameter $\boldsymbol{\theta}_{i}^{*}$ of the $i$-th unit is either equal to one of the other units or a new one from the base distribution $H_0$, i.e.:
\begin{equation}
\tilde{\boldsymbol{\theta}}_{ik}^{*}|\tilde{\boldsymbol{\theta}}_{1k}^{*},\ldots,\tilde{\boldsymbol{\theta}}_{i-1k}^{*}=\frac{i}{\psi-\nu+i}\sum_{h=1}^{i-1}\delta_{\boldsymbol{\tilde{\theta}}_{hk}^{*}}(\tilde{\boldsymbol{\theta}}_{ik}^{*})+\frac{\psi}{\psi-\nu+i}H_0(\tilde{\boldsymbol{\theta}}_{ik}^{*})\label{prior23}
\end{equation}
This sequential allocation procedure is generating clusters in the parameter space, where the number of clusters is random. The Pitman-Yor process induces the following prior distribution on the number of clusters $M_{k}$ 
\begin{equation*}
{P}(M_{k}=h)=\frac{\nu^{h-1}\Gamma(\psi/\nu+h)\Gamma(\psi+1)}{\Gamma(\psi/\nu+1)\Gamma(\psi+N)}S_{\nu}(N,h)
\end{equation*}
with $h\in\mathbb{N}$, where $S_{\nu}(N,h)$ is a generalized Stirling number of the first kind, and $\Gamma(x)$ is the one-parameter gamma function (e.g., see \cite{Pitman2006}, Ch. 1 and 3). The following formula is used to evaluate the prior mean of the number of clusters:
\begin{equation*}
\mathbb{E}(M_{k})=\left\{%
\begin{array}{ll}
\sum_{h=1}^{N}\frac{\psi}{\psi+h-1},& \hbox{if}\,\,\nu=0,\\
\frac{\Gamma(\psi+\nu+N)\Gamma(\psi+1)}{\nu\Gamma(\psi+\nu)\Gamma(\psi+N)}-\frac{\psi}{\nu},& \hbox{if}\,\,\nu\neq0.\\
\end{array}
\right.
\end{equation*}
We summarize our Bayesian nonparametric model in the Directed Acyclic Graph representation of Fig. \ref{DAG}.

\tikzset{
    obs/.style={rectangle, draw, minimum width = 5ex,
                minimum height = 5ex, inner sep = 0.1},
    latent/.style={circle, draw, minimum size =5ex, fill=gray!10,
                inner sep = 0.1},
    par/.style={circle, draw, minimum size =5ex, fill=white,
                inner sep = 0.1},
}
\begin{figure}[t]
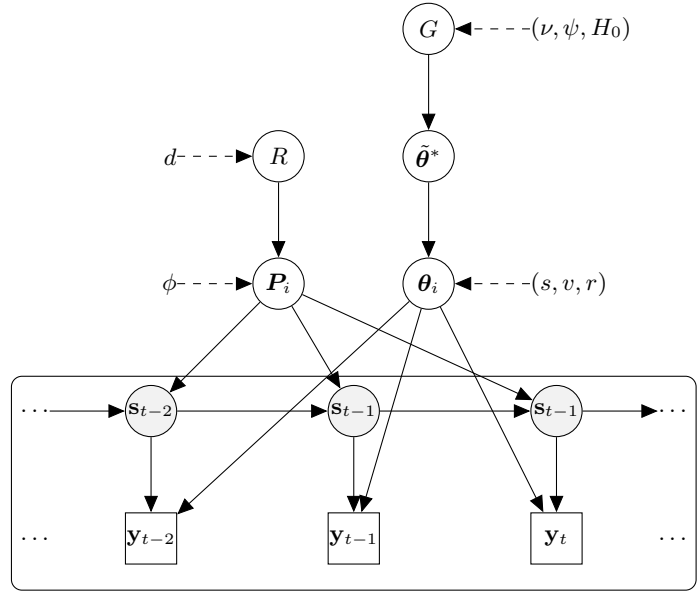

  \begin{center}
    \tikz[scale=3]{ %
%%%%%%%%%%%%%%%%%%%%%%%%%%%%%%%%%%%%%%%%%%%%%%%%%%%%%%%%         OBS
         \node[obs] (y1) {$\mathbf{y}_{t-1}$} ; %
         \node[obs, right = of y1,xshift = 1cm] (y2) {$\mathbf{y}_{t}$} ; %
         \node[const, right = of y2,xshift = 0cm] (y3) {$\ldots$} ; %
         \node[obs, left = of y1, xshift = -1cm] (y12) {$\mathbf{y}_{t-2}$} ;
         \node[const, left= of y12,xshift = 0cm] (y13) {$\ldots$} ; %
%%%%%%%%%%%%%%%%%%%%%%%%%%%%%%%%%%%%%%%%%%%%%%%%%%%%%%%         LAT
         \node[latent, above =of y1] (s1) {$\mathbf{s}_{t-1}$} ; %        
         \node[latent,right= of s1,xshift = 1cm] (s2) {$\mathbf{s}_{t-1}$} ; %        
         \node[const, right= of s2,xshift = 0cm] (s3) {$\ldots$} ; %        
         \node[latent,left= of s1,xshift = -1cm] (s12) {$\mathbf{s}_{t-2}$} ; 
         \node[const, left= of s12,xshift = 0cm] (s13) {$\ldots$} ; %        
 %%%%%%%%%%%%%%%%%%%%%%%%%%%%%%%%%%%%%%%%%%%%%%%%%%%%%%%         DAG
      \edge {s1}{y1}
      \edge {s2}{y2}
      \edge {s12}{y12}
      \edge {s1}{s2}
      \edge {s2}{s3}
      \edge {s12}{s1}
      \edge {s13}{s12}
 %%%%%%%%%%%%%%%%%%%%%%%%%%%%%%%%%%%%%%%%%%%%%%%%%%%%%%%         PAR
      \node[par,  above= of s1, xshift = -1cm, yshift = 0cm] (p) {$\boldsymbol{P}_{i}$};      
      \node[par,  above= of p] (R) {$R$};      
      \node[const,  left= of p] (f) {$\phi$};      
      \node[const,  left= of R] (d) {$d$};      
%%%%%%%%%%%%%%%%%%%%%%%%%
      \node[par,  right= of p, xshift = 0.3cm] (mu1) {$\boldsymbol{\theta}_{i}$};      
      \node[par,  above= of mu1, xshift = 0cm, yshift = 0cm] (theta) {$\tilde{\boldsymbol{\theta}}^{*}$};      
%%%%%%%%%%%%%%%%%%%%%%%%%
      \node[const,  right= of mu1] (s) {$(s,v,r)$};      
      \node[par,  above= of theta] (G) {$G$};      
      \node[const,  right= of G] (G0) {$(\nu,\psi,H_0)$};      
      \edge {mu1}{y1}
      \edge {mu1}{y2}
      \edge {mu1}{y12}
      \edge {p}{s1}
      \edge {p}{s2}
      \edge {p}{s12}
      \edge {R}{p}
      \edge{G}{theta}
      \edge{theta}{mu1}
%%%%%%%%%%%%%%%%%%% hyparpar
      \edge[style={dashed}]{G0}{G}
      \edge[style={dashed}]{s}{mu1}
      \edge[style={dashed}]{d}{R}      
      \edge[style={dashed}]{f}{p}      
\plate {sec} {(y13) (y12) (y1) (y2) (y3) (s13) (s12) (s1) (s2) (s3)} {} ;        
}
  \end{center}
  \caption{DAG of the Bayesian nonparametric MS-GARCH panel model. It exhibits the hierarchical structure of the observations $\mathbf{y}_t=(y_{1t},\ldots,y_{Nt})$ (boxes), the latent variables $\mathbf{s}_{t}=(s_{1t},\ldots,s_{Nt})$ (gray circles), the parameters $P_i=(p_{i,11},\ldots,p_{i,1K},\ldots,p_{i,K1},\ldots,p_{i,KK})$, $\boldsymbol{\theta}_i=(\mu_i,\gamma_i,\alpha_i,\beta_i)$, the hyperparameters of the first stage $R=(\mathbf{r}_1,\ldots,\mathbf{r}_K)$, $\tilde{\boldsymbol{\theta}}^{\ast}_i=(\tilde{\mu}^*_i,\tilde{\gamma}^*_i,\tilde{\alpha}^*_i,\tilde{\beta}^*_i)$ and  of the second stage $G$ (white circles). The directed arrows show the causal dependence structure of the model.} \label{DAG}
\end{figure}

\medskip

It is possible to show that the PYP clustering effects on the cross section of time series correspond to a probabilistic clustering of the parameters based on an infinite mixture distribution. The Pitman-Yor process prior can be written in a Sethuraman's like representation as a discrete random measure
\begin{equation}
G_k(\boldsymbol{\theta}^{\ast})=\sum_{h=1}^{\infty}W_{hk}\delta_{\boldsymbol{\theta}_{hk}^{\ast}}(d\boldsymbol{\theta}^{\ast})\label{rm1}
\end{equation}
where the atoms $\boldsymbol{\theta}_{hk}^{*}$ are i.i.d. random variables from the base measure $H_0$ and the random weights $W_{hk}$ have the stick-breaking representation 
\begin{equation}
W_{hk}=V_{hk}\prod_{l=1}^{h-1}(1-V_{lk})\label{rm2}
\end{equation}
with $V_{lk}\sim \mathcal{B}e(1-\nu,\phi+\nu l)$ i.i.d. $l=1,2,\ldots$ \cite[see][]{Arbel2019}.

By integrating out the discrete part of the hierarchical prior one obtains the following infinite mixture representation of the prior distribution on $\boldsymbol{\theta}$ 
\begin{eqnarray}
\boldsymbol{\theta}_{ik}|G_k &\overset{ind}{\sim}&\int \pi(\boldsymbol{\theta}_{ik}|\boldsymbol{\theta}^{*})G_{k}(d\boldsymbol{\theta}^{*})=\sum_{h=1}^{\infty}W_{hk}\pi(\boldsymbol{\theta}_{ik}|\boldsymbol{\theta}^{*}_{hk})\label{rm3}
\end{eqnarray}
where $\pi(\boldsymbol{\theta}_{ik}|\boldsymbol{\theta}^{*})$ is the joint parameter distribution at the first stage of hierarchical prior (see Eqs. \ref{prior11}-\ref{prior12}) and $G_k(\boldsymbol{\theta}^{*})$ is the distribution at the second stage. In conclusion the PYP prior allows for probabilistic clustering in the parameter space. 

\medskip

The predictive density induced by our prior assumptions can be written as
\begin{eqnarray}
y_{it}|G,s_{it} &\overset{ind}{\sim}&\sum_{h=1}^{\infty}W_{hs_{it}}\int f_t(y_{it}|s_{it},\boldsymbol{\Theta})\pi(d\boldsymbol{\theta}|\boldsymbol{\theta}^{*}_{hs_{it}})
\end{eqnarray}
where $f_t(y_{it}|s_{it}=k,\boldsymbol{\Theta})=f(y_{it}|\mu_{ik},\sigma_{ik,t})$ is the transition kernel of the MS-GARCH with $\sigma_{ik,t}^{2}=\gamma_{ik}+\alpha_{ik}\varepsilon_{it-1}^2+\beta_{ik}\sigma_{it-1}^{2}$ for $k=1,\ldots,K$ and $\boldsymbol{\Theta}=(\boldsymbol{\theta}_1,\ldots,\boldsymbol{\theta}_{K})$, $\boldsymbol{\theta}_{k}=(\boldsymbol{\theta}_{1k},\ldots,\boldsymbol{\theta}_{Nk})$ and $\boldsymbol{\theta}_{ik}=(\mu_{ik},\gamma_{ik},\alpha_{ik},\beta_{ik})$. This prior predictive densities accounts for various forms of possible heterogeneity in the data such as asymmetry, excess of kurtosis and multimodality. 

\section{Posterior approximation}\label{CompDet}
Let $\Theta=(\boldsymbol{\theta}_{1},\ldots,\boldsymbol{\theta}_{K})$ be the collection of the unit- and regime-specific parameters $\boldsymbol{\theta}_{k}=(\boldsymbol{\theta}_{1k},\ldots,\boldsymbol{\theta}_{Nk})$ and $\boldsymbol{\theta}_{ik}=(\mu_{ik},\gamma_{ik},\alpha_{ik},\beta_{ik})$, and $P=(P_1,\ldots,P_N)$ the collection of transition probabilities. Let $Y=(\mathbf{y}_1,\ldots,\mathbf{y}_T)$ be the collection over time of the observation vector $\mathbf{y}_t=(y_{1t},\ldots,y_{Nt})$ and $S=(\mathbf{s}_1,\ldots,\mathbf{s}_T)$ be the collection over time of the latent vectors $\mathbf{s}_{t}=(s_{1t},\ldots,s_{Nt})$. The likelihood function of the proposed MS-GARCH panel model is
\begin{equation}
L(Y|\Theta,P)=\sum_{s_{1},\ldots,s_{T}\in\{1,\ldots,K\}}\prod_{t=1}^{T}\prod_{i=1}^{N}f(y_{it}|\theta_{i},s_{it})f(s_{it|s_{it-1}},P_i)
\end{equation}
where
\begin{equation}
f(s_{it}|s_{it-1},P_i)=\prod_{k=1}^{K}\prod_{l=1}^{k}p_{i,kl}^{\mathbb{I}(s_{it}=l)\mathbb{I}(s_{it-1}=k)}
\end{equation}
which is not tractable since it is written in integral form as usually in latent variable models. Nevertheless a data-augmentation principle  can be applied \citep{TanWon87} in order to develop efficient posterior simulation methods. Following a common strategy in panel Markov-switching literature \citep[e.g., see][]{Bil13,CFMR2019,BiaBilCasGui15}, we introduce the set of auxiliary allocation variables $\xi_{ikt}=\mathbb{I}(s_{it}=k)$ which allow us to write the complete-data likelihood function as follows
\begin{equation}
L(Y,\Xi|\Theta,P)=\prod_{t=1}^{T}\prod_{i=1}^{N}f(y_{it}|\theta_{i},s_{it})\prod_{k=1}^{K}\prod_{l=1}^{k}p_{i,kl}^{\xi_{ikt-1}\xi_{ilt}}
\end{equation}
where $\Xi=(\Xi_1,\ldots,\Xi_T)$ is the collection over time of the latent vectors $\Xi_t=(\boldsymbol{\xi}_{1t},\ldots,\boldsymbol{\xi}_{Nt})$ with $\boldsymbol{\xi}_{it}=(\xi_{i1,t},\ldots,\xi_{iKt})$.

The joint hierarchical prior distribution is
\begin{eqnarray}
\pi(\Theta,G)=\prod_{k=1}^{K}\left(\prod_{i=1}^{N}\pi(\boldsymbol{\theta}_{ik}|G_k)\prod_{l=1}^{k}p_{i,kl}^{r_{l}-1}\right)\pi(V_{k})\pi(\Theta_{k}^{*})
\end{eqnarray}
where 
\begin{eqnarray}
\pi(\boldsymbol{\theta}_{ik}|G_k)=\sum_{h=1}^{\infty}W_{hk}\pi(\boldsymbol{\theta}_{ik}|\boldsymbol{\theta}_{hk}^{*})\label{infmix}
\end{eqnarray}
is the infinite mixture prior where we recall
$\pi(\boldsymbol{\theta}_{ik}|\boldsymbol{\theta}_{hk}^{*})=$ $\mathcal{N}(\mu_{ik}|\mu_{hk}^{*},s)$ $\mathcal{B}e(\alpha_{ik}|\alpha_{hk}^{*},r)$ $\mathcal{B}e(\beta_{ik}|\beta_{hk}^{*},r)$ $\mathcal{B}e(\gamma_{ik}/a|\gamma_{hk}^{*}/a,r)$ is the first-stage joint prior distribution given in Eqs. \ref{prior11}-\ref{prior12} and
\begin{eqnarray}
\pi(\Theta_{k}^{*})&=&\prod_{h=1}^{\infty}\mathcal{N}(\mu_{hk}^{*};m^{*},s^{*})\mathcal{U}(\alpha_{hk}^{*};0,1)\mathcal{U}(\beta_{hk}^{*};0,1)\mathcal{U}(\gamma_{hk}^{*};0,a)\label{infatom}\\
\pi(V_{k})&=&\prod_{l=1}^{\infty}\mathcal{B}e(V_{lk};1-\nu,\psi+\nu l)\label{infweights}
\end{eqnarray}
is joint distribution of the infinite collection of stick-breaking variables and atoms, $V_k=(V_{1k},V_{2k},\ldots)$ and $\Theta_{k}^{*}=(\boldsymbol{\theta}_{1k}^{*},\boldsymbol{\theta}_{2k}^{*},\ldots)$, respectively, which are involved in the definition of the random measures $G_k(\boldsymbol{\theta}_{k}^{*})$ $k=1,\ldots,K$.

The joint prior distribution in a Bayesian nonparametric framework is usually not tractable since its support is the space of the discrete random measures which are infinite-dimensional objects (see Eqs. \ref{infmix}-\ref{infweights}). Nevertheless, the data-augmentation principle can be applied in order to make the inference problem more tractable. Following the recent Bayesian nonparametrics literature \citep[e.g., see][]{Bas13,Bas18,Bil19}, we introduce a set of slice variables $U_{ik}\sim \mathcal{U}(0,1)$ and define the index set $\mathcal{A}_{ik}=\{h|U_{ik}<W_{hk}\}$. Then the infinite mixture can be demarginalized as follows
\begin{eqnarray}
\pi(\boldsymbol{\theta}_{ik}|U_k, V_k,\theta_{k}^{*})&=&\sum_{h=1}^{\infty}\mathbb{I}(U_{ik}<W_{hk})\pi(\boldsymbol{\theta}_{ik}|\boldsymbol{\theta}_{hk}^{*})\\
&=&\sum_{h\in \mathcal{A}_{ik}}\mathbb{I}(U_{ik}<W_{hk})\pi(\boldsymbol{\theta}_{ik}|\boldsymbol{\theta}_{hk}^{*})\nonumber
\end{eqnarray}
which is a almost-surely finite mixture since $\hbox{Card}(\mathcal{A}_{ik})<\infty$ a.s., where $U_k=(U_{1k},\ldots,U_{Nk})$ is the collection of slice variables. 

Following the standard practice in finite mixture modelling we introduce the latent allocation variable $D_{ik}\in\mathcal{A}_{ik}$ and obtain
\begin{eqnarray}
\pi(\boldsymbol{\theta}_{ik}|U_k, D_k, V_k,\theta_{k}^{*})&=&\mathbb{I}(U_{ik}<W_{D_{ik}k})\pi(\boldsymbol{\theta}_{ik}|\boldsymbol{\theta}_{D_{ik}k}^{*})
\end{eqnarray}
where $D_k=(D_{1k},\ldots,D_{Nk})$. Let us denote with $V=(V_1,\ldots,V_K)$, $U=(U_1,\ldots,U_K)$ and $\Theta^{*}=(\boldsymbol{\theta}_{1}^{*},\ldots,\boldsymbol{\theta}_{K}^{*})$ the collections of regime-specific auxiliary variables and atoms. The joint posterior distribution $\pi(\Xi,\Theta,P,U,D,V,\Theta^{*}|Y)$ is proportional to 
\begin{eqnarray}
\quad\quad L(Y,\Xi|\Theta,P)&=&\! \prod_{k=1}^{K}\left(\prod_{i=1}^{N}\pi(\boldsymbol{\theta}_{ik}|U_k, D_k, V_k,\Theta_{k}^{*})\prod_{l=1}^{k}p_{i,kl}^{r_{l}-1}\right)\pi(V_{k})\pi(\Theta_{k}^{*}).
\end{eqnarray}

Note that the allocation variables allows to reconcile the notations used in the hierarchical model of Eqs. \ref{prior11}-\ref{prior22} and the random measure representation in Eqs. \ref{rm1}-\ref{rm3} as follows: 
\begin{equation}
\tilde{\boldsymbol{\theta}}_{ik}^{*}=\boldsymbol{\theta}_{D_{ik}k}^{*}\label{ClustPar}
\end{equation}

A Gibbs sampler is used to generate random samples from the joint posterior and to approximate the Bayesian estimator. The Gibbs sampler iterates the following steps
\begin{enumerate}
\item Sample slice and stick-breaking variables $U$ and $V$ given $\Xi,\Theta,P,D,\Theta^{*},Y$
\item Sample the transition probabilities $P$ given $\Xi,\Theta,U,D,V,\Theta^{*},Y$
\item Sample the atoms $\Theta^{*}$ given $\Xi,\Theta,P,U,D,V,Y$
\item Sample the MS-GARCH parameters $\Theta$ given $\Xi,P,U,D,V,\Theta^{*},Y$
\item Sample the switching allocation variables $\Xi$ given $\Theta,P,U,D,V,\Theta^{*},Y$
\item Sample the mixture allocation variables $D$ given $\Xi,\Theta,P,U,V,\Theta^{*},Y$
\end{enumerate}
The derivation of the full conditional distributions is given in Appendix \ref{app}.

%The identification issue relates to the invariant property of the posterior distribution of the Markov-switching parameters when the labels of the parameters are permuted. As a consequence, an identical set of marginal posterior distributions is obtained for each switching component of the parameters. In this paper, we follow a commonly used approach by imposing a restriction on the intercept of each of the measurement equations i.e.  $\mu_{i1}>\mu_{i2} >\ldots >\mu_{iK}$  \citep[e.g., see][]{Celeux1998,Fruhwirth2001,Fruhwirth2006}.

\section{Numerical illustrations}\label{EmpApp}
\subsection{Simulation results}\label{simMain}
For inference and model validation, we run a set of simulation experiments on synthetic datasets. In this section we report the results for one of the experiments, in which we examine the efficiency and effectiveness of our MCMC sampling scheme in estimating the number of clusters in each regime. 

We generate a panel of 30 time series with length 300 each from the data generating process (DGP) corresponding to the model defined by Eqs. \ref{MeaEqn}-\ref{prior12} for two regimes ($K = 2$), including their time-invariant transition probabilities and switching conditional mean and variance. The DGP is assumed to be as realistic as possible for illustrative purposes. In particular, the number of groups in the clusters across the two regimes is being kept relatively small. 

In the first regime (i.e., $s_{it}=1$), we assume that the units are clustered into two groups with equal probability. In formulas:
$$
\begin{aligned}
\mu_{i1} &= % 1 + 0.001\eta_{i1}, &\eta_{i1}\sim\mathcal{N}(0,1)\\
\left\{
\begin{array}{ll}
1 + 0.01\eta_{i1}, &\hbox{with probability}\, p_1=0.5,\\
1.5+0.01\eta_{i1},& \hbox{with probability}\, (1- p_1)\\
\end{array}
\right.\\
\gamma_{i1} & = 
\left\{
\begin{array}{ll}
0.1+0.01\zeta_{i1}^2, & \hbox{with probability}\, p_1=0.5,\\
0.2+0.01\zeta_{i2}^2,& \hbox{with probability}\,   (1-p_1),\\
\end{array}
\right.\\
(\alpha_{i1},\beta_{i1},x) &\sim \mathcal{D}ir(1000(0.05, 0.8, 1-0.85))
\end{aligned}
$$
In the second regime (i.e, $s_{it}=2$), the units are clustered into three groups. In formulas:
$$
\begin{aligned}
    \mu_{i2}&=
\left\{
\begin{array}{ll}
-1.1+0.01\eta_{i2},& \hbox{with probability}\, p_1=0.3,\\
-1.5+0.01\eta_{i2},& \hbox{with probability}\, p_2=0.3,\\
-1.0+0.01\eta_{i2},& \hbox{with probability}\, (1-p_1-p_2),\\
\end{array}
\right.\\
    \gamma_{i2}&=
\left\{
\begin{array}{ll}
0.5+0.01\zeta_{i2}^2,& \hbox{with probability}\, p_1=0.3,\\
0.8+0.01\zeta_{i2}^2,& \hbox{with probability}\, p_2=0.3,\\
0.1+0.01\zeta_{i2}^2,& \hbox{with probability}\, (1-p_1-p_2),\\
\end{array}
\right.\\
(\alpha_{i2},\beta_{i2},x) &\sim \mathcal{D}ir(1000(0.05, 0.8, 1-0.85))
\end{aligned}
$$
where $\eta_{i1}\sim\mathcal{N}(0,1)$, $\eta_{i2}\sim\mathcal{N}(0,1)$, $\zeta_{i1}\sim\mathcal{N}(0,1)$ and $\zeta_{i2}\sim\mathcal{N}(0,1)$.

The transition probabilities are $p_{i,11}\sim\mathcal{B}e(1000 p,1000(1-p))$ and $p_{i,22}\sim\mathcal{B}e(1000 p,1000(1-p))$ i.i.d. for $i=1,\ldots,N$, where $p=0.98$. 

\begin{figure}[h!]
\begin{center}
\setlength{\tabcolsep}{5pt}
\renewcommand{\arraystretch}{4.1}
\begin{tabular}{cc}
$\mu_{i}$ & $\gamma_{i}$\\
\includegraphics[scale=0.4]{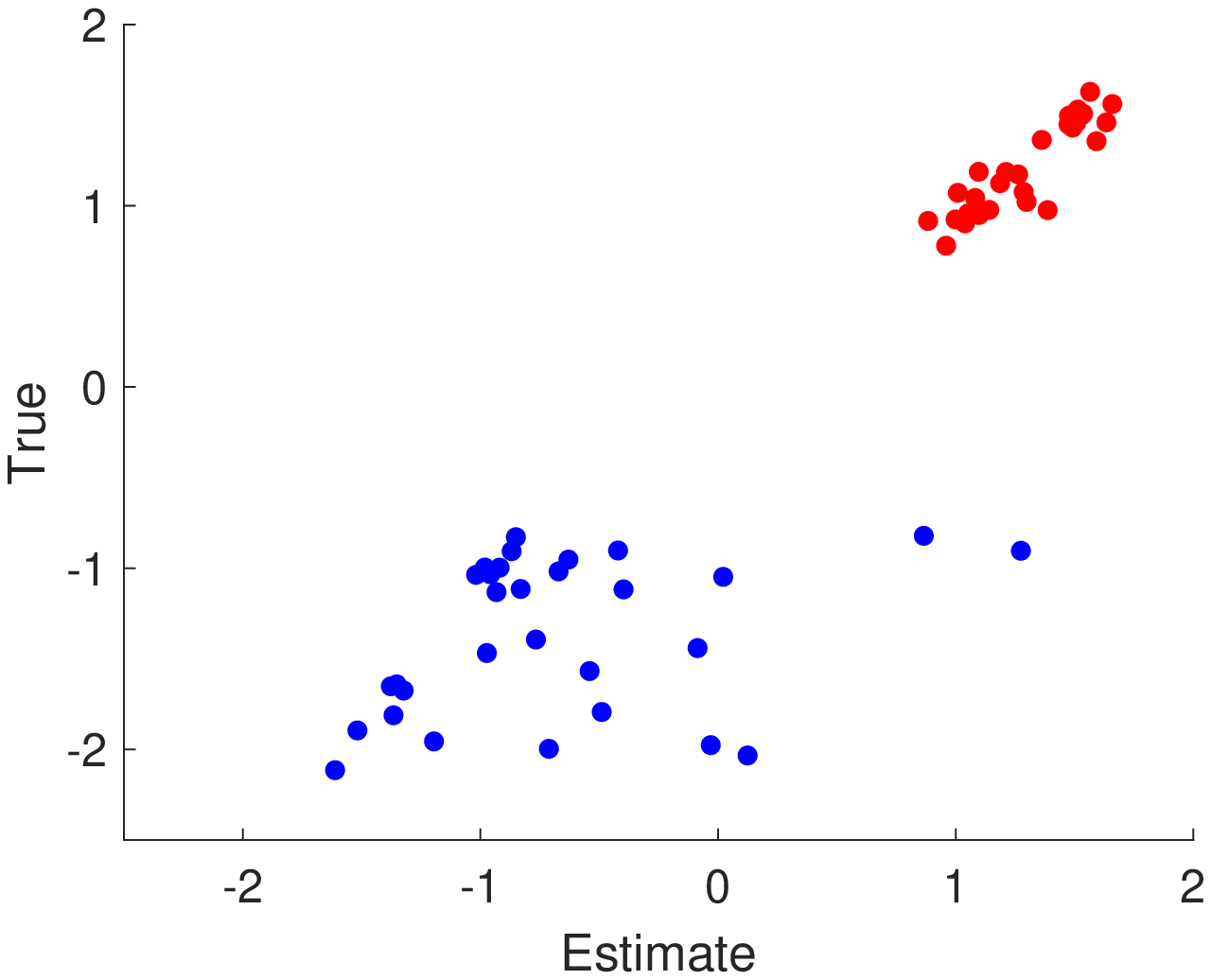}&
\includegraphics[scale=0.4]{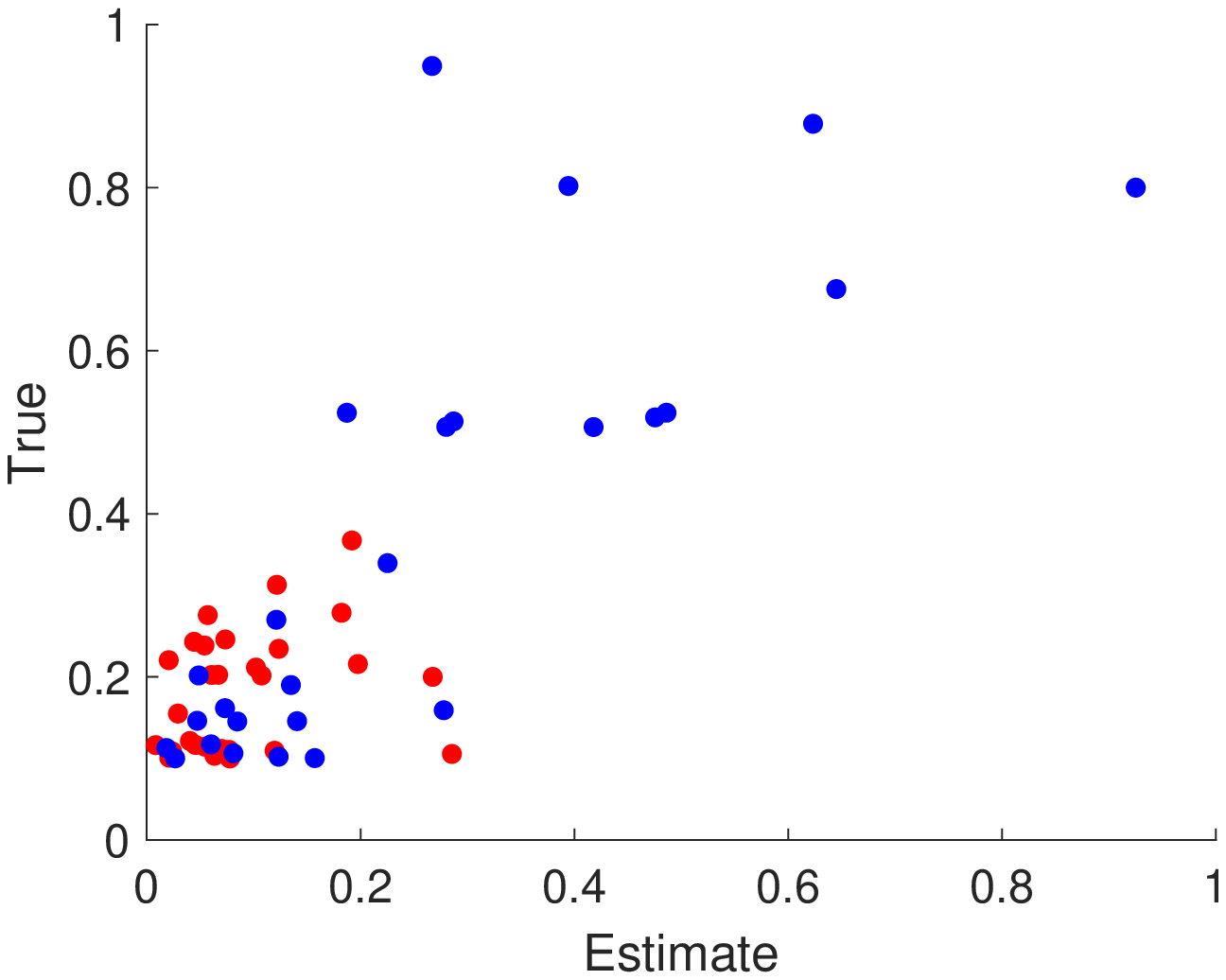}
\end{tabular}
\end{center}
\caption{Simulation results. True (vertical axis) and estimated (horizontal axis) values of the intercept ($\mu_{i}$) of the measurement equation and GARCH parameter $\gamma_{i}$ for each unit i  in regime 1 (red) and 2 (blue).} \label{Nclusparameters}
\end{figure}

\begin{figure}[h!]
\begin{center}
\setlength{\tabcolsep}{5pt}
\begin{tabular}{cc}
%\hspace{-20pt}\includegraphics[scale=0.45]{Figures/FigSimulatedData/Nclust1.eps}&
%\hspace{-20pt}\includegraphics[scale=0.45]{Figures/FigSimulatedData/Nclust2.eps}\\
%\includegraphics[scale=0.5]{Figures/FigSimulatedData/AllocationRegime1.eps}&
%\includegraphics[scale=0.5]{Figures/FigSimulatedData/AllocationRegime2.eps}
\hspace{-20pt}\includegraphics[scale=0.4]{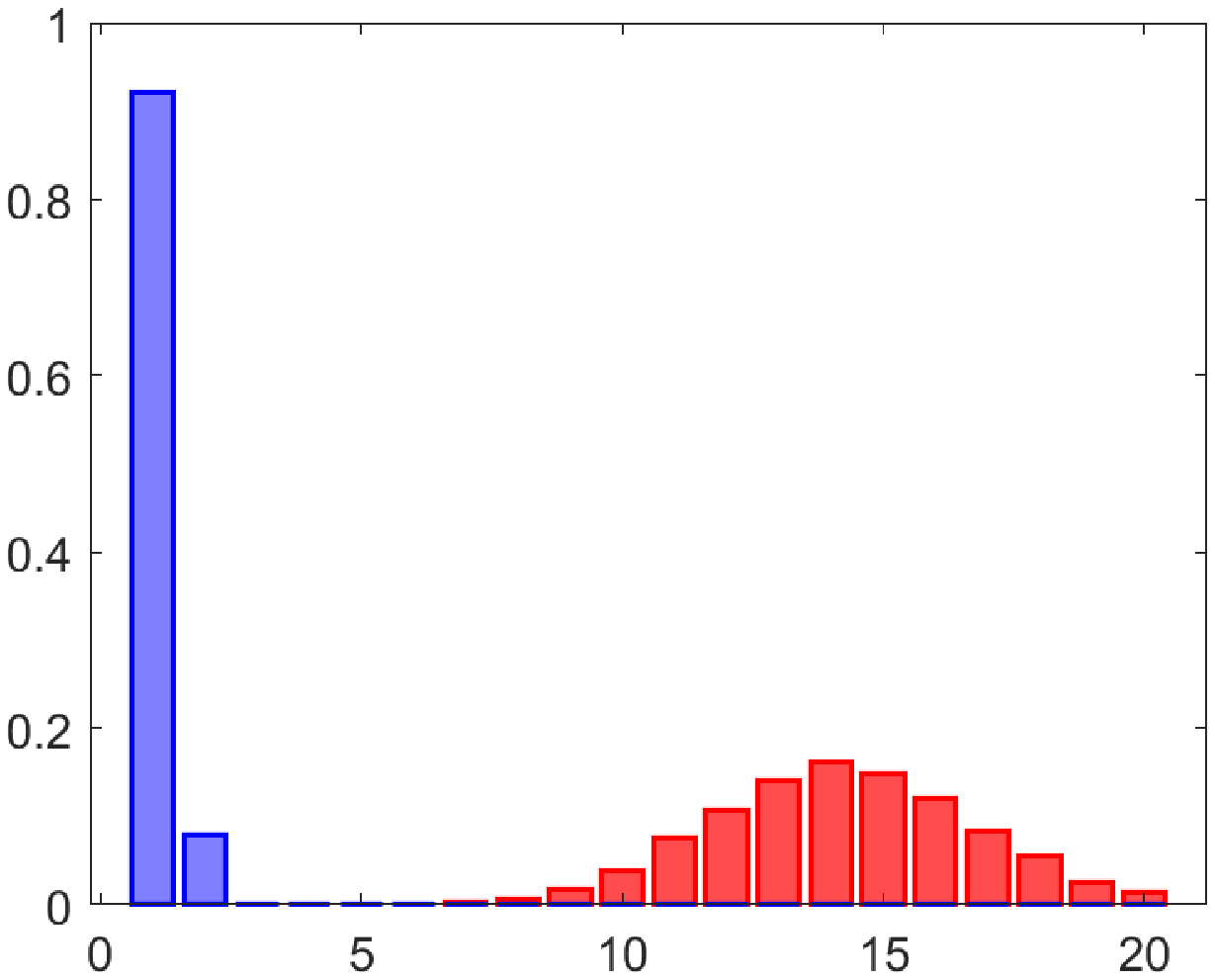}&
\hspace{-20pt}\includegraphics[scale=0.4]{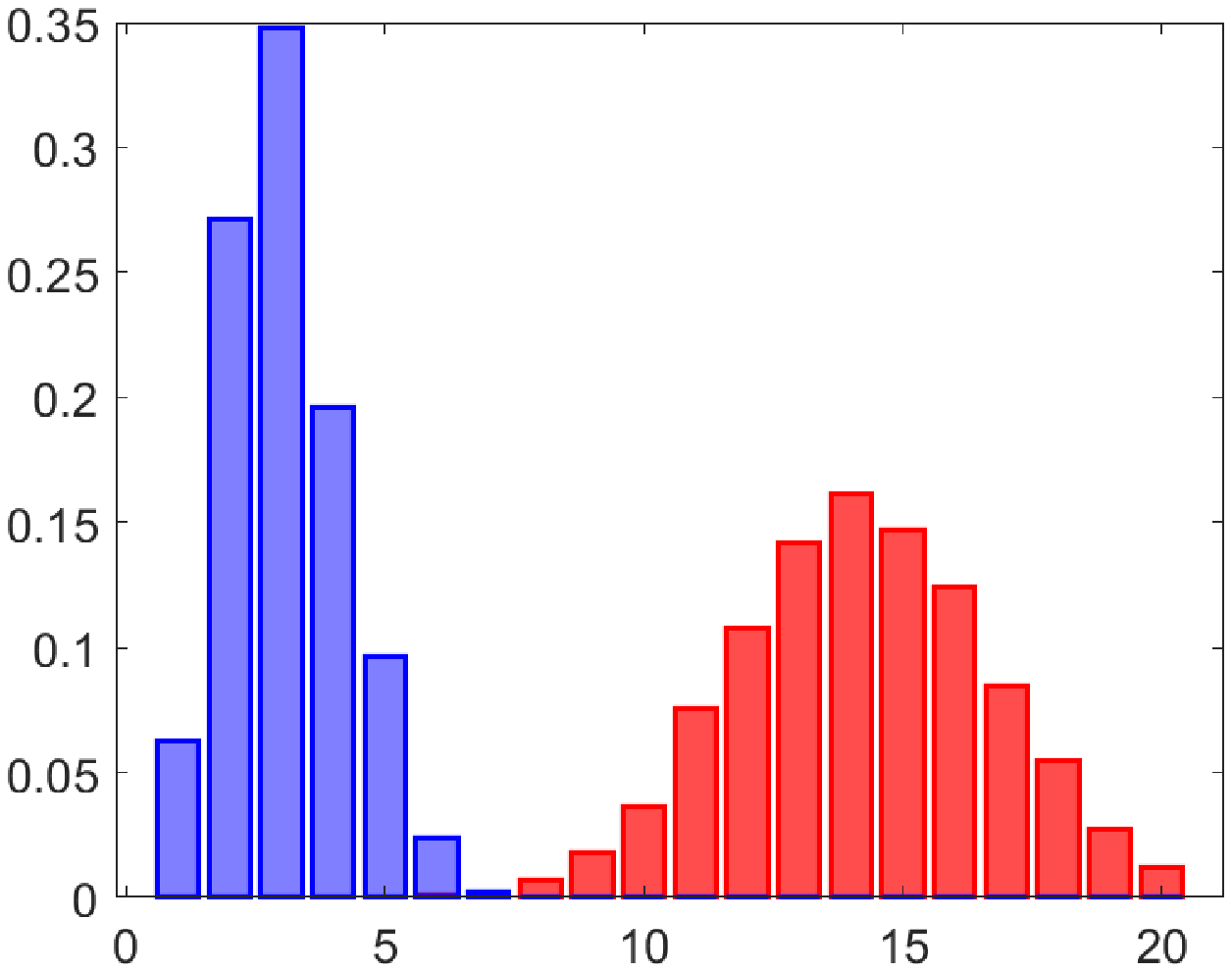}
\\
\end{tabular}
\end{center}
\caption{Simulation results. Prior (red) and posterior (blue) distribution of the number of clusters in regime 1 (left) and 2 (right). 
%Bottom: posterior co-clustering matrix in regime 1 (left) and 2 (right).
}\label{NclustCoclust}
\end{figure}

In Fig. \ref{Nclusparameters}, the true and the estimated values of the intercept ($\mu_{i}$) of the measurement equation and GARCH parameter $\gamma_{i}$ (see Eqs. \ref{MeaEqn} and \ref{GARCHEqn}) in regime 1 (red) and 2 (blue) are illustrated.{\footnote{
For other parameters and the trajectory of the Markov chain see Appendix \ref{SimuPlor}.}}
The findings seems to reveal that the inference is able to recover the true value of the parameters.

Figure \ref{NclustCoclust} shows that data are informative about the number of clusters in each regimes, and there is a substantial revision of the prior distributions (red) and the posterior distributions (blue) concentrate about the true number of clusters in the two regimes. For our simulated dataset, the Maximum a Posteriori (MAP) estimation of the number of cluster is 1 for the first regime and 3 for the second regime.  

\subsection{Volatility clusters in the S\&P 100}
We consider 78 assets of the 101 constituents of the SP\&100 index and collect the percentage log-returns at a weekly frequency. We do this to have a balanced panel of observations from $6^{th}$ January to $3^{rd}$ October 2020 (the sectorial classification of these assets is reported in Tab. \ref{tab} of Appendix \ref{emp}).
\begin{figure}[t]
\begin{center}
\setlength{\tabcolsep}{10pt}
\begin{tabular}{cc}
\includegraphics[scale=0.4]{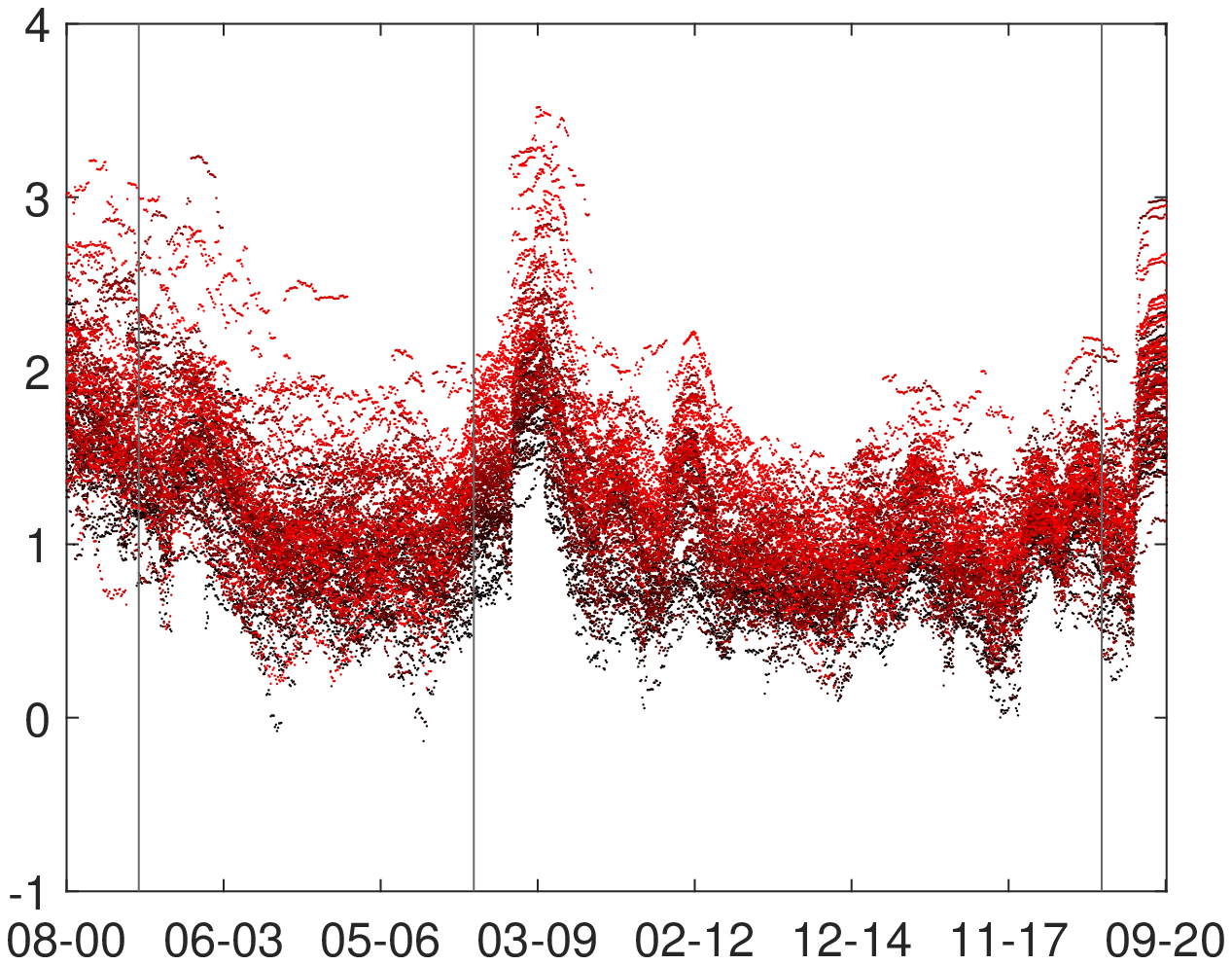}&
\includegraphics[scale=0.4]{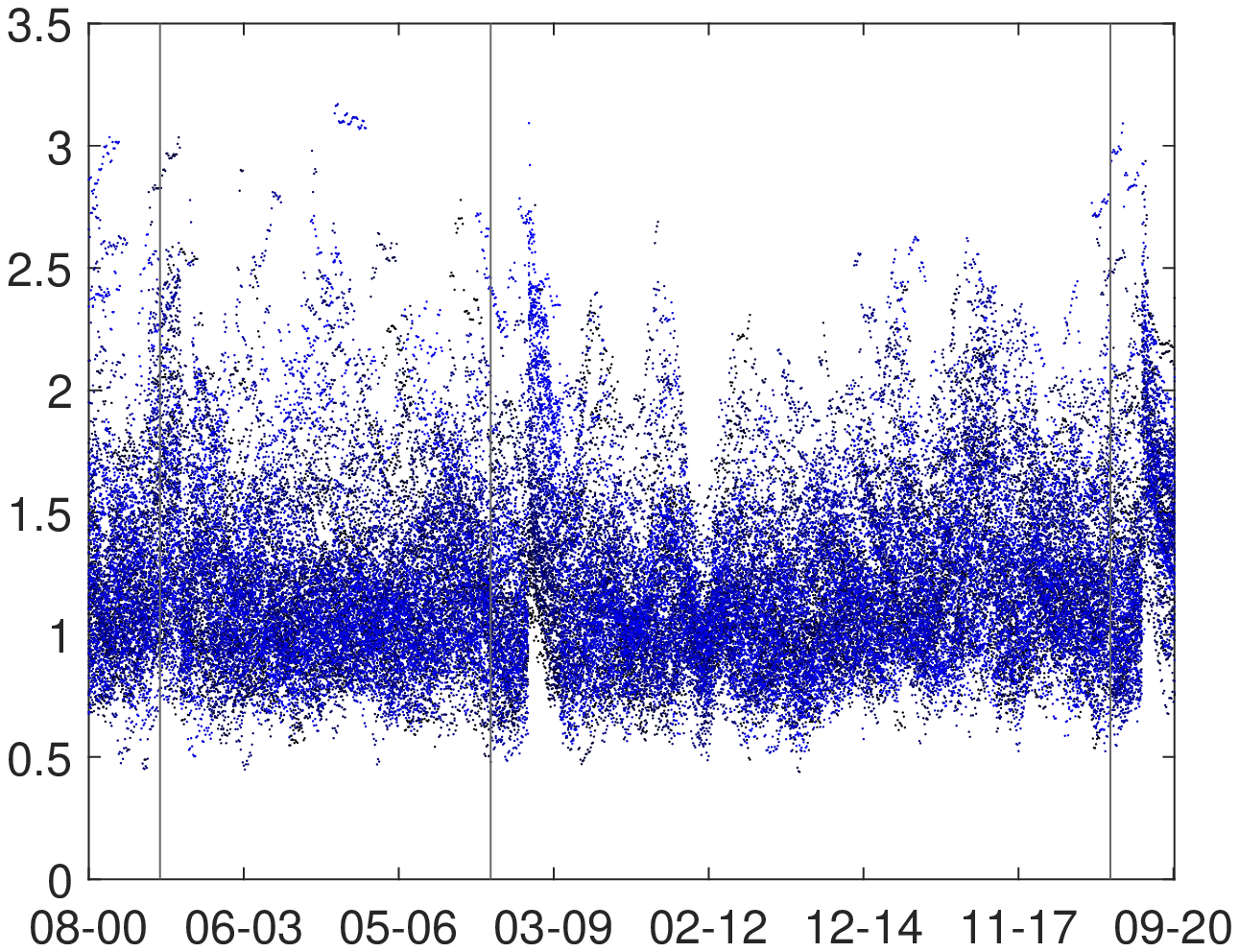}\\
&\\
\includegraphics[scale=0.4]{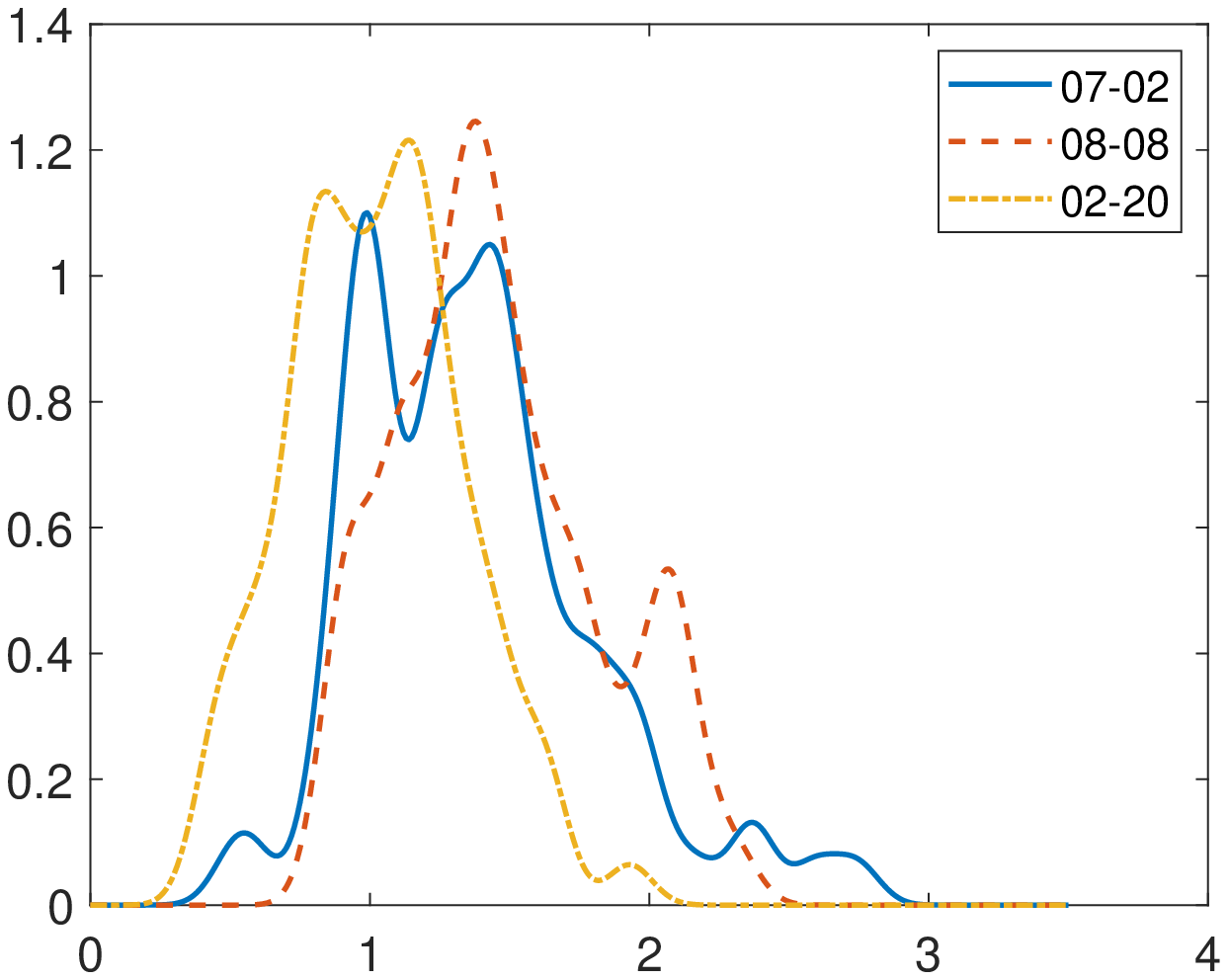}&
\includegraphics[scale=0.4]{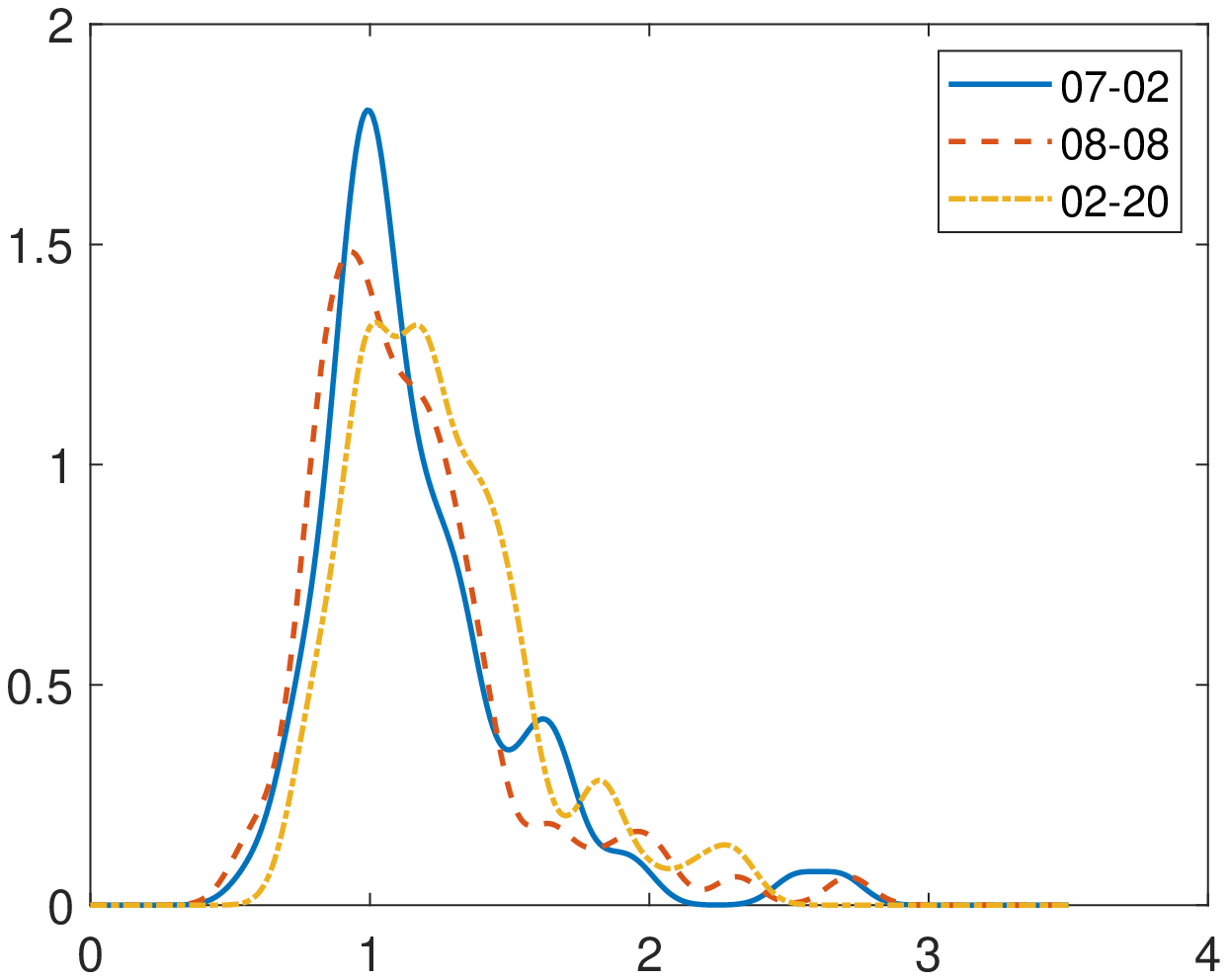}
\end{tabular}
\end{center}
\caption{Top: rolling window estimates of the log-volatility (left) and log-kurtosis (right) for the SP\&100's constituents for the period $6^{th}$ January 2000 to $3^{rd}$ October 2020 (30 weeks window). Vertical bars indicate three reference dates: $6^{th}$ July 2002, $23^{rd}$ August 2008 and $22^{nd}$ February 2020. Bottom: cross-sectional distribution of the log-volatility (left) and log-kurtosis (right) in three reference dates.}\label{Styleized}
\end{figure}

The empirical analysis aims to identify regimes of under-performance and over-performance of expected returns. Moreover, we use the sector classification and some fundamental financial ratios to study the composition of the clusters.  

As a preliminary analysis, we plot in Fig. \ref{Styleized} the estimates of the log-volatility and log-kurtosis of the 78 constituents considered in the analysis. This figure also shows the cross-sectional distribution of the log-volatility and log-kurtosis. The figure indicates that volatility and kurtosis change over time with time series clustering effects (see top plots of Fig. \ref{Styleized}). This seems to suggests the use of GARCH and Markov-switching models. Furthermore, the cross-sectional distribution of the volatility and kurtosis exhibits multiple modes and long tails (see bottom plots of Fig. \ref{Styleized}).\footnote{ In Fig. \ref{rob} of Appendix \ref{emp}, we also report the estimates of the cross-sectional distribution of the log-volatility (left) and log-kurtosis (right) of the SP\&100's constituents log-returns in the three dates ($6^{th}$ July 2002, $23^{rd}$ August 2008 and $22^{nd}$ February 2020) for three different sizes of the rolling window. The results show that the preliminary evidence on cross-sectional heterogeneity is robust with respect to the choice of the window size.} This fact seems to imply cross-section heterogeneity in the data with possible clustering effects in the parameters of the GARCH process. These effects cannot be captured only by a Markov-switching (MS) model, therefore there is a need of combining the MS-GARCH with a probabilistic clustering mechanism.

\begin{figure}[t]
\begin{center}
\setlength{\tabcolsep}{5pt}
\begin{tabular}{cc}
\hspace{-20pt}\includegraphics[scale=0.4]{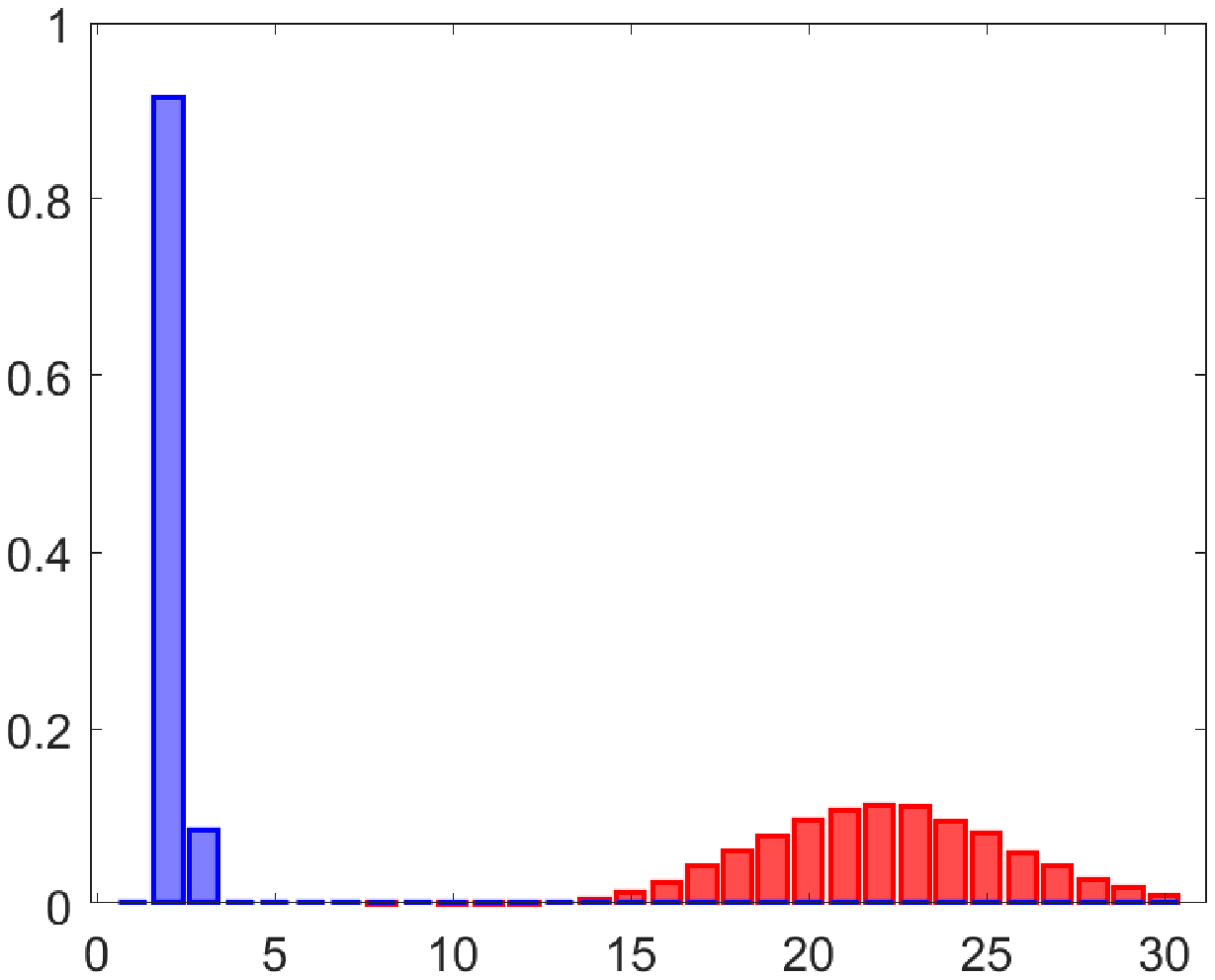}&
\hspace{-20pt}\includegraphics[scale=0.4]{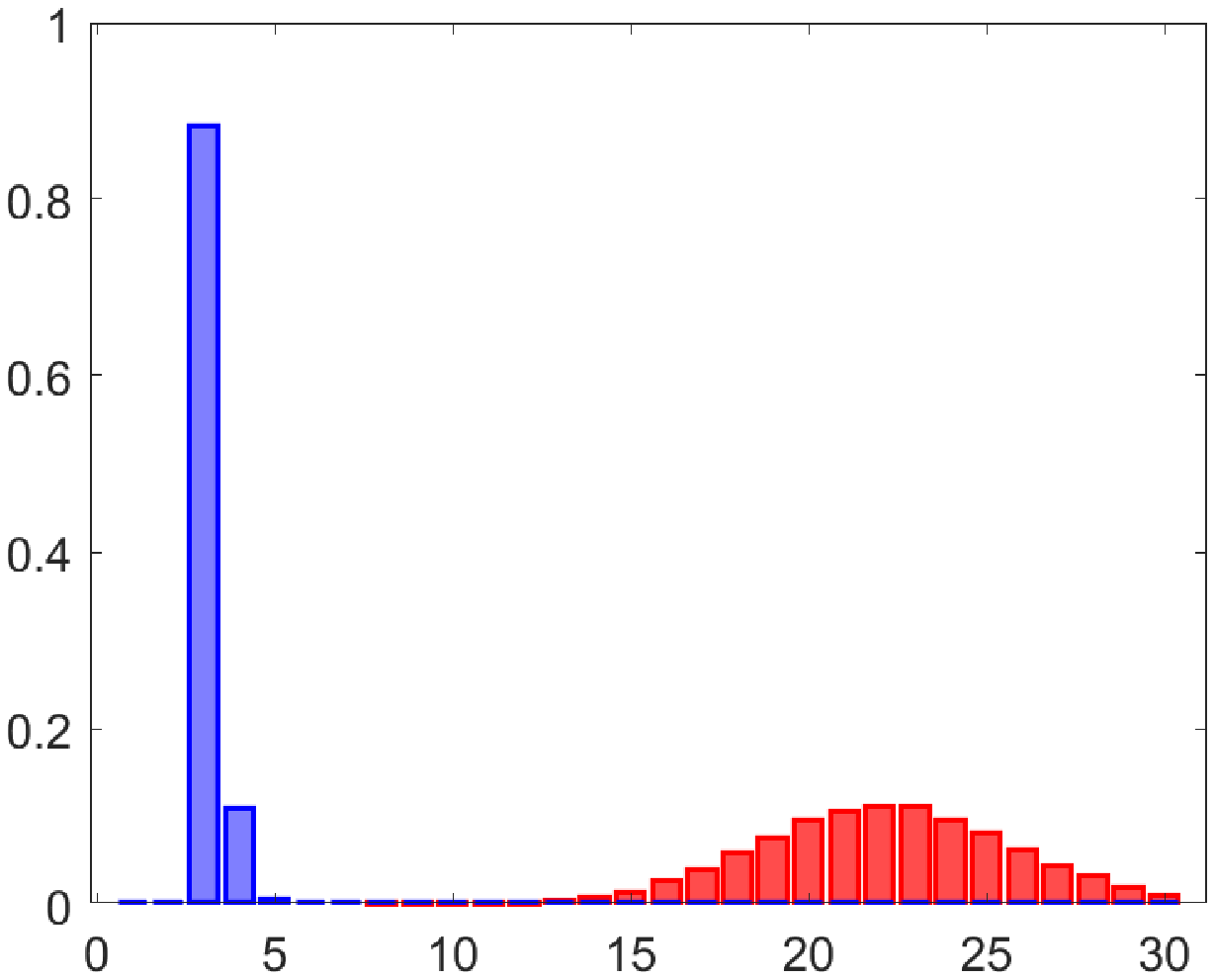}
\end{tabular}
\end{center}
\caption{Prior (red) and posterior (blue) distribution of the number of clusters in the over-performing regime 1 (left) and under-performing regime 2 (right). } \label{NclustCoclust2}
\end{figure}

\begin{figure}[t]
\begin{center}
\setlength{\tabcolsep}{10pt}
\begin{tabular}{cc}
\includegraphics[scale=0.4]{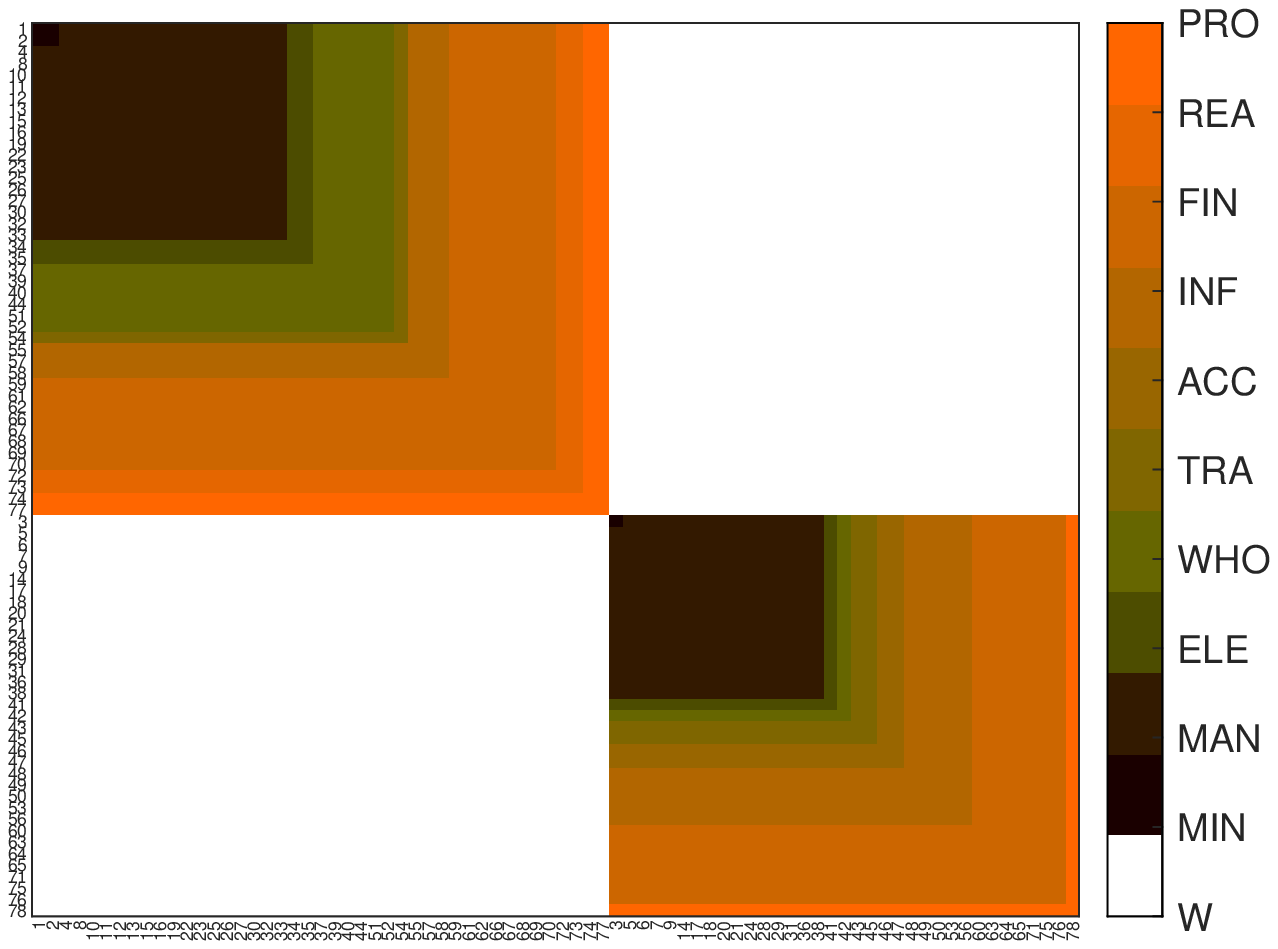}&
\includegraphics[scale=0.4]{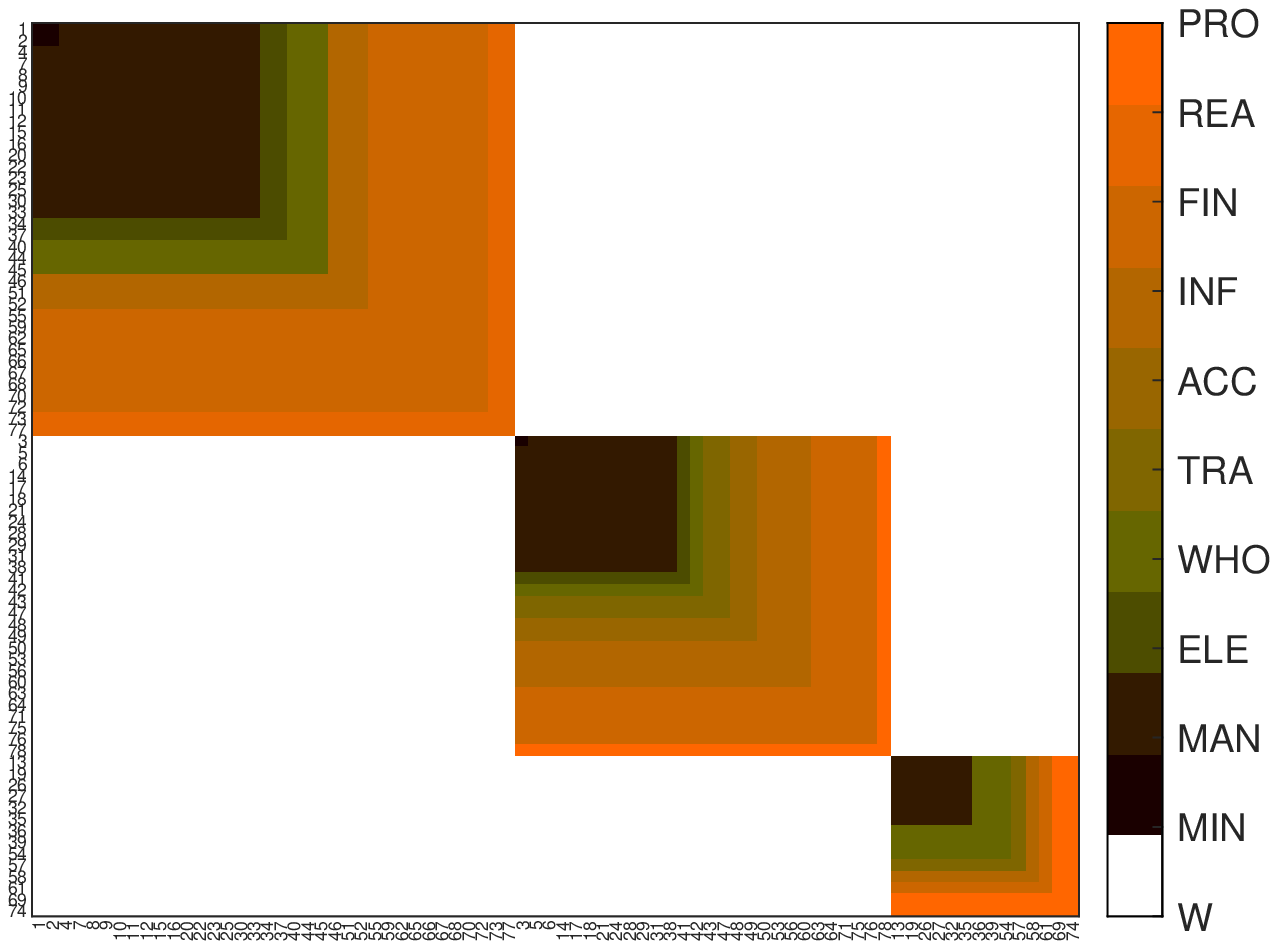}\vspace{10pt}\\
\multicolumn{2}{c}{\includegraphics[scale=0.4]{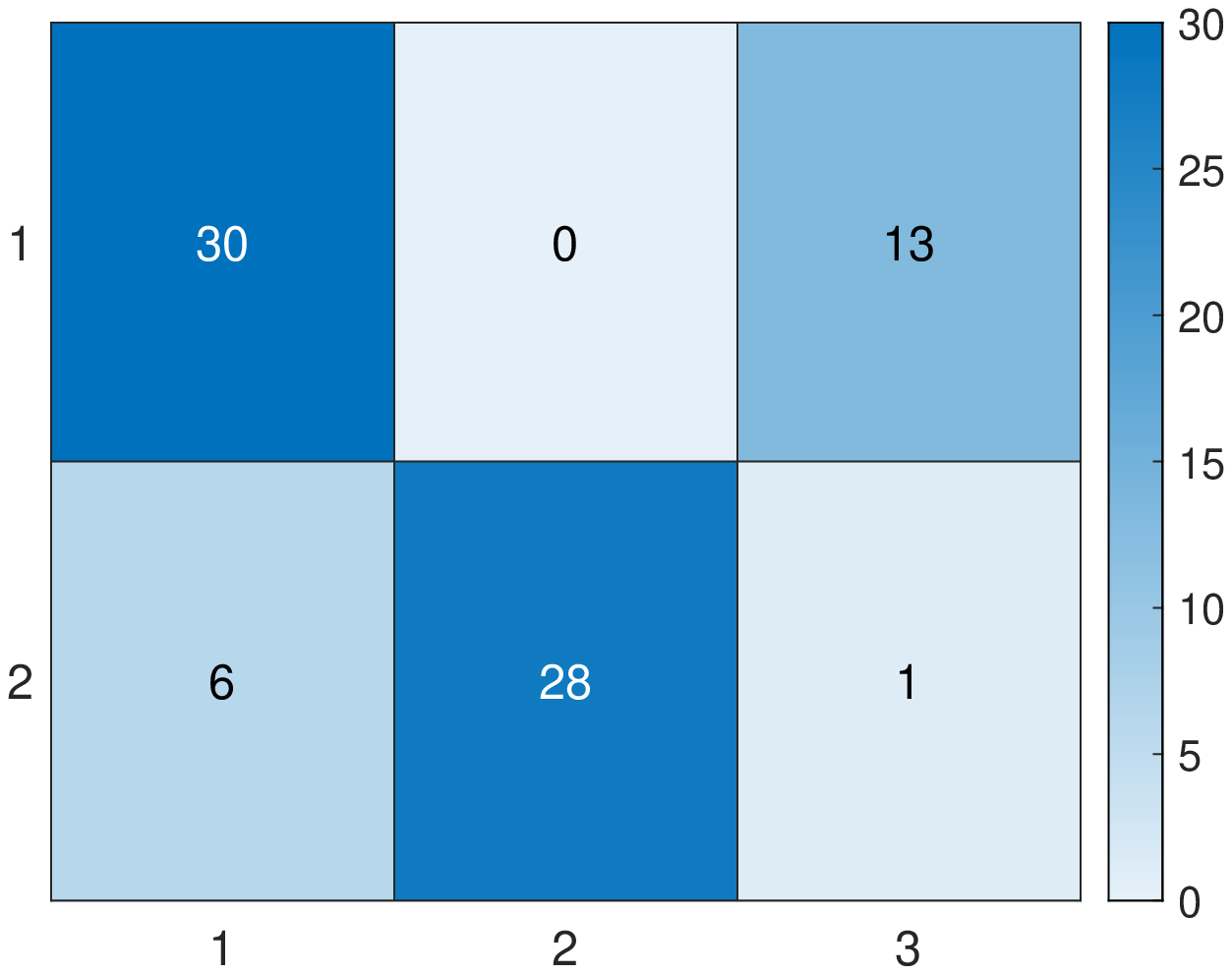}}
\end{tabular}
\end{center}
\caption{Top: posterior co-clustering matrix in regime 1 (left) and 2 (right). In each block, colors represent the sector labels of the units. Bottom: the number of assets (cell entries) shared by two clustering structures in regime 1 (vertical axis) and regime 2 (horizontal axis).}\label{NclustCoclustCompare}
\end{figure}

In our analysis, we first identify the two regimes, and then the clusters of assets within each regime. Lastly, we use the sector classification and some fundamental financial ratios to study the composition of the clusters. 

Regime identification is achieved by ordering the expected returns $\mu_{i1}>\mu_{i2}$, such that regime 1 corresponds to a relative over-performance state and regime 2 to an under-performance one. This identification constraint is strongly supported by the data and allows us to separate the assets returns in two performance regimes (see Fig. \ref{ParamBNPFina} in Appendix \ref{emp}). 

Regarding the cluster identification, Fig. \ref{NclustCoclust2} reports the prior (red) and posterior (blue) distribution of the number of clusters in regime 1 (left) and 2 (right). We set $\nu=0$ and $\phi=10$ in the PYP prior in order to have quite diffuse prior distributions. The posterior distribution is concentrated suggesting a substantial revision of the prior information and the MAP estimates of the number of clusters is 2 and 3 for regime 1 and 2, respectively.

\begin{figure}[h!]
\begin{center}
\setlength{\tabcolsep}{10pt}
\renewcommand{\arraystretch}{1.2}
\begin{tabular}{cc}
\includegraphics[scale=0.4]{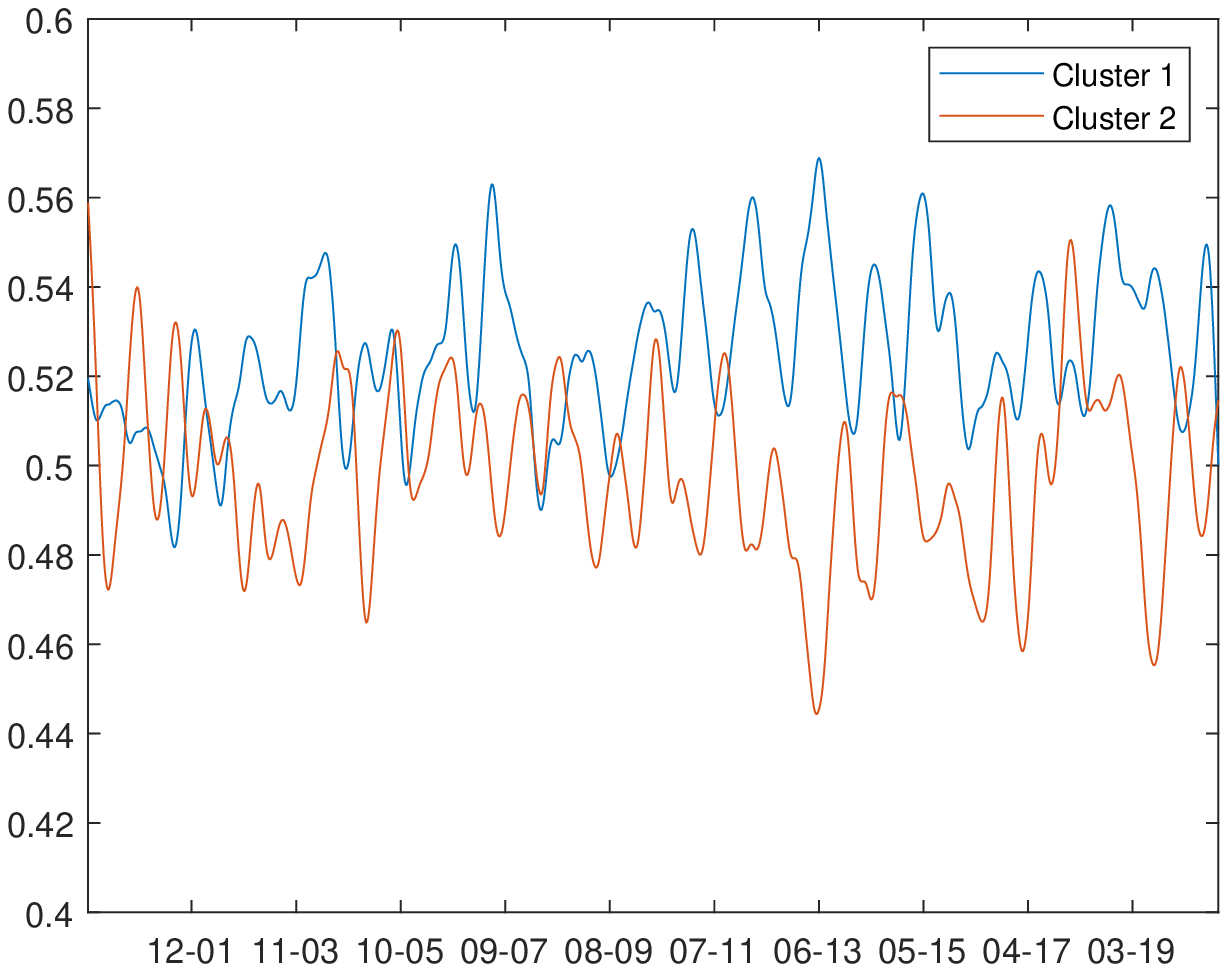}&
\includegraphics[scale=0.4]{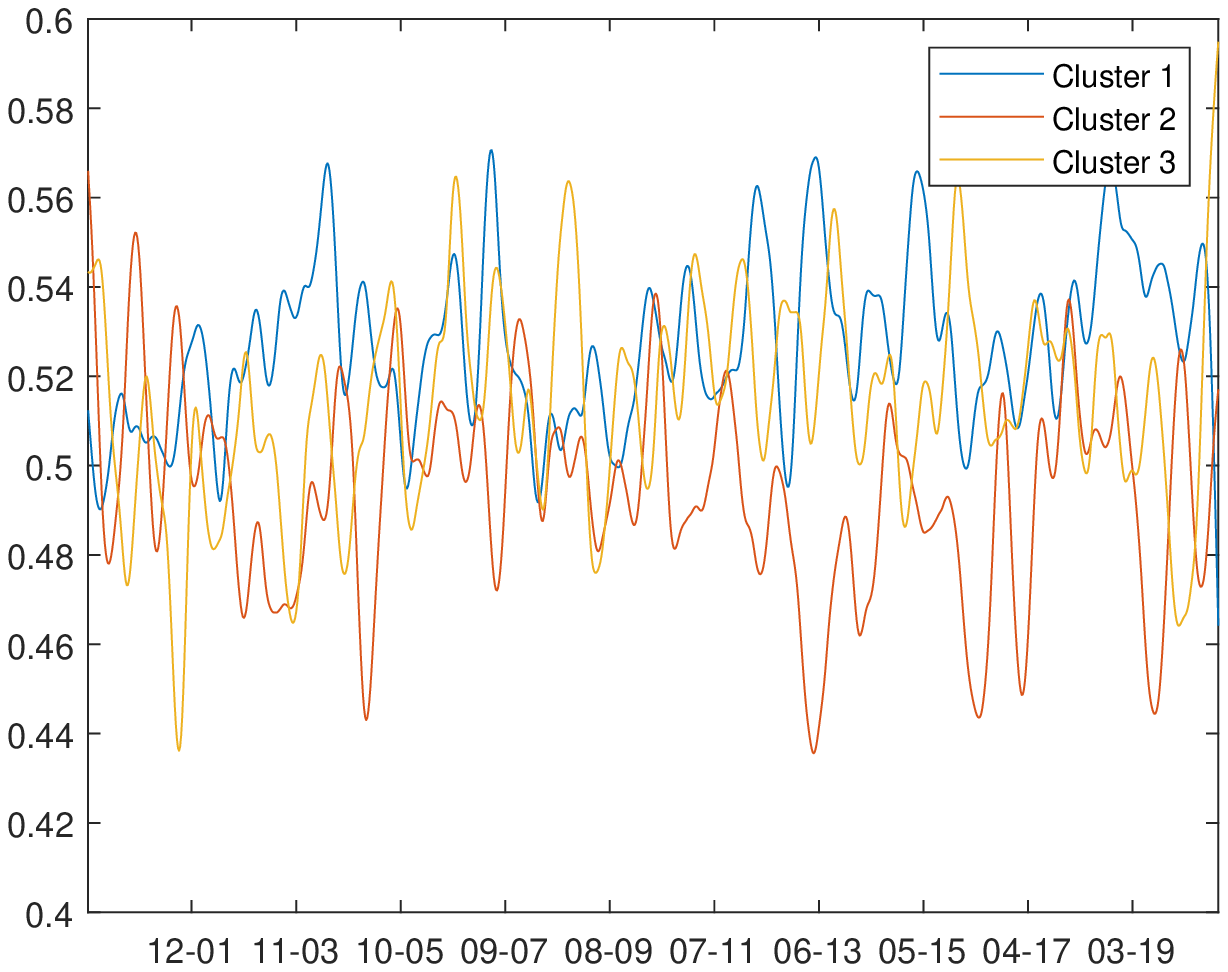}\\
\includegraphics[scale=0.4]{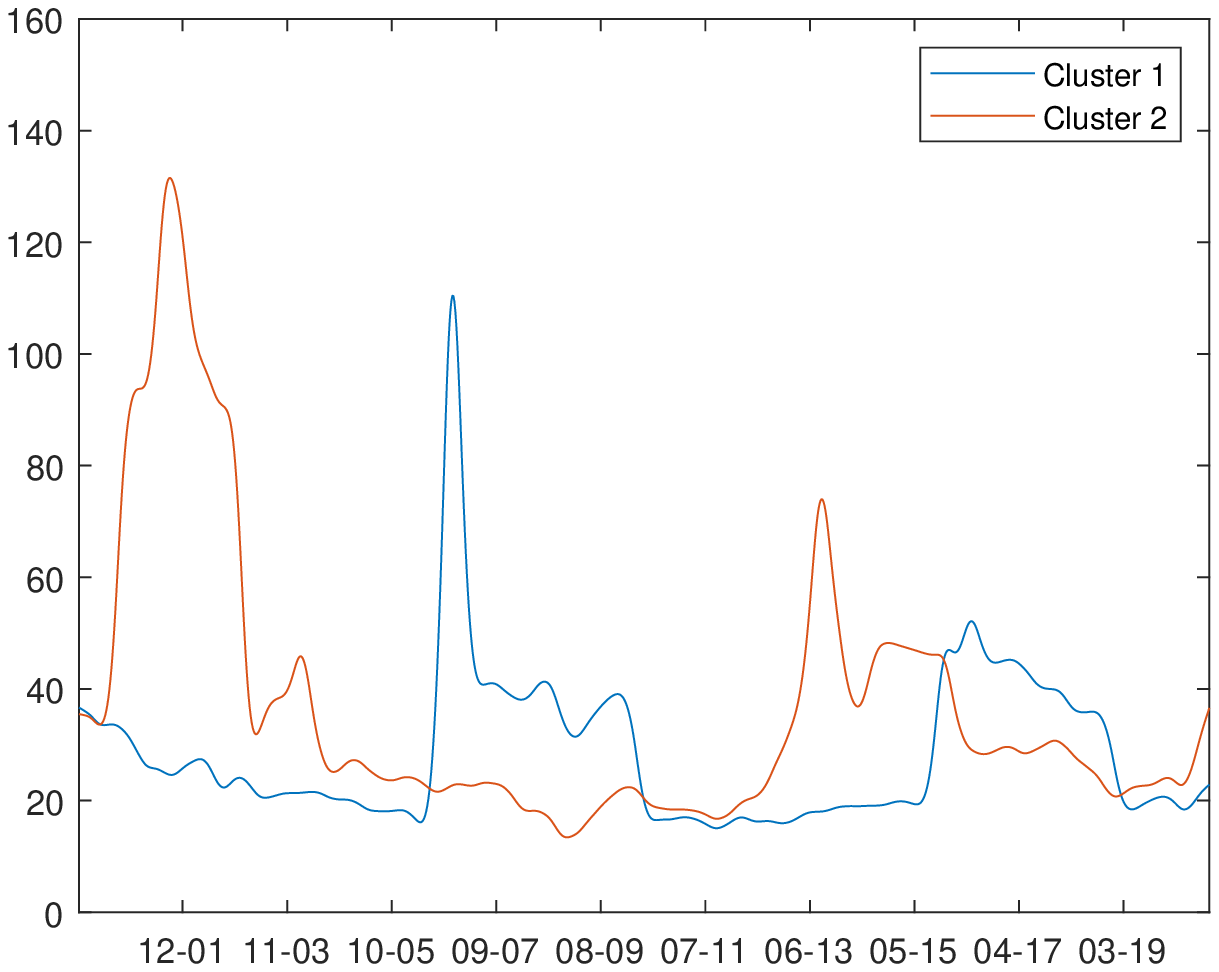}&
\includegraphics[scale=0.4]{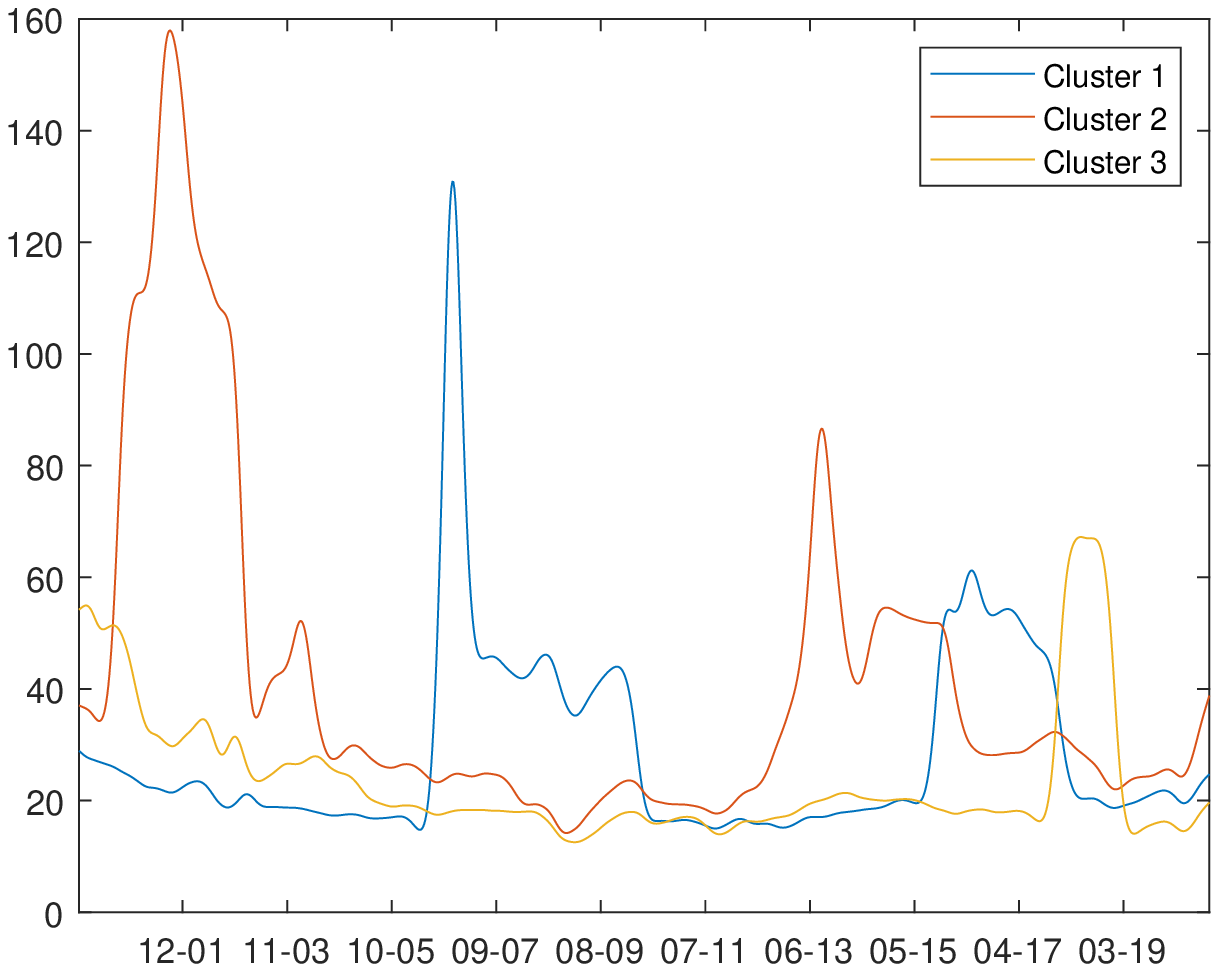}\\
\includegraphics[scale=0.4]{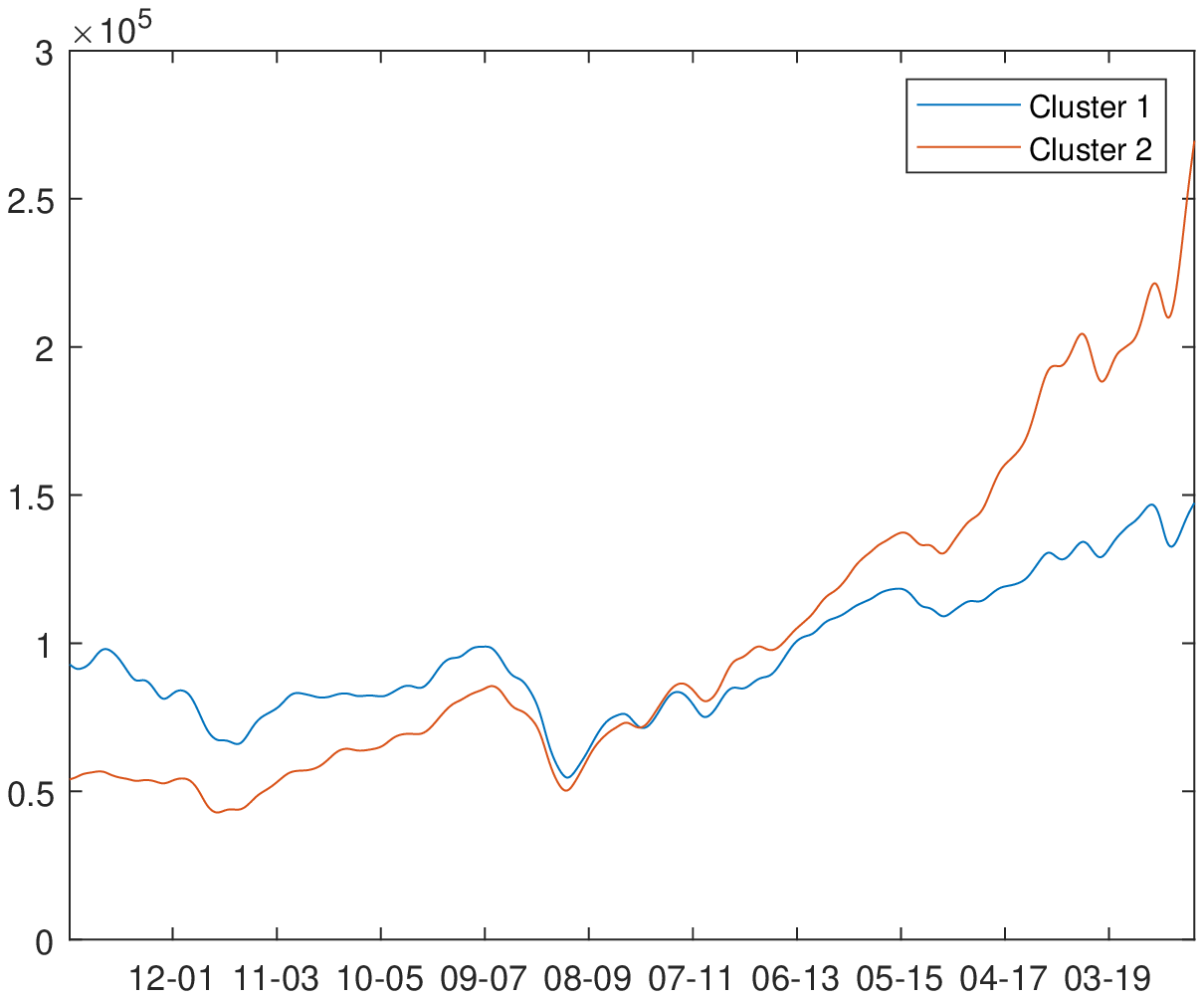}&
\includegraphics[scale=0.4]{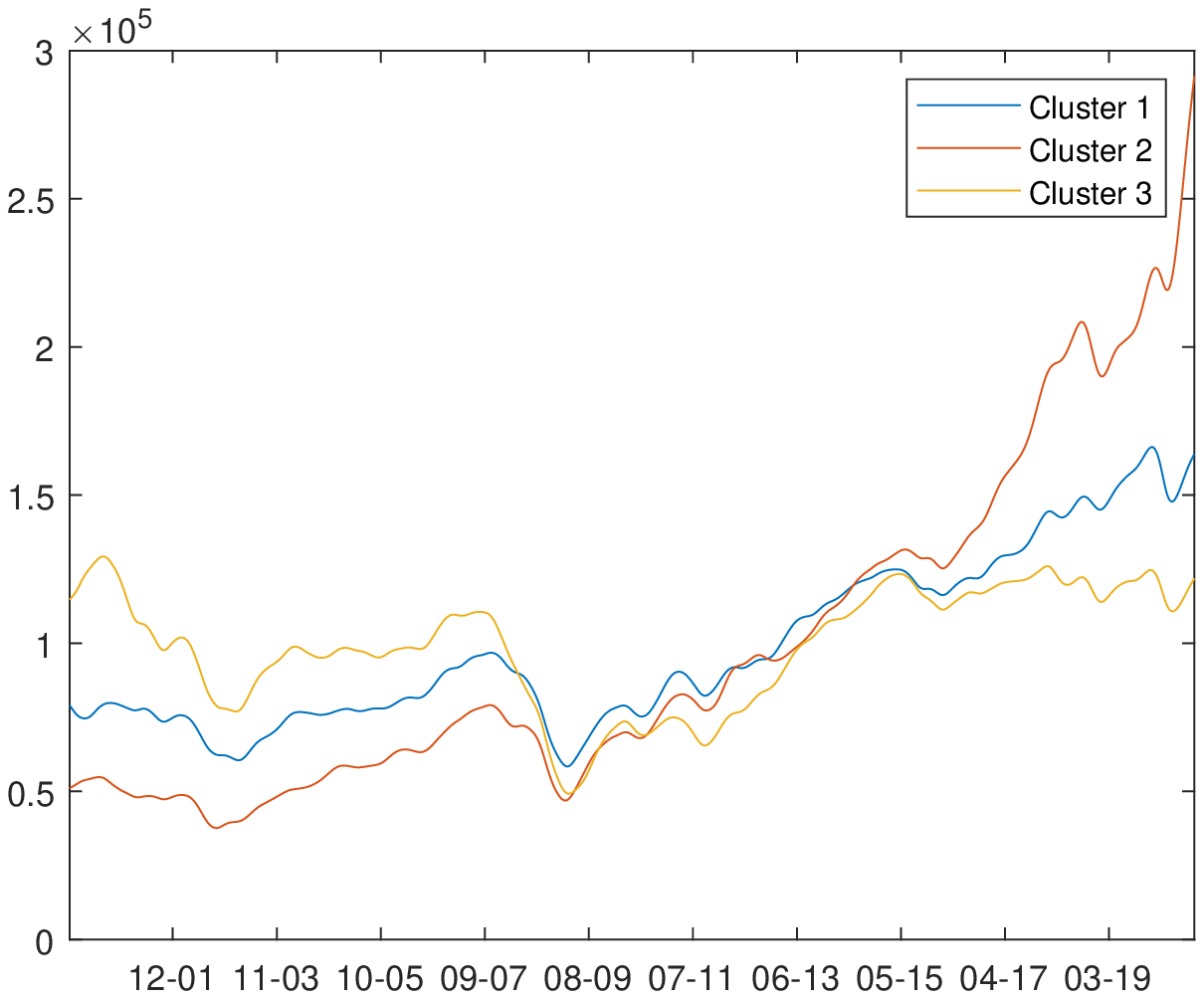}\\
\end{tabular}
\end{center}
\caption{Average overperforming probability of the assets in the clusters (first row), Price-to-Earning (second row), and Market Capitalization (third row) for the assets in the clusters of regime 1 (left) and of regime 2 (right).}\label{NclustCoclust1}
\end{figure}
We also estimate the co-clustering probability matrix to delve into the composition of the clusters. In this way, the  probability $\mathbb{P}(\{D_{ik}=D_{jk}\}|M_k,Y)$ that the parameters $\boldsymbol{\theta}_{ik}^{*}$ and $\boldsymbol{\theta}_{jk}^{*}$ are in the same cluster is given. This probability can be easily approximate by using the MCMC samples as follow
\begin{equation}
\frac{1}{\hbox{Card}(\mathcal{R}_k)}\sum_{r\in \mathcal{R}_k}\delta(D_{jk}^{(r)}-D_{ik}^{(r)})
\end{equation}
where $D^{(r)}_{ik}$ is a sample of the allocation variable for the $i$-unit parameters in the regimes $k$ and $\mathcal{R}_k=\{r=1,\ldots,R|N_{k}^{(r)}=M\}$ contains the values of MCMC iterations such that the parameters of the panel units have been allocated to exactly $M$ mixture components. Note that a spectral clustering algorithm have been applied to re-order the series and to provide better graphical representation of the clusters.

The top panel of Fig. \ref{NclustCoclustCompare} report the co-clustering matrices for the two regimes. In each block matrix, the algorithm identifies a constituent (asset) belonging to a cluster with the label 1 (color patch) and 0 (white patch) otherwise. The colors represent the sectors in the clusters. Following \cite{wade2018}, we use the variation of information (VI) metric proposed by \cite{MEILA2007873} to compare the two regimes (in terms of clusters). This measure compares the information in the two regimes with the information shared between the two regimes. We compute the normalized value of VI (0.20), which suggests a substantial difference between the clustering and composition in the two regimes.{\footnote{VI lies in the interval  $0 - \log_2(N)$ and a normalize value is obtained dividing VI by $\log_2(N)$.}} 

The bottom panel of Fig. \ref{NclustCoclustCompare} shows the relationship between the clustering structures of the two regimes. We order the clusters following the numbers of constituents from the largest to the smallest. Most of the assets in cluster 1 in the first regime belong to cluster 1 in the second regime, whereas many assets of the first group in regime 1 belong to the third group in regime 2. In particular, for cluster 1 in both regimes, we observe that the majority of the sectors representing the assets are: manufacturing (about 40\% in both regimes), financial and insurance (19\% in the first regime), and  wholesale and retail (25\% in the second regime). Similar results for the sectors are found for cluster 2 in the two regimes. More specifically, the manufacturing sector represents about 40\% of the assets in the two regimes, while the financial and insurance sector is about 18\% for regime 2, and information and communication is around 20\% (for details on sectors see Tabs. \ref{tab:addlabel} and \ref{tab:addlabe2} in Appendix \ref{emp}).

Further, in order to characterize the clusters in terms of the market size of the constituents, we first rank the companies by computing the average size of each of them using the last year of the sample period. Then, we classify the assets into three groups, namely small (bottom 30\%), medium (middle 40\%) and big (top 30\%) companies (see Tab. \ref{tab:addlabe3}). We also compute the percentage of companies belonging to the clusters in each regime in terms of size. In regime 1, companies with the medium size represent the largest majority about 40\%, and a similar outcome is also observed for regime 2. More specifically, the following emerge.
%\footnote{The constituents of different sizes are reported in Tabs. \ref{tab:addlabel} and \ref{tab:addlabe2} in Appendix \ref{emp}.} 

%\textcolor{red}{GIVEN THAT WE HAVE REPORTED THE RESULT IN THE TABLE IN THE TEXT ....SHOULD WE KEEP THE TABLE OR NOT? Tab. \ref{tab:clust} reports the average of the market size, Price-to-Earnings ratio along with the returns and their standard deviations. Looking at Fig. \ref{NclustCoclustCompare} and Tab. \ref{tab} the following emerges. }

In regime 1, we have: 
\begin{enumerate}
\item Cluster 1 is characterized by 40\% and 20\% small and large size companies, respectively. This cluster also shows values on average of the market size, returns and their standard deviation equal to $1.15\times 10^5, 0.21$ and $4.72$, respectively. \footnote{We computed the average return and standard deviation using the whole sample, while the average market size is calculated using the last 10 years of the sample period.} 
\item Cluster 2 consists of 17\% small and 42\% medium size companies with an average market size of $1.50\times 10^5$, a return equal to $0.34$ and standard deviation of $4.18$.
\end{enumerate}

The clusters composition in Regime 2 is as follows:
\begin{enumerate}
\item Cluster 1 comprises of small (39\%) and medium size firms (36\%) with an average market size of $1.26\times 10^5$, return of $-0.02$ and standard deviation of $4.27$.
\item 17\% and 42\% of the assets in cluster 2 are small and medium size companies, respectively. While, on average the market capitalization, return and standard deviation are $1.48\times 10^5, 0.15$ and $4.50$, respectively.
\item Cluster 3 is characterized by 29\% small and 50\% medium size companies. Moreover, the average market capitalization, returns and standard deviation are $1.09\times 10^5, 0.08$ and $4.78$.
\end{enumerate}

For the cluster composition, we also compute the value of the Price-to-Earnings ratio (PE) for all the clusters\footnote{Following the standard practice in style analysis the average PE is computed over the last 10 years.}. For regime 1, clusters 1 and 2 have values of PE equal to 26.97 and 32.38, respectively. For regime 2, these values are 27.08, 35.69 and 23.72 for clusters 1, 2 and 3, respectively. These results seem to indicate that in both regimes cluster 2 is overvalued. 
 
To provide additional information on the composition of the clusters, we plot the dynamics over time of the average probabilities of the assets in each cluster, the market size and the PE in Fig. \ref{NclustCoclust1}. Regarding the average probabilities, cluster 1 shows the highest probability in regime 1, while in regime 2 clusters 1 and 3 display similar probabilities. As for PE, the dynamics in clusters 1 and 2 in regime 1 resemble those in regime 2 (see second line of Fig. \ref{NclustCoclust1}), and the market capitalization pattern indicates that relatively to cluster 1 in the over-performance state, assets in cluster 2 seems to show lowest values before the 2008/09 Global financial crisis and 
larger afterwards. The same dynamics for this two clusters is also observed in the under-performance state. 

The following summarizes the results. There is evidence of time-varying clustering structures in our panel of time series. Regime 1 (over-performance phase) and regime 2 (under-performance phase) comprise 2 and 3 clusters, respectively. The composition of the clusters varies across regimes while some similarities in terms capitalization and financial ratios are observed. In regime 1 and 2 medium size companies represent the largest majority (about 40\%) and assets in cluster 2 seem to be over valued by their PE ratio. The pattern of the market capitalization is similar across the two regimes as cluster 2 shows lower (larger) values compared to cluster 1 before (after) the 2008/09 Global financial crisis.

\section{Conclusion}\label{Concl}
The increase of interest in the study of volatility of large panels of asset returns and the evidence of regimes in volatility of financial returns has suggested to adapt Markov switching models to GARCH effect. In this respect, this paper introduces a new model for panel data with Markov-switching GARCH effects. 

In particular, we propose to model cross-sectional clustering effects with a Bayesian nonparametric technique that considers a hierarchical Pitman-Yor process prior for the Markov Switching GARCH parameters. The Bayesian nonparametric approach is a two-stage procedure. In the first stage, the hierarchical prior allows for cross-unit heterogeneity, while shrinking all unit-specific parameters towards a common mean. In the second stage, the hierarchical procedure allows for mixed effects in the common mean. 

This paper makes a contribution in some respects. First, the new model allows us to make 
inference on the number of mixture components in the cross-sectional clustering. Second, the model is sufficiently flexible to embody different shapes of the prior and  posterior predictive distributions. Third, uncertainty and the number of mixture components
are incorporated in the predictive distribution. Lastly, through a data-augmentation strategy, this paper makes the inference more tractable for our high dimensional model. A simulation exercise is carried out for inference and model validation. 

We apply the new model to 78 assets of the SP\&100 index from $6^{th}$ January 2000 to $3^{rd}$ October 2020. Our results may have some implications for portfolio making and for style investing decisions. The evidence shows that regime 1 (over-performance phase) and regime 2 (under-performance phase) differ in terms of clustering structures comprising 2 and 3 clusters, respectively. Within each regime the clusters differ substantially in terms of over-performance probability and in terms of style features, when considering capitalization and Price-to-Earnings. The heterogeneity of the clusters in terms of sectors and styles allows for portfolio diversification. Across regimes, the composition of the clusters changes, nevertheless some clusters share some similarities in terms of style features, allowing for the implementation of rotating style strategies. 

Further research may consider the choice of the number of performance regimes, the sensitivity with respect to nonparametric prior specification and some forecasting comparisons with exogenous clustering models.

\section*{Acknowledgements}
This research used the SCSCF multiprocessor cluster system provided by the Venice Centre for Risk Analytics (VERA) at University Ca' Foscari of Venice.

%\begin{supplement}
%\sname{Supplement B}\label{suppB}
%\stitle{Computational details}
%\slink[url]{http://www.e-publications.org/ims/support/dowload/supplementB.pdf}
%\sdescription{This document contains the derivation of the full conditional distributions of the Gibbs sampler and a description of the sampling methods used.}
%\end{supplement}
%
%\begin{supplement}
%\sname{Supplement C}\label{suppC}
%\stitle{Simulation results}
%\slink[url]{http://www.e-publications.org/ims/support/dowload/supplementC.pdf}
%\sdescription{This document reports the results of the simulation experiments used to check the efficiency of the proposed MCMC procedure.}
%\end{supplement}
%
%\begin{supplement}
%\sname{Supplement D}\label{suppD}
%\stitle{Data description}
%\slink[url]{http://www.e-publications.org/ims/support/dowload/supplementD.pdf}
%\sdescription{This document contains a description and a preliminary analysis of the data used in the empirical application.}
%\end{supplement}
%
%
%\begin{supplement}
%\sname{Supplement E}\label{suppE}
%\stitle{Further empirical results}
%\slink[url]{http://www.e-publications.org/ims/support/dowload/supplementE.pdf}
%\sdescription{This document provides further empirical results and robustness checks.}
%\end{supplement}

\bibliographystyle{imsart-nameyear}
\bibliography{BiblioNew}

%%%%%%%%%%%%%%%%%%%%%%%%%%%%%%%%%%%%%%%%%%%%%%%%%%%%%%%%%%%%%%%%%
%%%%%%%%%%%%%%%%%%%%%%%%%%%%%%%%%%%%%%%%%%%%%%%%%%%%%%%%%%%%%%%%%
%%%%%%%%%%%%%%%%%%%%%%%%%%%%%%%%%%%%%%%%%%%%%%%%%%%%%%%%%%%%%%%%%
%%%%%%%%%%%%%%%%%%%%%%%%%%%%%%%%%%%%%%%%%%%%%%%%%%%%%%%%%%%%%%%%%

\appendix
\section{Proof of the results in Section 3}\label{app}
We introduce for $h\geq 1$ the set of parameters allocated to the $h$-th mixture component in the regime $k$, $\mathcal{D}_{hk}=\{i=1,\ldots,N|D_{ik}=h\}$ and the set of the non-empty mixture components $\mathcal{D}^{*}_{k}=\{h|\mathcal{D}_{hk}\neq \emptyset\}$. The number of stick-breaking components needed for the finite mixture representation is $D^{*}_k=\max\{D_{ik},i=1,\ldots,N\}$. When sampling from the full conditional distribution of $\Theta^*_k$ and $V_k$ only $N^*_k$ element are sampled where $N^{*}_k$ is the smallest integer such that $\sum_{h=1}^{N^*_k}W_{hk}>1-U_{k}^*$ where $U_{k}^*=\min\{U_{ik},i=1,\ldots,N\}$. 
\subsection{Full conditional distribution of $V$ and $U$}
Let us split $V_{k}$ in three blocks: $V_k^*=\{V_{lk}: l\in\mathcal{D}_{k}^*\}$, $V_k^{**}=\{V_{kD^*_{k}+1},\ldots, V_{kD^*_{k}+N^*_k}\}$ and $V_k^{***}=\{V_{lk}:l>N^*_k\}$. The samples are generated from a collapsed Gibbs step
\begin{enumerate}
\item the full conditional of the elements in $V^*_k$ given $\Xi,\Theta,P,D,\Theta^{*},Y$
\begin{equation}
f(V_{lk}|\cdots)\propto \mathcal{B}e\left(1-\nu+\sum_{i=1}^{N}\mathbb{I}(D_{ik}=l),\psi+\nu l+ \sum_{i=1}^{N}\mathbb{I}(D_{ik}>l)\right)
\end{equation}
for $l\leq D^*_k$, 
\item the full conditional of the elements elements of $V^{**}_k$ and $V^{***}_k$ given $\Xi,\Theta,P,D,V^*,\Theta^{*},Y$, which coincide with the prior distributions $\mathcal{B}e(1-\nu,\psi+\nu l)$ for $l>D^*_k$
\item the full conditional of the elements of $U_k$ given $V$ and $\Xi,\Theta,P,D,\Theta^{*},Y$
\begin{equation}
f(U_{ik}|\cdots)\propto \mathbb{I}(U_{ik}<W_{D_{ik}k})
\end{equation}
which is uniform on the interval $(0,W_{D_{ik}})$
\end{enumerate}
\subsection{Full conditional distribution of $P$ and $R$}
We apply a collapsed-Gibbs step and sample $\mathbf{r}_{k}$ $k=1,\ldots,K$ given $\Xi, \Theta,D,V, Y$ and $\mathbf{p}_{k}$ $k=1,\ldots,K$ from its conditional given $R$ and $\Xi, \Theta_{-p},D,V, Y$.
As regards the transition probabilities, from standard calculations in Markov-switching regression models we obtain
\begin{eqnarray}
&&f(\mathbf{p}_{i,k}|\cdots)\propto \prod_{h=1}^{K}p_{i,kh}^{\phi r_{kh}+n_{i,kh}-1}\propto \mathcal{D}(\phi r_{k1}+n_{i,k1},\ldots,\phi r_{kK}+n_{i,kK})
\end{eqnarray}
where 
\begin{equation}
n_{i,kh}=\sum_{t=1}^{T}\xi_{ikt-1}\xi_{iht}
\end{equation}
The marginal distribution is
\begin{eqnarray}
&&f(\mathbf{r}_{k}|\cdots)\propto \int_{\Delta_{[0,1]^{K}}^{N}}\prod_{i=1}^{N}\prod_{h=1}^{K}p_{i,kh}^{\phi r_{kh}+n_{i,kh}-1}\frac{\Gamma(\phi)}{\Gamma(\phi r_{kh})}dp_{i,kh} \pi(\mathbf{r}_k)\\
&&\propto \left(\prod_{h=1}^{K}r_{kh}^{d-1}\right)\left(\prod_{i=1}^{N}\prod_{h=1}^{K}\frac{\Gamma(\phi r_{kh}+n_{i,kh})}{\Gamma(\phi+n_{i,k})}\frac{\Gamma(\phi)}{\Gamma(\phi r_{kh})}\right)\nonumber
\end{eqnarray}
where $n_{i,k}=n_{i,k1}+\ldots+n_{i,kK}$ and $\Delta_{[0,1]^{K}}=\{(p_1,\ldots,p_K)\in\mathbb{R}^K|p_k>0\,\forall k,\, p_1+\ldots+p_K=1\}$ is the $K$-dim standard simplex. From the properties of the gamma functions
\begin{eqnarray*}
\Gamma(\phi r_{kh}+n_{i,kh})&=&\prod_{l=1}^{n_{i,kh}}(\phi r_{kh}+l-1)\Gamma(\phi r_{kh})\\
\Gamma(\phi+n_{i,k})&=&\prod_{l=1}^{n_{i,k}}(\phi+l-1)\Gamma(\phi)
\end{eqnarray*}
we obtain
\begin{eqnarray}
&&f(\mathbf{r}_{k}|\cdots)\propto\mathcal{D}ir(d+m_{k1},\ldots,d+m_{kK})g(\mathbf{r}_{k})
\end{eqnarray}
where
\begin{equation*}
g(\mathbf{r}_{k})=\prod_{i=1}^{N}\left(\prod_{l=1}^{n_{i,k}}(\phi+l-1)\right)^{-1}\prod_{h=1}^{K}\prod_{l=2}^{n_{i,kh}}(\phi r_{kh}+l-1)
\end{equation*}
and $m_{kh}=\hbox{Card}(\mathcal{M}_{kh})$, $\mathcal{M}_{kh}=\{i=1,\ldots,N|n_{i,kh}>0\}$. Samples from this full conditional distribution are obtain by a Metropolis-Hastings algorithm with independent proposal distribution $\mathcal{D}ir(d+m_{k1},\ldots,d+m_{kK})$.
\subsection{Full conditional distribution of $\Theta^*$}
The full conditional distribution of $\mathbf{\theta}_{hk}^{*}=(\mu_{hk}^*,\gamma_{hk}^*,\alpha_{hk}^*,\beta_{hk}^*)$ can be sampled by simulating iteratively from the following conditional distributions.
The full conditional of $\mu_{hk}^{*}$:
\begin{eqnarray}
f(\mu_{hk}^{*}|\cdots)&\propto& \mathcal{N}(\mu_{hk}^*|m^*,s^{*})\prod_{i\in \mathcal{D}_{hk}}\mathcal{N}(\mu_{ik}|\mu_{hk}^*,s)\\
&\propto& \mathcal{N}\left(\mu_{hk}^*|\overline{m}_{hk},\overline{s}_{hk}\right)\nonumber
\end{eqnarray}
where, 
\begin{equation*}
\overline{m}_{hk} = \overline{s}_{hk}^{2}\left(\frac{m^{*}}{s^{*2}} +\frac{\sum_{i\in \mathcal{D}_{hk}} \mu_{ik}}{s^2}  \right)
,~~~{\text{and}}~~\overline{s}_{hk}= \left( \frac{1}{s^{*2}} + \frac{\hbox{Card}(\mathcal{D}_{hk})}{s^{2}}\right)^{-1/2} 
\end{equation*}
The full conditional distribution of $\gamma_{hk}^{*}$
\begin{eqnarray}
&&f(\gamma_{hk}^{*}|\cdots)\propto \mathbb{I}_{[0,a]}(\gamma_{hk}^{*})\prod_{i\in \mathcal{D}_{hk}}\mathcal{B}e(\gamma_{ik}/a|r\gamma_{hk}^*/a,r(1-\gamma_{hk}^*/a))
\end{eqnarray}
\begin{eqnarray*}
&&\propto\mathbb{I}_{[0,a]}(\gamma_{hk}^{*})\prod_{i\in\mathcal{D}_{hk}} \dfrac{\exp\{(r\gamma_{hk}^{*}/a-1)\log(\gamma_{ik}/a)+(r(1-\gamma_{hk}^{*}/a)-1)\log(1-\gamma_{ik}/a)\}}
{\Gamma{(r\gamma_{hk}^*/a)} \Gamma{(r(1-\gamma_{hk}^*/a))}}
\\
&&\propto\exp\{- \kappa_{hk}\gamma_{hk}^{*}\}\left(\dfrac{1}{\Gamma{(r\gamma_{hk}^*/a)} \Gamma{(r(1-\gamma_{hk}^*/a))}}\right)^{\scriptsize{\hbox{Card}}(\mathcal{D}_{hk})} \mathbb{I}_{[0,a]}(\gamma_{hk}^{*})
\end{eqnarray*}
where 
\begin{equation*}
\kappa_{hk}=\dfrac{r}{a}\sum_{i\in\mathcal{D}_{hk}}\log((a-\gamma_{ik})/\gamma_{ik})
\end{equation*}
which can be simulated exactly by the inverse cdf method where the cdf is
\begin{eqnarray*}
\left(1-\exp\{- \kappa_{hk}\gamma_{hk}^{*}\}\right)\frac{1}{1-\exp\{-a\kappa_{hk}\}}\mathbb{I}_{[0,a]}(\gamma_{hk}^{*}).
\end{eqnarray*}

The full conditional distribution of $\alpha_{hk}^{*}$
\begin{eqnarray}
&&f(\alpha_{hk}^{*}|\cdots)\propto \mathbb{I}_{[0,1]}(\alpha_{hk}^{*})\prod_{i\in \mathcal{D}_{hk}}\mathcal{B}e(\alpha_{ik}|r\alpha_{hk}^*,r(1-\alpha_{hk}^*))
\end{eqnarray}
\begin{eqnarray*}
&&\propto\mathbb{I}_{[0,1]}(\alpha_{hk}^{*})\prod_{i\in\mathcal{D}_{hk}} \dfrac{\exp\{(r\alpha_{hk}^{*}-1)\log(\alpha_{ik})+(r(1-\alpha_{hk}^{*})-1)\log(1-\alpha_{ik})\} }
{\Gamma{(r\alpha_{jk}^*)} \Gamma{(r(1-\alpha_{hk}^*))}}
\\
&&\propto\exp\{- \tau_{hk}\alpha_{hk}^{*}\}\left(\dfrac{1}{\Gamma{(r\alpha_{hk}^*)} \Gamma{(r(1-\alpha_{hk}^*))}}\right)^{\scriptsize{\hbox{Card}}(\mathcal{D}_{hk})} \mathbb{I}_{[0,1]}(\alpha_{hk}^{*})
\end{eqnarray*}
where 
\begin{equation*}
\tau_{hk}=r\sum_{i\in\mathcal{D}_{hk}}\log((1-\alpha_{ik})/\alpha_{ik})
\end{equation*}
which can be simulated exactly by the inverse cdf method where the cdf is
\begin{eqnarray*}
\left(1-\exp\{- \tau_{hk}\alpha_{hk}^{*}\}\right)\frac{1}{1-\exp\{-\tau_{hk}\}}\mathbb{I}_{[0,1]}(\alpha_{hk}^{*}).
\end{eqnarray*}
Similar argument is applied to the full conditional distributions of $\beta_{hk}^{*}$. 
\subsection{Full conditional distribution of $\Theta$}
The full conditional distribution of the elements of $\boldsymbol{\theta}_{ik}$ $k=1,\ldots,K$ are discussed. Let $\boldsymbol{\mu}_i=(\mu_{i1},\ldots,\mu_{iK})$, its full conditional distribution 
\begin{equation}
f(\boldsymbol{\mu}_{i}|\cdots)\propto \left(\prod_{t=1 }^{T}\mathcal{N}(  y_{it}|\mu_{i}(s_{it}),\sigma_{it})\right) \prod_{k=1}^{K}\mathcal{N}(  \mu_{ik}|\tilde{\mu}_{ik}^{*},s)
\end{equation}
which is not tractable due to the recursive form of $\sigma^2_{it}$. Thus we sample from the full conditional by Metropolis-Hastings with proposal distribution obtained through the approximation $\sigma_{it}^{*2}$ of $\sigma_{it}^{2}$. It can easily be shown, by the completing of the square argument, that the joint full conditional distribution of $\boldsymbol{\mu}_i$ can be approximated by a normal distribution with mean and covariance
\begin{equation}
\mathbf{m}_{i} = 
S_{i} \left( 
\begin{array}{c}
m_{i1}/ s_{i1}^{2}  \\
m_{i2}/s_{i2}^{2}\\
\vdots\\
m_{iK}/s_{iK}^{2}\\
\end{array} 
\right),~~~
S_{i} = 
\left( 
\begin{array}{cccccc}
s_{i1}^{2}  &    0                  &\ldots & 0\\
          0            & s_{i2}^{2} &   0   &\vdots\\
       \vdots          &    0                  &\ddots &  0\\
       0               &    0                  &\ldots & s_{iK}^{2}\\
\end{array}
\right)
\end{equation}
where
\begin{equation*}
m_{ik} = s_{ik}^2 \left(\frac{\tilde{\mu}_{ik}^{*}}{s^{2}}+ \sum_{t\in\mathcal{T}_{y,ik}} \frac{y_{it}}{\sigma_{it}^{*2}}   \right),~~~{\text{and}}~~~
s_{ik}^2 = \left(\frac{1}{s^{2}}+\sum_{t\in\mathcal{T}_{y,ik}}\frac{1}{\sigma_{it}^{*2}}\right)^{-1}
\end{equation*}
with $\mathcal{T}_{y,ik} = \lbrace t=1,\ldots,T \vert s_{it} = k \rbrace $ and 
$$\sigma_{it}^{*2}=\gamma_{i}(s_{it})+\alpha_{i}(s_{it})(y_{t-1}-\mu_{i}(s_{it-1}))^{2}+(\beta_{i}(s_{it}))\sigma_{t-1}^{*2}.$$ 
The mean and variance thus constructed are used in defining the parameters of the
normal  mixture proposal distribution for $\boldsymbol{\mu}_{i}$.
$$
f({\boldsymbol\mu}_{i}|\ldots ) = 0.05{\mathcal{N}}({\boldsymbol\mu}_{i}; {\bf{m}}_{i}, S_{i}) + 
0.95{\mathcal{N}}({\boldsymbol\mu}_{i}; {\boldsymbol{\mu}}_{i}^{(r-1)}, S_{i})
$$
As regards the parameters of the volatility process the full conditional probability distribution is \\
let $\boldsymbol{\gamma}_{i}=(\gamma_{i1},\ldots,\gamma_{iK})$, $\boldsymbol{\alpha}_{i}=(\alpha_{i1},\ldots,\alpha_{iK})$, $\boldsymbol{\beta}_{i}=(\beta_{i1},\ldots,\beta_{iK})$, 
\begin{equation}
\begin{aligned} 
f(\boldsymbol{\gamma}_{i},\boldsymbol{\alpha}_{i},\boldsymbol{\beta}_{i}|\cdots)\propto &\prod_{t=1 }^{T}\mathcal{N}(  y_{it}|\mu_{i}(s_{it}),\sigma_{it})\prod_{k=1}^{K}
\mathcal{B}e(\gamma_{ik}/a|r\tilde{\gamma}_{ik}^{*}/a,r(1-\tilde{\gamma}_{ik}^{*}/a))\\
&\mathcal{B}e(\alpha_{ik}|r\tilde{\alpha}_{ik}^{*},r(1-\tilde{\alpha}_{ik}^{*}))
\mathcal{B}e(\beta_{ik}|r\tilde{\beta}_{ik}^{*},r(1-\tilde{\beta}_{ik}^{*}))
\end{aligned} 
\end{equation}

We follow the ARMA approximation of the MS-GARCH process, that is

\begin{eqnarray}
&&\sigma_{it}^{2} =\gamma_{i}(s_{it})\! +\! \alpha_{i}(s_{it})\epsilon_{it-1}^{2} + \beta_{i}(s_{it})\sigma_{it-1}^{2}\\&&\epsilon_{it}^{2}= \gamma_{i}(s_{it}) + (\alpha_{i}(s_{it})\!+\!\beta_{i}(s_{it}))\epsilon_{it-1}^{2} \!-\! \beta_{i}(s_{it})(\epsilon_{it-1}^{2}\!-\!\sigma_{it-1}^{2}) + (\epsilon_{it}^{2}\!-\!\sigma_{it}^{2}).
\end{eqnarray}

Let
$$
w_{it} = \epsilon_{it}^{2}-\sigma_{it}^{2} = \left(\dfrac{\epsilon_{it}^{2}}{\sigma_{it}^{2}}-1\right)\sigma_{it}^{2} = (\chi^{2}(1)-1)\sigma_{it}^{2}
$$
with 
$$E_{t-1}[w_{it}]=0; \quad {\text{and}}\quad Var_{t-1}[w_{it}]=2\sigma_{it}^{4}.$$
Subject to the above and following \cite{Nak98} suggestion, we assume that $w_{it}\approx w_{it}^{*} \sim \mathcal{N}(0,2\sigma_{it}^{4})$. Then we have the following auxiliary ARMA model for the squared error term $\epsilon_{it}^{2}$
\begin{equation}
\epsilon_{it}^{2}= \gamma_{i}{s_{it}} + (\alpha_{i}(s_{it})+\beta_{i}(s_{it}))\epsilon_{it-1}^{2} - \beta_{i}(s_{it})w_{it-1}^{*} + w_{it}^{*}
\end{equation}
with $w_{it}^{*} \sim \mathcal{N}(0,2\sigma_{it}^{4})$, which returns
\begin{equation}
w_{it}^{*} = \epsilon_{it}^{2} - \gamma_{i}(s_{it}) -  \alpha_{i}(s_{it})\epsilon_{it-1}^{2} - \beta_{i}(s_{it})(\epsilon_{it-1}^{2}-w_{it-1}^{*}).
\end{equation}
Following \cite{Ardia08} we further express $w_{it}^{*}$ as a linear function of the $(3K\times1)$ vector $\boldsymbol{\theta}_{i\sigma}=(\gamma_{i1},\ldots,\gamma_{iK},\alpha_{i1},\ldots,\alpha_{iK},\beta_{i1},\dots,\beta_{iK})'$. To do this, we approximate the function $w_{t}^{*}$ by the first order Taylor's expansion about $\theta_{i\sigma}^{(r-1)}=(\gamma_{i1}^{(r-1)},\dots,\gamma_{iK}^{(r-1)},\alpha_{i1}^{(r-1)},\dots,\alpha_{iK}^{(r-1)},\beta_{i1}^{(r-1)},\dots,\beta_{iK}^{(r-1)})'$.
\begin{equation} 
w_{it}^{*} \approx w_{it}^{**} =w_{it}^{*}(\boldsymbol{\theta}_{i\sigma}^{(r-1)}) + \nabla_{it}'(\theta_{i\sigma}-\boldsymbol{\theta}_{i\sigma}^{(r-1)}),
\end{equation}
where
\begin{equation}
\nabla_{it}= 
vec \left(
\begin{array} {c}
\nabla_{it1}' \\
\nabla_{it2}' \\
\vdots\\
\nabla_{itK}' \\
\end{array}
\right),~~
\nabla_{itk} =
\left(
\begin{array} {c}
\dfrac{\partial w_{it}^{*}}{\partial\gamma_{ik}} \\  
\dfrac{\partial w_{it}^{*}}{\partial\alpha_{ik}} \\
\dfrac{\partial w_{it}^{*}}{\partial\beta_{ik}} \\
\end{array}
\right)
\end{equation}
with 
\begin{equation}
\left(
\begin{array} {c}
\nabla_{it1}' \\
\nabla_{it2}' \\
\vdots\\
\nabla_{itK}' \\
\end{array}
\right) =
{\boldsymbol\xi}_{it}'E_{t}  +
({\boldsymbol\xi}_{it}\beta_{i}')
\left(
\begin{array} {c}
\nabla_{it-1,1}' \\
\nabla_{it-1,2}' \\
\vdots\\
\nabla_{it-1,K}' \\
\end{array}
\right)
\end{equation}
$E_{t} = (-1, -\epsilon_{it-1}^{2}, -(\epsilon_{it-1}^{2} - w_{it-1}^{*}))$, $\beta_{i} = (\beta_{i1},\beta_{i2},\ldots,\beta_{iK})$, $\nabla_{i0k} = \boldsymbol{0}$ and ${\boldsymbol\xi}_{it}$ is a row  vector. 
%and 
%\begin{equation}
%\begin{aligned}
%\dfrac{\partial w_{it}^{*}}{\partial\gamma_{k}} &= - \xi_{it,k} + (\boldsymbol{\xi}_{t}'\beta)\dfrac{\partial w_{it-1}^{*}}{\partial\gamma_{ik}}\\
%\dfrac{\partial w_{it}^{*}}{\partial\alpha_{ik}} &= - \xi_{i,tk}\epsilon_{i,t-1}^{2} + (\boldsymbol{\xi}_{t}'\beta)\dfrac{\partial w_{it-1}^{*}}{\partial\alpha_{ik}}\\
%\dfrac{\partial w_{it}^{*}}{\partial\beta_{ik}}  &= - \xi_{i,tk}(\epsilon^{2}_{it-1}-w_{it-1}^{*}) + (\boldsymbol{\xi}_{it}'\beta)\dfrac{\partial w_{it-1}^{*}}{\partial\beta_{ik}}
%\end{aligned}
%\end{equation}
%for $k=1,\dots, K$, evaluated at $\theta_{i\sigma}^{(r-1)}$. In matrix form we can  
%\begin{equation}
%\left(
%\begin{array} {c}
%\nabla_{it1}' \\
%\nabla_{it2}' \\
%\vdots\\
%\nabla_{itK}' \\
%\end{array}
%\right) =
%{\boldsymbol\xi}_{it}\left({\bf{E}}  +
%{\boldsymbol\beta} 
%\left(
%\begin{array} {c}
%\nabla_{it-1,1}' \\
%\nabla_{it-1,2}' \\
%\vdots\\
%\nabla_{it-1,K}' \\
%\end{array}
%\right)\right),~~
%\end{equation}

Upon defining $r_{it}^{*}=w_{it}^{*}(\boldsymbol{\theta}_{-i\pi}^{(r-1)})-\nabla_{it}'\boldsymbol{\theta}_{i\sigma}^{(r-1)}$, it turns out that $w_{it}^{**} = r_{it}^{*} + \nabla_{it}'\boldsymbol{\theta}_{i\sigma}$. Furthermore, by defining 
$\mu_{i} = (\mu_{i1},\mu_{i2},\ldots, \mu_{iK})$, $\alpha_{i}=(\alpha_{i1},\alpha_{i2},\ldots,\alpha_{iK})$, $\gamma_{i}=(\gamma_{i1},\gamma_{i2},\ldots, \gamma_{iK})$,
the $T\times 1$ vectors $\mathbf{w}_{i}=(w_{i1}^{**},\dots,w_{iT}^{**})'$,  $\mathbf{r}^{*}_{i}=(r_{i1}^{*},\dots,r_{iT}^{*})'$, a $T\times 3K$ matrix $\nabla_{i} = (\nabla_{i1},\nabla_{i2},\dots,\nabla_{iT})'$ as well as a $T\times T$ matrix 
\begin{equation}
\Upsilon_{i} = 2
\begin{pmatrix}
\sigma_{i1}^{**4} &\cdots & 0\\
\vdots         &\ddots & \vdots\\
0              &\cdots & \sigma_{iT}^{**4}\\
\end{pmatrix},
\end{equation}
with $\sigma_{it}^{**2}=({\boldsymbol{\xi}_{it}}\gamma_{i}^{(r-1)'})+({\boldsymbol{\xi}_{it}}\alpha_{i}^{(r-1)'})(y_{t-1}-{\boldsymbol{\xi}_{t-1}}\mu_{i}^{(r)'})^{2}+({\boldsymbol{\xi}_{it}}\beta_{i}^{(r-1)'})\sigma_{it-1}^{**2}$, we end up with $\mathbf{w}_{i}=\mathbf{r}^{*}_{i} +\nabla_{i}\boldsymbol{\theta}_{i\sigma}$. Using this linear approximation, we can approximate the full conditional distribution of the volatility parameters as
\begin{equation}
\begin{aligned}
&f(\boldsymbol{\theta}_{i\sigma}|\boldsymbol{\xi}_{i,1:T}^{(r-1)},\mu_{i}^{(r)},y_{1:T})\propto \dfrac{1}{|\Upsilon_{i}|^{\frac{1}{2}}} \exp{\left(-\dfrac{{\bf{w}}'_{i}\Upsilon_{i}^{-1}{\bf{w}}_{i}}{2}\right)}\mathbb{I}_{\Theta}(\boldsymbol{\theta}_{i\sigma})\\
&\propto {\mathcal{N}}_{3K}(m_{i\sigma},S_{i\sigma})\mathbb{I}_{\Theta}(\boldsymbol{\theta}_{i\sigma}),
\end{aligned}\label{eqGARCHprop}
\end{equation}
where $\Theta=\{\gamma_{i1}>0,\ldots,\gamma_{iK}>0,0<\alpha_{i1}<1,\ldots,0<\alpha_{iK}<1,0<\beta_{i1}<1\dots,0<\beta_{iK}<1\}$ and
\begin{equation}
\begin{aligned}
S_{i\sigma} &= (\nabla_{i}'\Upsilon_{i}^{-1}\nabla_{i})^{-1}\\
m_{i\sigma}   &= -S_{i\sigma}\nabla_{i}'\Upsilon_{i}^{-1}{\bf{r}}_{i}^{*}.
\end{aligned}
\label{propV}
\end{equation}
The mean and variance defined above are used to characterize proposal distribution for $\theta_{i\sigma}$, that is a mixture of truncated normal distributions. In our MCMC exercise, we sample from the normal mixture and check that each sample satisfies the constraints. 
$$
f(\theta_{i\sigma}|\ldots ) = 0.05{\mathcal{N}}(\theta_{i\sigma}; {\bf{m}}_{i\sigma}, S_{i\sigma}) + 
0.95{\mathcal{N}}(\theta_{i\sigma}; \theta_{i\sigma}^{(r-1)}, S_{i\sigma})
$$
\subsection{Full conditional distribution of $\Xi$}
The full joint conditional distribution of the state variables, $\boldsymbol{\xi}_{i,1:T} = \left(\boldsymbol{\xi}_{i1},\dots, \boldsymbol{\xi}_{iT}\right)$ with 
$\boldsymbol{\xi}_{it} = \left(\xi_{i1,t},\ldots, \xi_{iK,t} \right)$
, given the parameter values and return series
\begin{equation}
p(\boldsymbol{\xi}_{i,1:T}|\ldots) \propto \prod_{t=1}^{T}f(y_{it}|\theta_{i},s_{it})\prod_{k=1}^{K}\prod_{l=1}^{k}p_{i,kl}^{\xi_{ikt-1}\xi_{ilt}}
\end{equation}
is a non-standard distribution. For this reason, following \cite{Antho14}, we propose a Metropolis-Hastings algorithm with proposal distribution given by an approximation of the  smoothed probability $p(\boldsymbol{\xi}_{i,1:T}|\ldots)$. Precisely, the algorithm involves running a Forward Filtering Backward Sampling (FFBS) on a auxiliary model to generate proposals at each iteration step. Among the several alternative MS-GARCH models based on collapsing procedure (see \cite{Antho14}), we adopt the \cite{klas02} MS-GARCH model as our auxiliary model because it accounts for the highest amount of information in its construction. We denote the proposal distribution by 
\begin{equation}
q(\boldsymbol{\xi}_{i,1:T}|\theta_{i},y_{i,1:T})=q(\boldsymbol{\xi}_{iT}|\theta_{i},y_{i,1:T})\prod_{t=1}^{T-1}q(\boldsymbol{\xi}_{it}|\boldsymbol{\xi}_{it+1},\theta_{i},y_{i,1:t}),\label{eq9}
\end{equation}
where 
$$
q(\boldsymbol{\xi}_{it}|\boldsymbol{\xi}_{it+1},\theta_{i},y_{i,1:t})=\dfrac{ q(\boldsymbol{\xi}_{it}|y_{i,1:t},\theta_{i})q(\boldsymbol{\xi}_{it+1}|\boldsymbol{\xi}_{it},\theta_{i})} {q(\boldsymbol{\xi}_{it+1}|y_{i,1:t},\theta_{i})}
$$ 
with $q(\boldsymbol{\xi}_{it}|y_{i,1:t},\theta_{i})$ representing filtered probability.

At time $t$, given $\theta_{i}$ and $y_{i,1:t}$ the prediction and filtering densities are respectively given by 
\begin{equation}
q(\boldsymbol{\xi}_{it}|\theta_{i},y_{i,1:t-1})=\sum_{k=1}^{K}\left(\prod_{l=1}^{K}p_{i,lk}^{\xi_{il,t}}\right)q(\boldsymbol{\xi}_{it-1}=e_{k}|\theta_{i},y_{i,1:t-1}),
\end{equation}
and 
\begin{equation}
q(\boldsymbol{\xi}_{it}|\theta_{i},y_{i,1:t})=\dfrac{g(y_{it}|\boldsymbol{\xi}_{it},\theta_{i},y_{i,1:t-1})q(\boldsymbol{\xi}_{it}|\theta_{i},y_{i,1:t-1})}{\sum_{k=1}^{K}g(y_{it}|\boldsymbol{\xi}_{it}=e_{k},\theta_{i},y_{i,1:t-1})q(\boldsymbol{\xi}_{it}=e_{k}|\theta_{i},y_{i,1:t-1})},
\end{equation}
where $e_{k}$ is the $k-$th row of a K-by-K identity matrix and $g(y_{it}|\boldsymbol{\xi}_{it},\theta_{i},y_{i,1:t-1})$ is the conditional density of unit $i$ return process under the auxiliary model
\begin{equation}
g(y_{it}|\boldsymbol{\xi}_{it},\theta_{i},y_{i,1:t-1}) \propto \prod_{\tau=1}^{t} \dfrac{1}{h_{i\tau}} \exp\left(-\dfrac{(y_{i\tau} - \mu_{i}(s_{i\tau}))^{2} }{2h_{i\tau}^{2}}\right)
\end{equation}
where 
$$
h_{it}^{2} = \gamma_{i}(s_{it}) + \alpha_{i}(s_{it})\epsilon^{2}_{(y)it-1} + \beta_{i}(s_{it})\sigma_{(y)i,kt-1}^{2}
$$
with
\begin{equation}
\begin{aligned}
\epsilon_{(y)it-1}&=y_{it-1}-\mu_{(y)i,kt-1}\\
\mu_{(y)ik,t-1}&=E[\mu_{i}(s_{it})|y_{i,1:t-1},\boldsymbol{\xi}_{it}=e_{k}]\\            
\sigma_{(y)i,kt-1}^{2}&= E[\sigma_{it-1}^{2}(y_{i,1:t-2},\boldsymbol{\xi}_{it-1},\boldsymbol{\xi}_{it-2})|y_{i,1:t-1},\boldsymbol{\xi}_{it}=e_{k}].
\end{aligned}
\end{equation}

Using the output of the FF, we compute  $q(\boldsymbol{\xi}_{iT}|\theta_{i},y_{i,1:T})$ and
\begin{equation}
q(\boldsymbol{\xi}_{it}|\boldsymbol{\xi}_{it+1},\theta_{i},y_{i,1:t})=\dfrac{\prod_{l=1}^{K}\left(\sum_{k=1}^{K}p_{i,lk}\xi_{i1,t}\right)^{\xi_{il,t+1}} q(\boldsymbol{\xi}_{it}|\theta_{i},y_{i,1:t})}{q(\boldsymbol{\xi}_{it+1}|\theta_{i},y_{i,1:t})},\label{eqBS}
\end{equation} 
for $t=T-1,T-2,\dots,2,1$. Then at each time step we sample $\boldsymbol{\xi}_{T}$ from $q(\boldsymbol{\xi}_{T}|\theta_{i},y_{i,1:T})$ and $\boldsymbol{\xi}_{it}$ from $q(\boldsymbol{\xi}_{it}|\boldsymbol{\xi}_{it+1},\theta_{i},y_{i,1:t})$ iteratively for $t=T-1,T-2,\dots,2,1$. This is the BS step. The BS procedure is implemented by first noting that $\boldsymbol{\xi}_{it+1}$ is the most recent value sampled for the hidden Markov chain at $t+1$ and since $\boldsymbol{\xi}_{it}$ can take one of $e_{1},\dots,e_{K}$, we compute the expression in equation (\ref{eqBS}) for each of these values. Sampling $\boldsymbol{\xi}_{it}$ from $q(\boldsymbol{\xi}_{it}|\boldsymbol{\xi}_{it+1},\theta_{i},y_{i,1:t})$ may be compared to multinomial sampling, provided that the probability of $\boldsymbol{\xi}_{ik}=e_{k}$, $k=1,\dots,K$, are known. 

\subsection{Full conditional distribution of $D$}
The full conditional of $D_{ik}$ is  $P(D_{ik}=h|\cdots)=c_h/c$ for $h\in\mathcal{A}_{ki}$, with $c_h=\mathcal{N}(\mu_{ik}|\mu^{*}_{hk},s)$ $\mathcal{B}e(\alpha_{ik}|r\alpha_{hk}^*,r(1-\alpha_{hk}^*))$ $\mathcal{B}e(\beta_{ik}|r\beta_{hk}^*,r(1-\beta_{hk}^*))\mathcal{B}e(\gamma_{ik}/a|r\gamma_{hk}^*/a,r(1-\gamma_{hk}^*/a))/a$ where $c=\sum_{h\in\mathcal{A}_{ki}}c_h$ is the normalizing constant and $a$ a real positive constant.

$\,$

\vfill

\newpage

\section{Further details on the simulation exercise}\label{sim}
\begin{figure}[h!]
\begin{center}
\setlength{\tabcolsep}{10pt}
\renewcommand{\arraystretch}{0.8}
\begin{tabular}{cc}
&\\
$p_{i,kk}$ & $\mu_{i,k}$\vspace{8pt}\\
\includegraphics[scale=0.4]{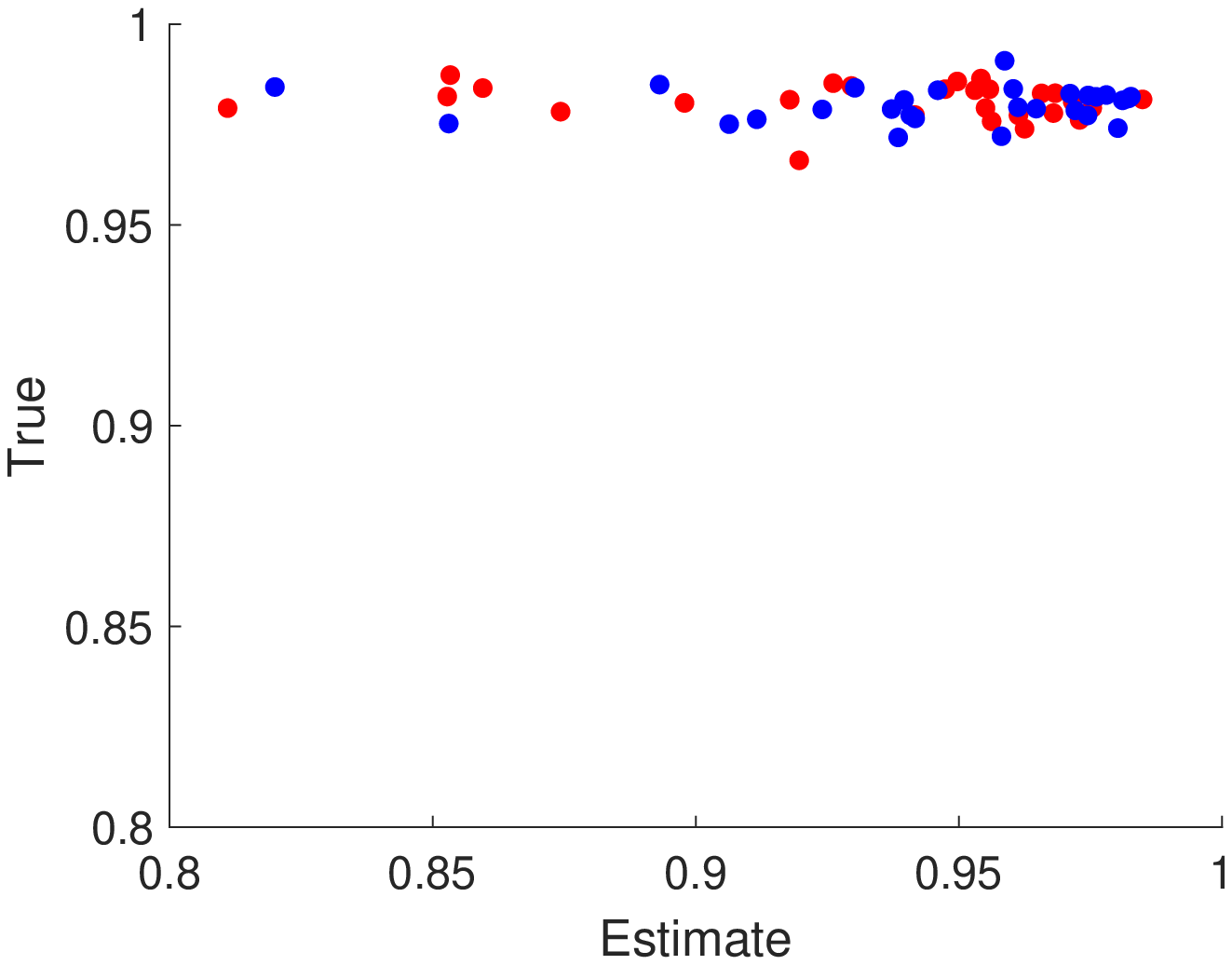}&
\includegraphics[scale=0.4]{Code/Figures/FigSimulatedBNP_ANTr10min2cluster/Mu.eps}\\
$\gamma_{i,k}$ &$\alpha_{i,k}$\vspace{8pt}\\
\includegraphics[scale=0.4]{Code/Figures/FigSimulatedBNP_ANTr10min2cluster/Gamma.eps}&
\includegraphics[scale=0.4]{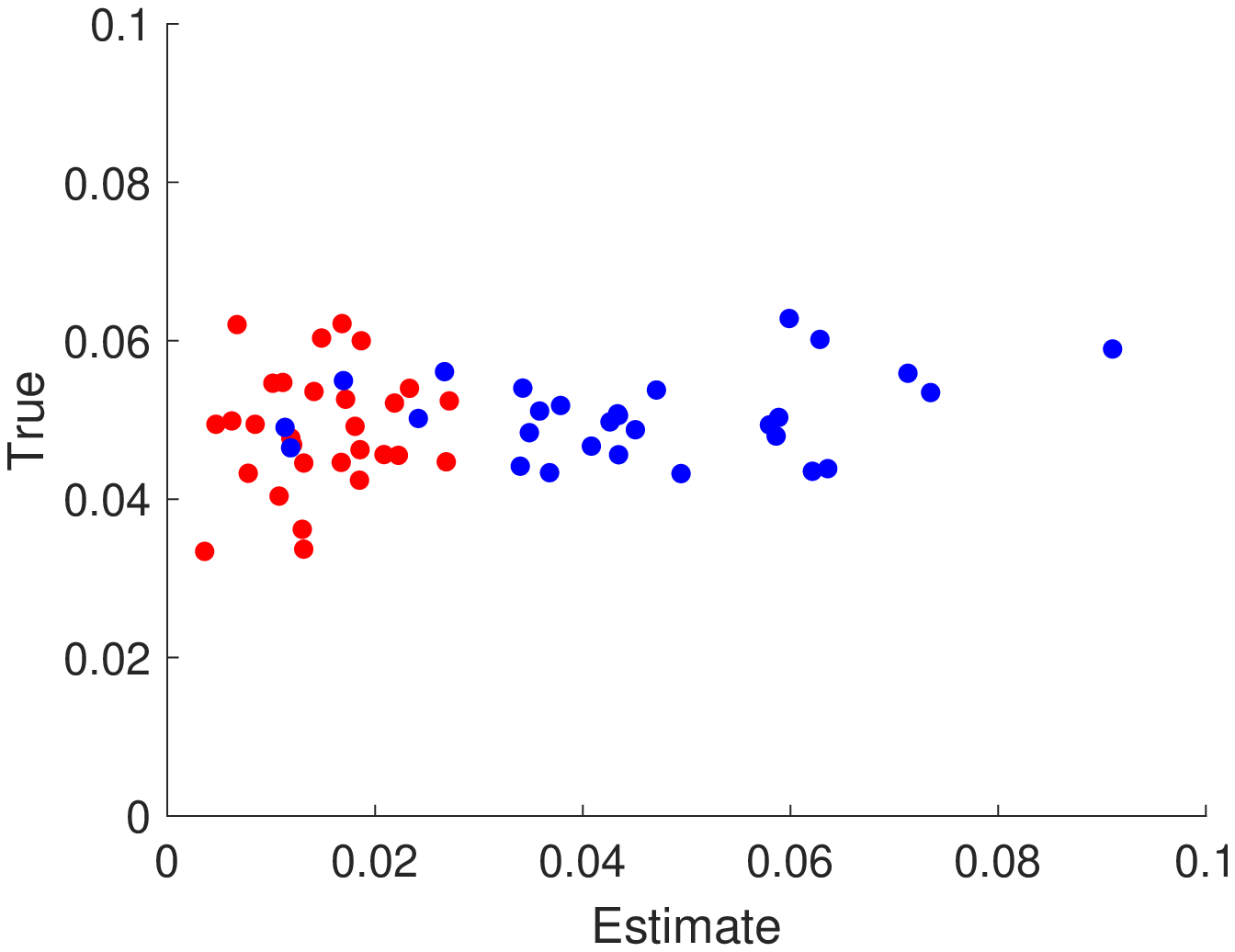}\\
$\beta_{i,k}$ &\vspace{8pt}\\
\includegraphics[scale=0.4]{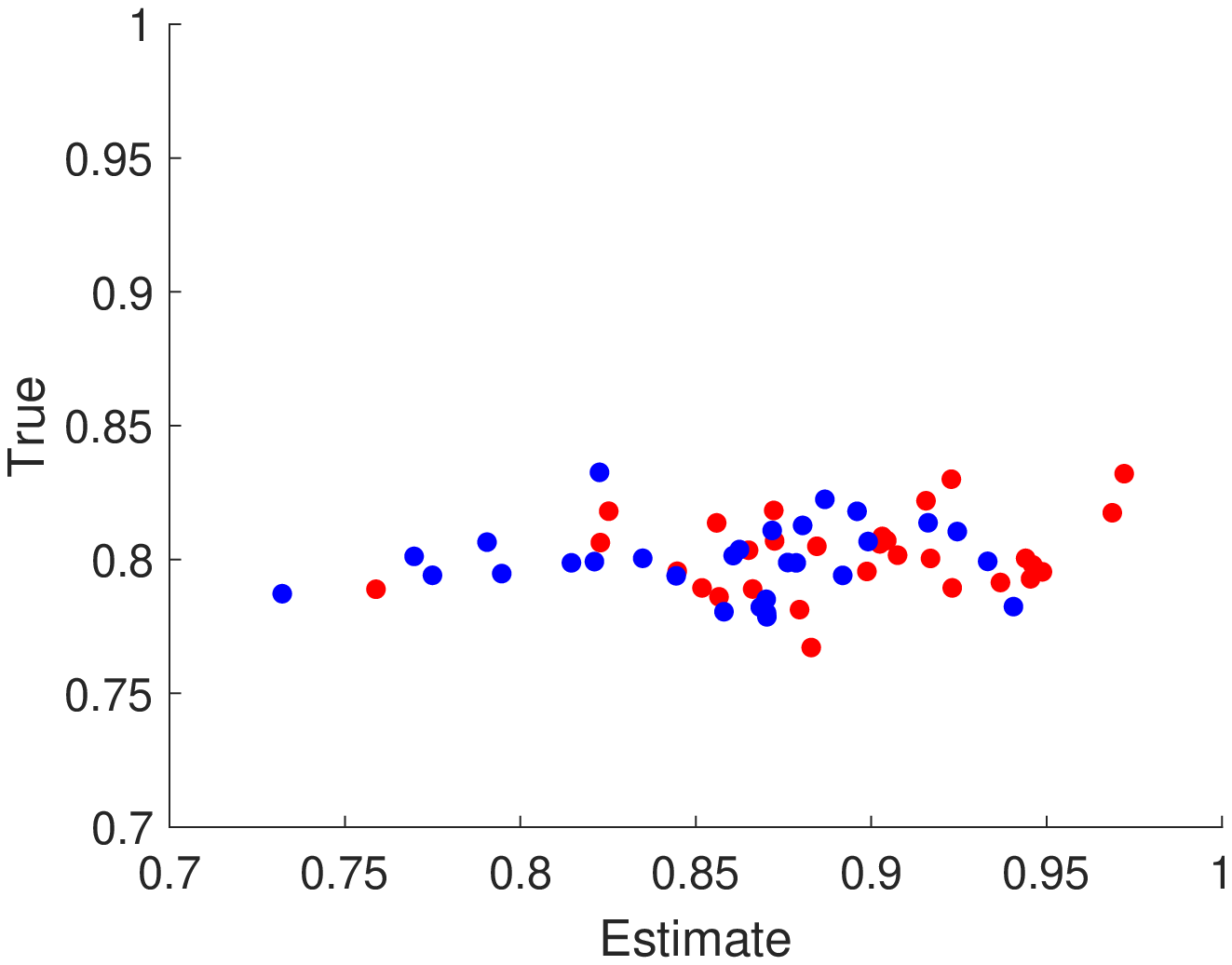}&
\end{tabular}
\end{center}
\caption{True (vertical axis) and estimated (horizontal axis) values of the parameters $\boldsymbol{\theta}_{ik}$ for each unit $i$ (dots) in regime $k=1$ (red) and $k=2$ (blue).} %\label{TransProb}
\label{SimuPlor}
\end{figure}

\begin{figure}[h!]
\begin{center}
\setlength{\tabcolsep}{2pt}
\begin{tabular}{ccccc}
\includegraphics[scale=0.18]{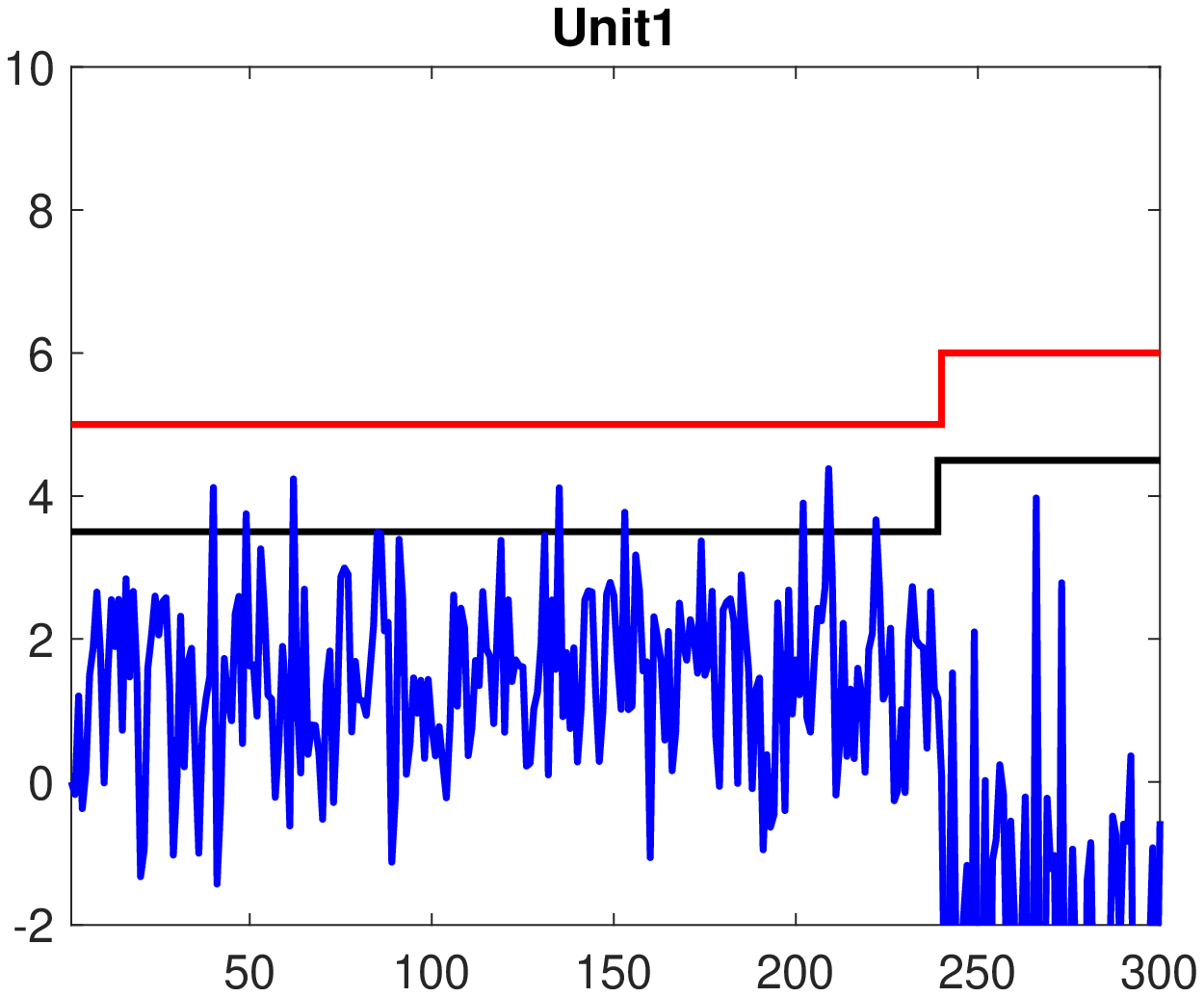}&
\includegraphics[scale=0.18]{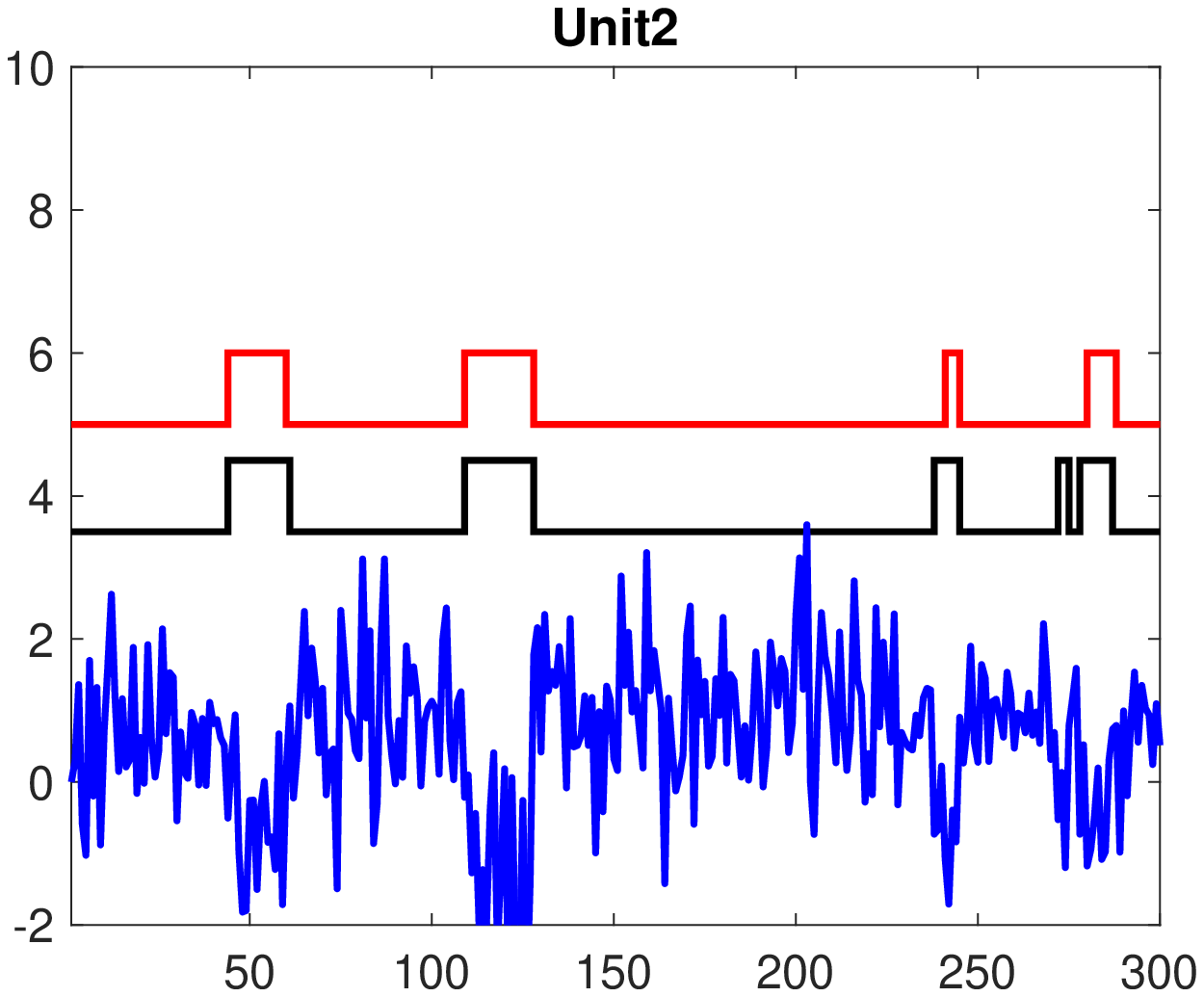}&
\includegraphics[scale=0.18]{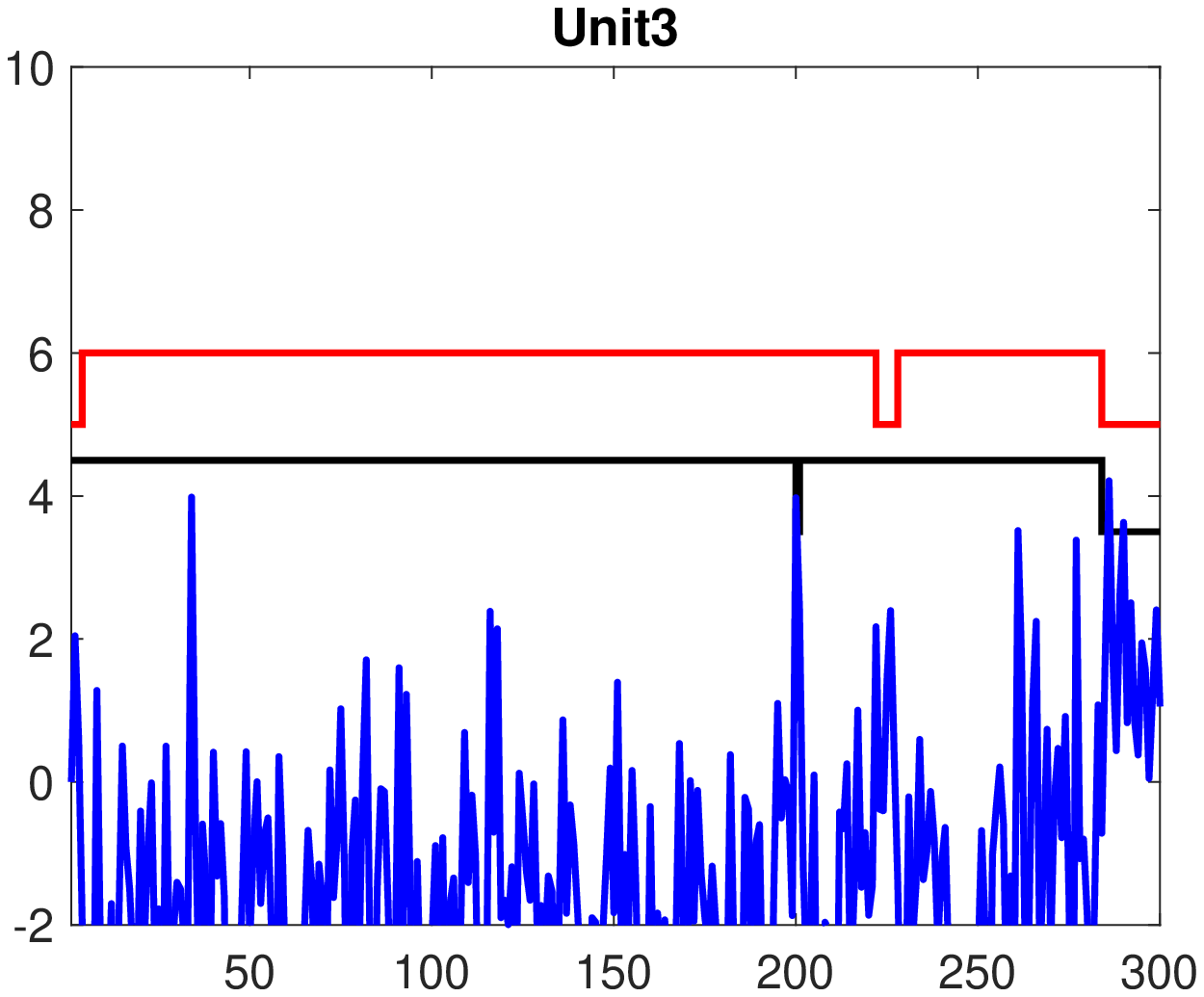}&
\includegraphics[scale=0.18]{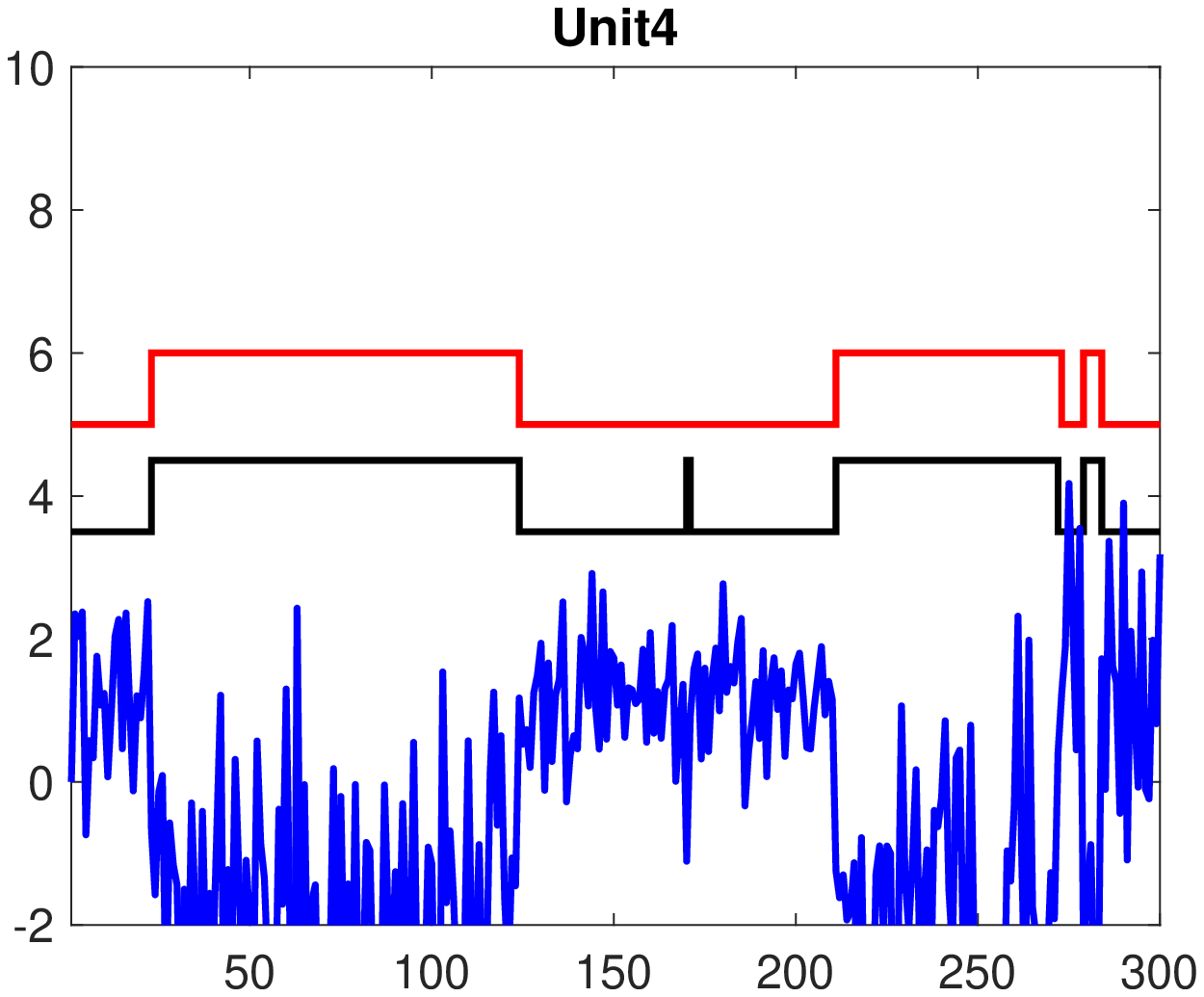}&
\includegraphics[scale=0.18]{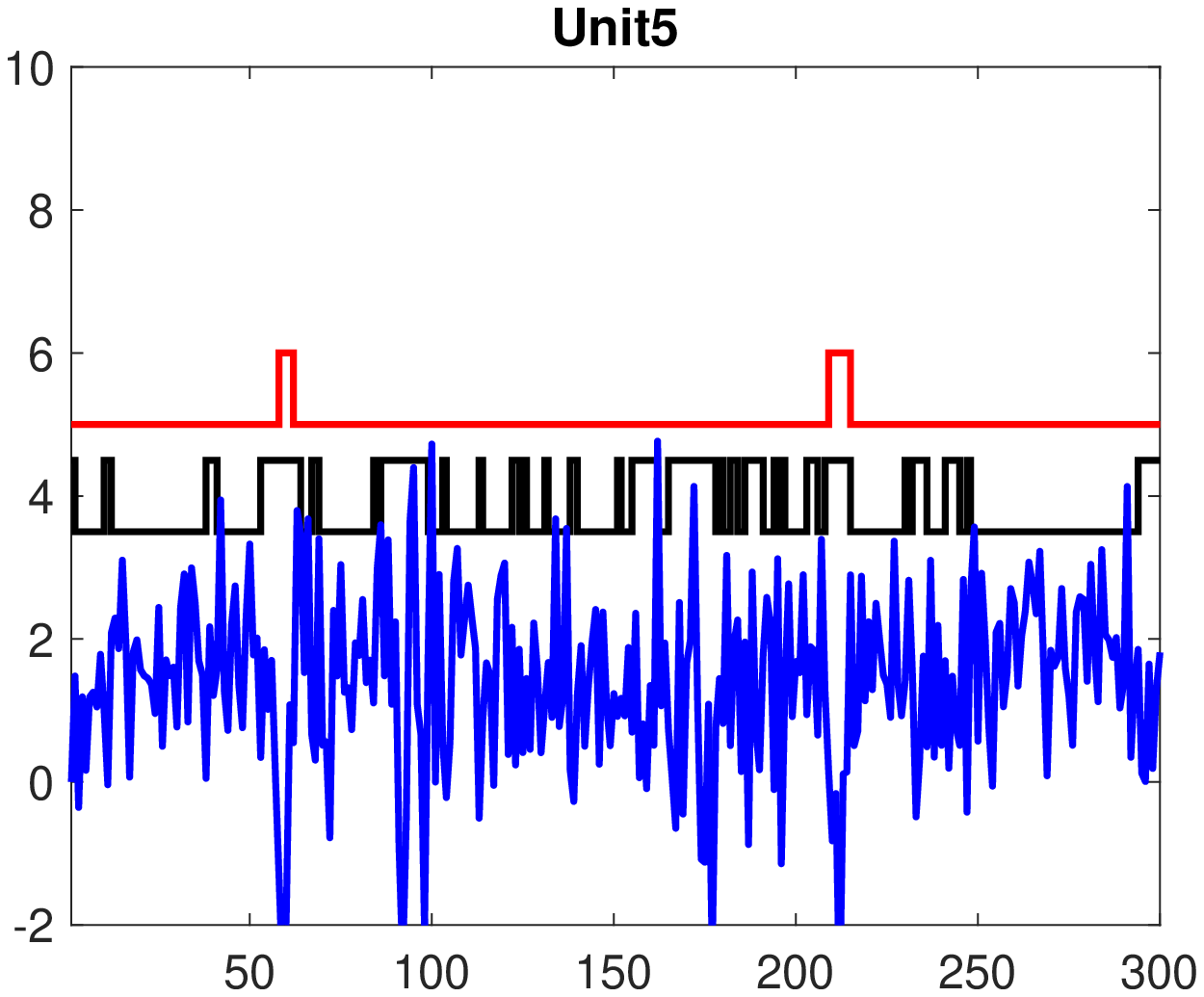}
\\\\\
\includegraphics[scale=0.18]{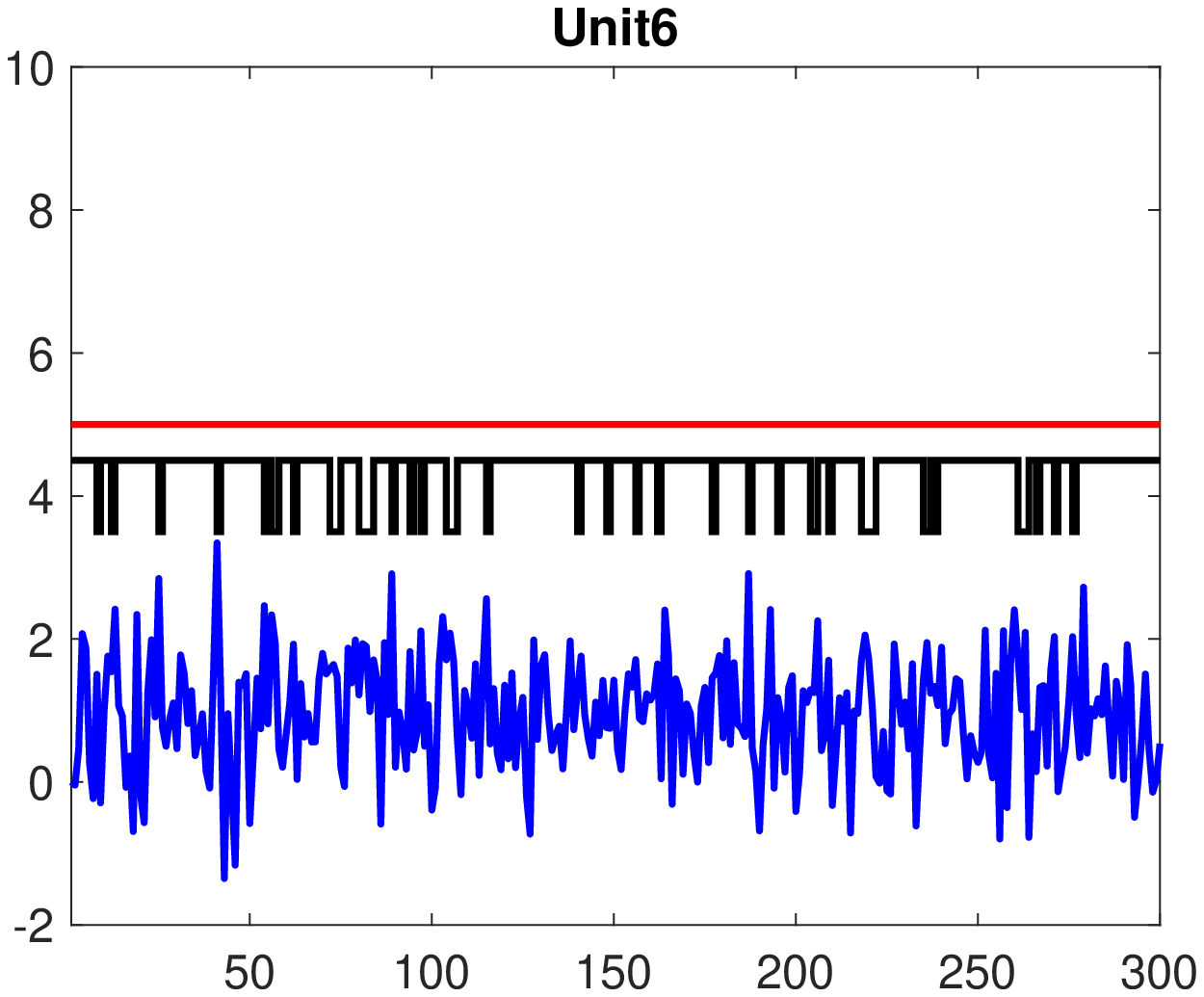}&
\includegraphics[scale=0.18]{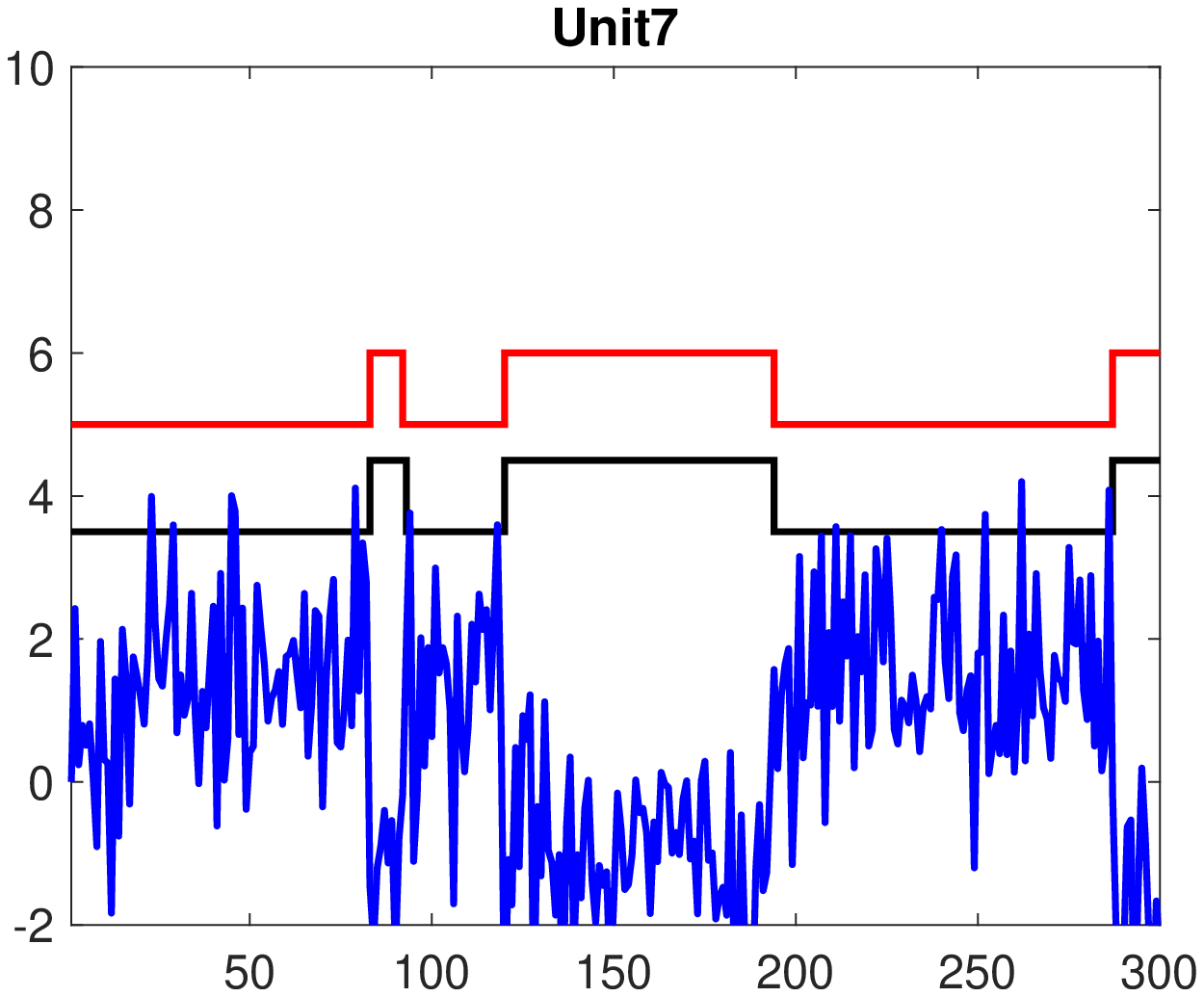}&
\includegraphics[scale=0.18]{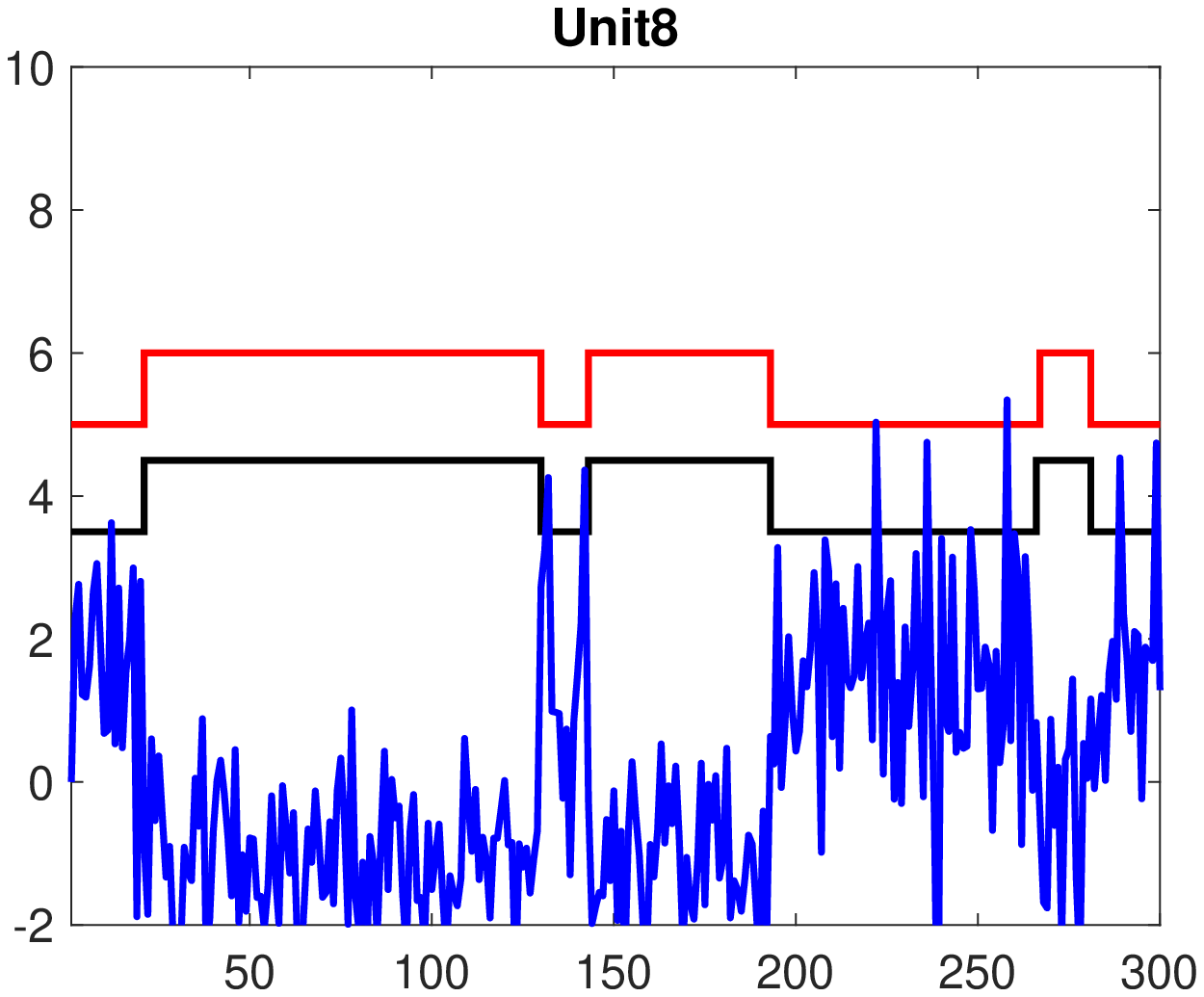}&
\includegraphics[scale=0.18]{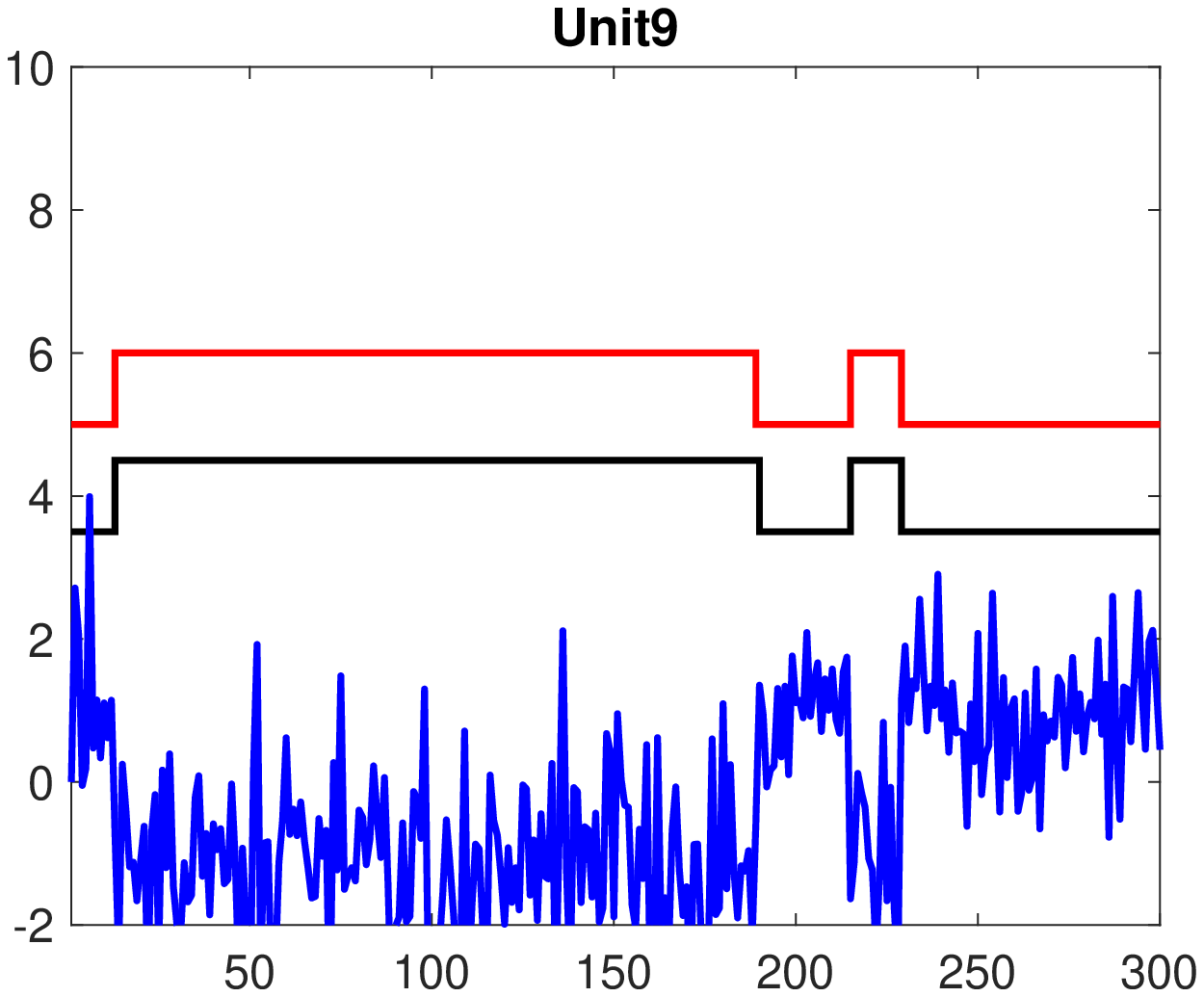}&
\includegraphics[scale=0.18]{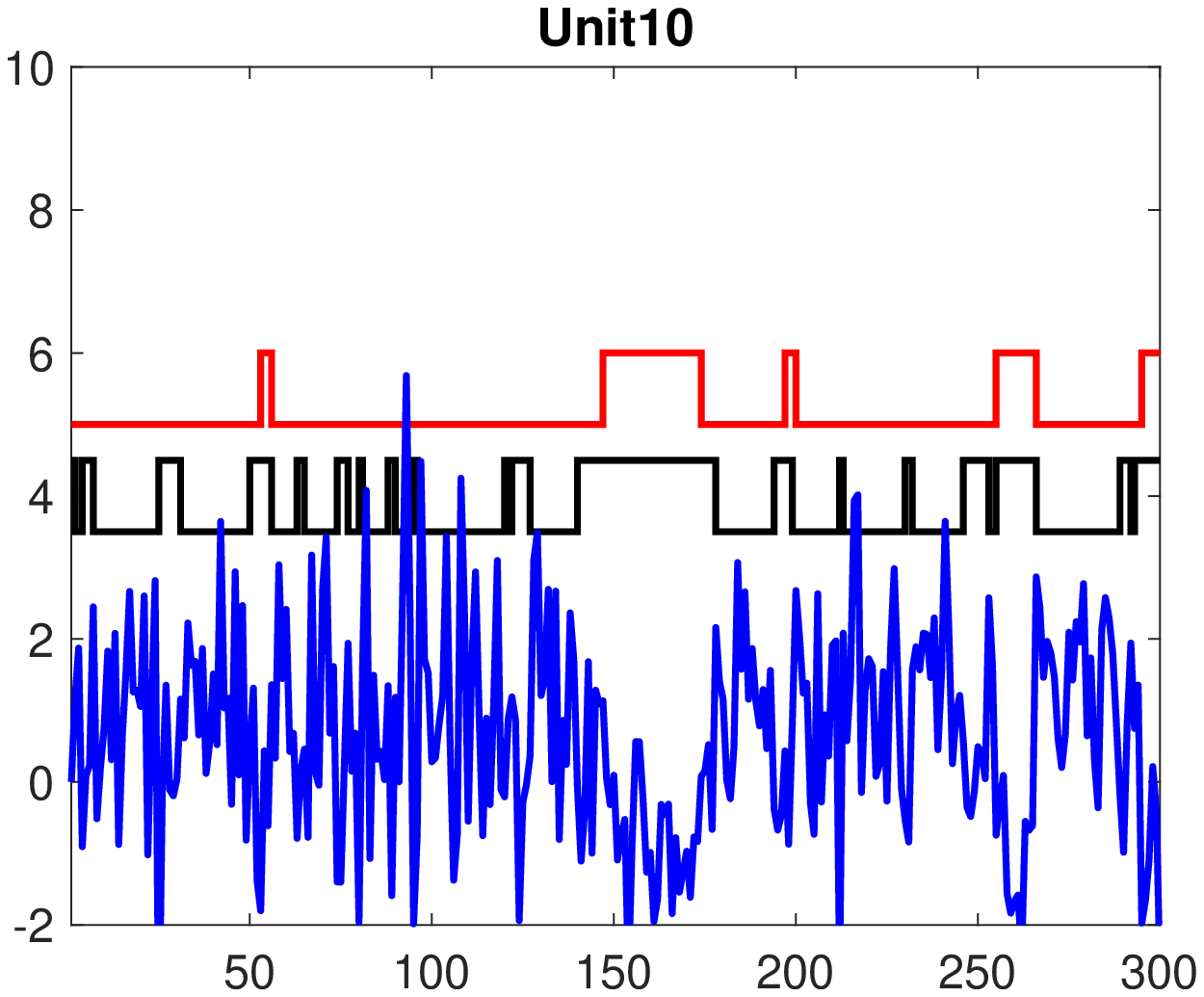}
\\\\\
\includegraphics[scale=0.18]{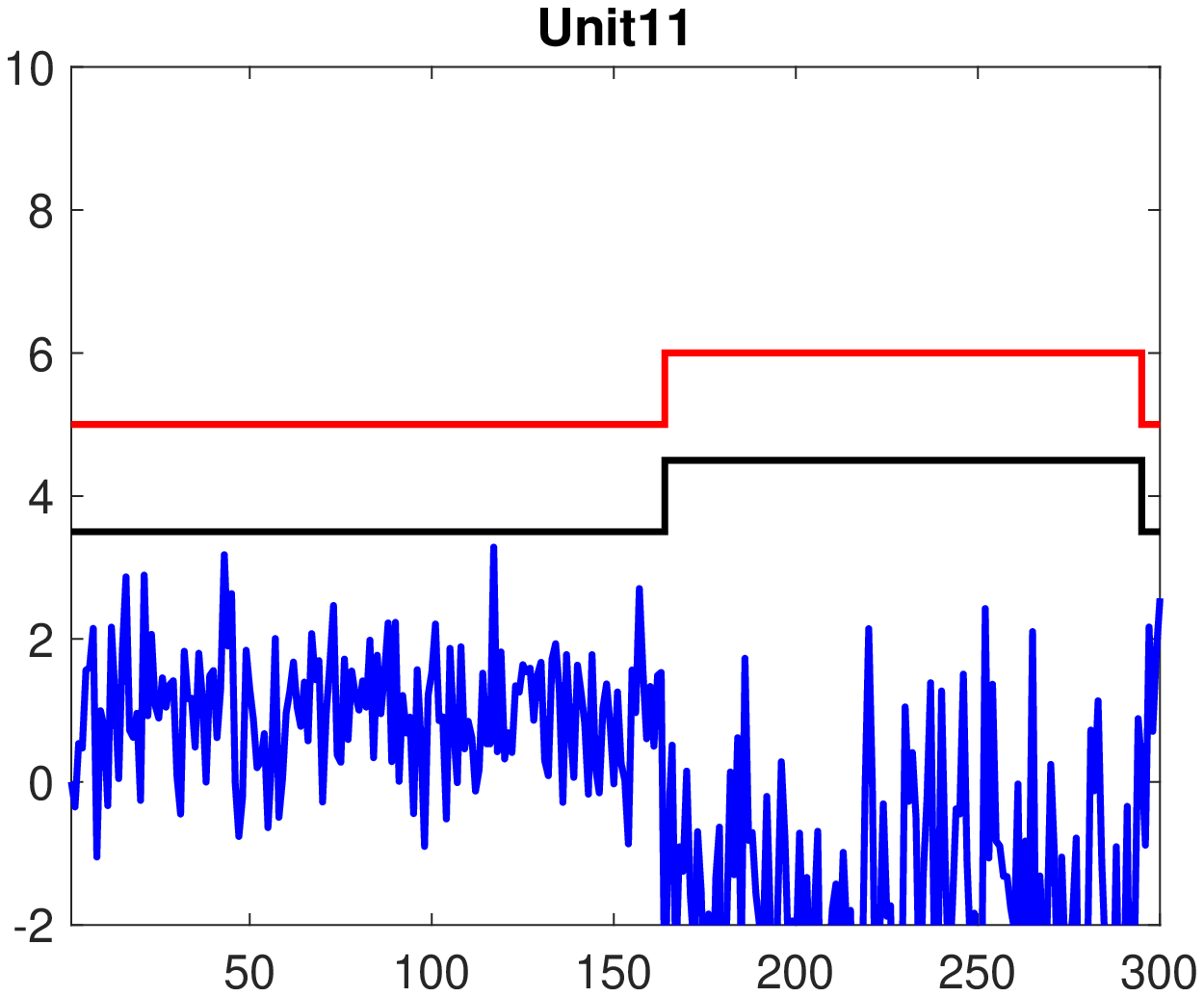}&
\includegraphics[scale=0.18]{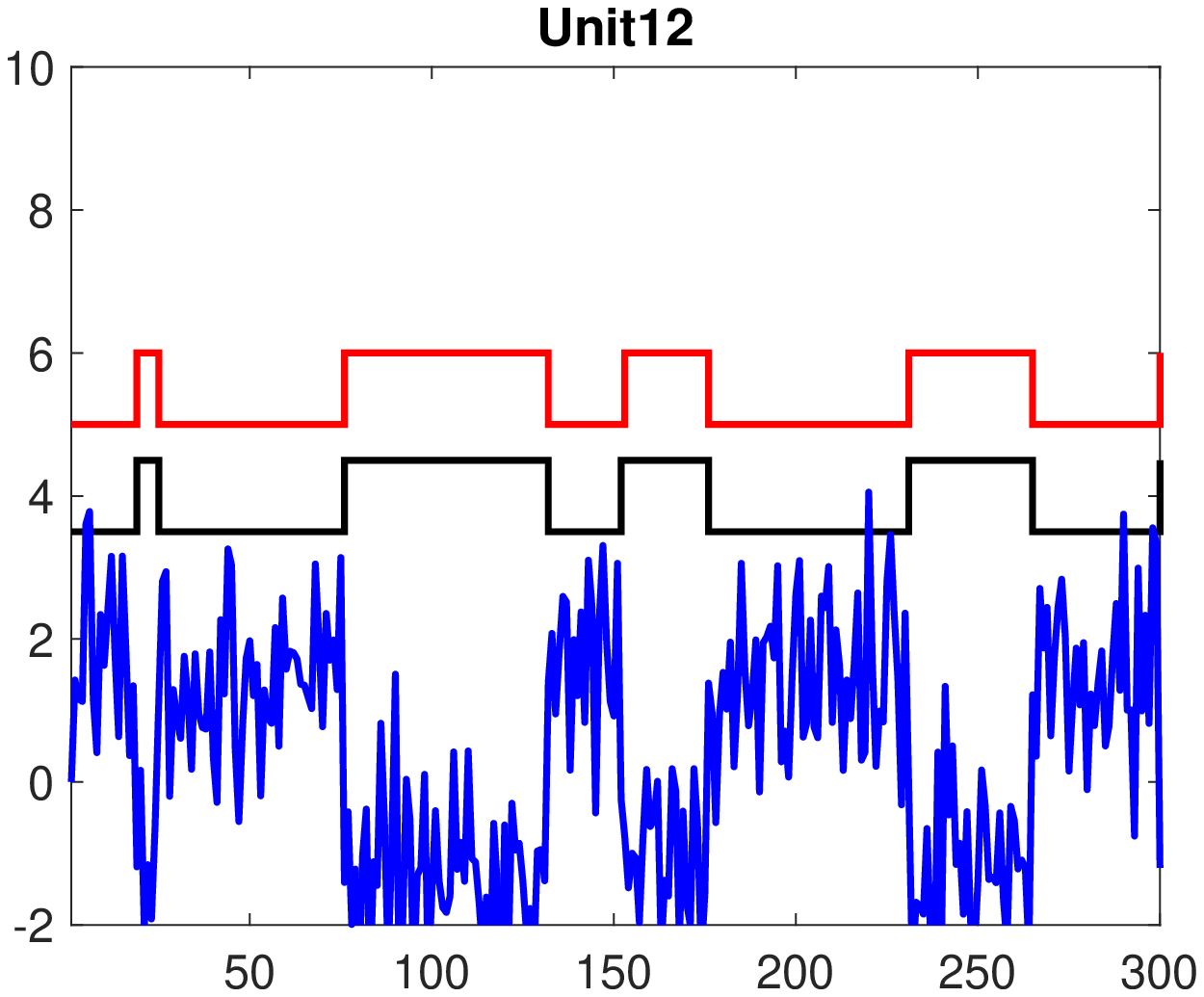}&
\includegraphics[scale=0.18]{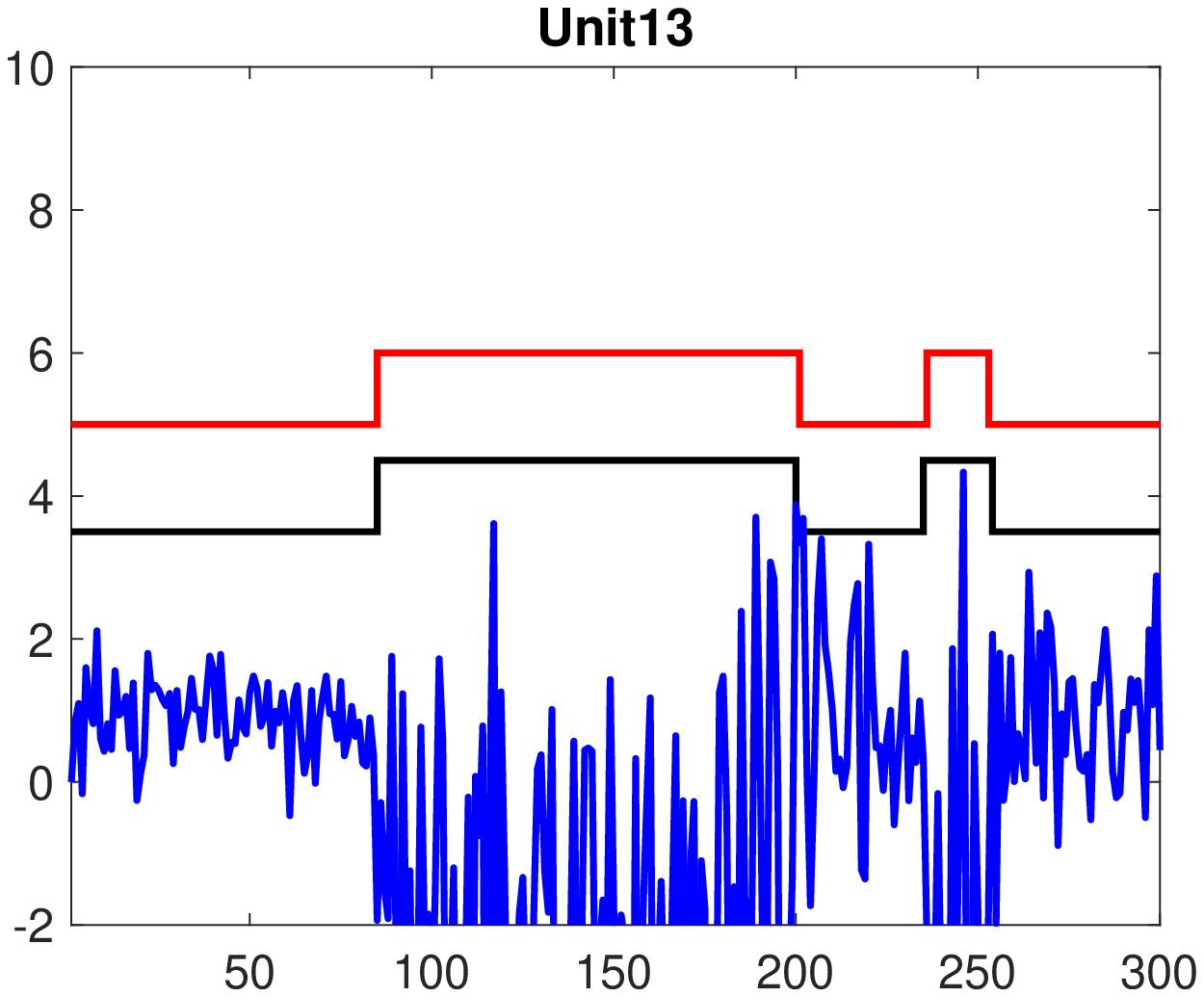}&
\includegraphics[scale=0.18]{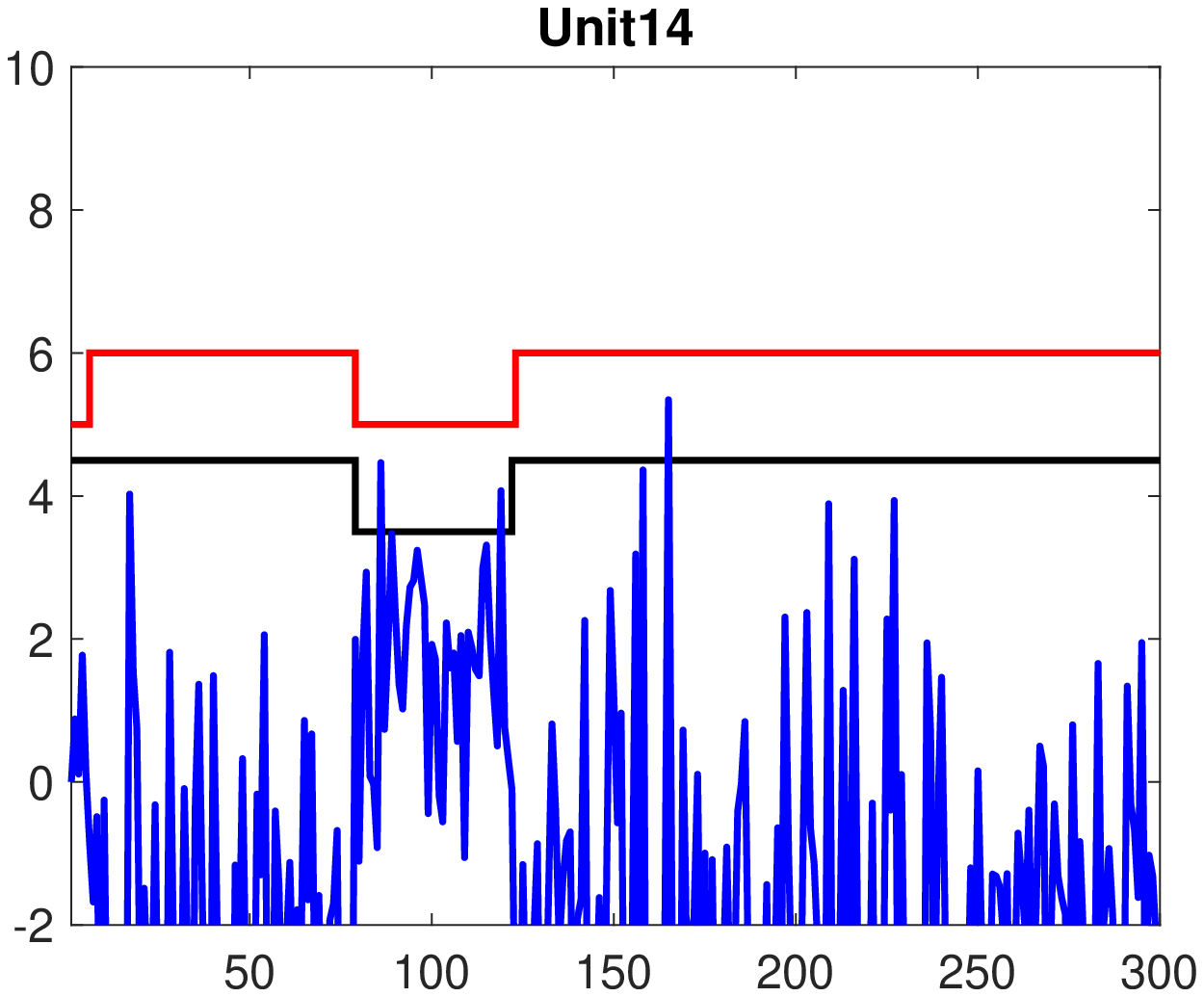}&
\includegraphics[scale=0.18]{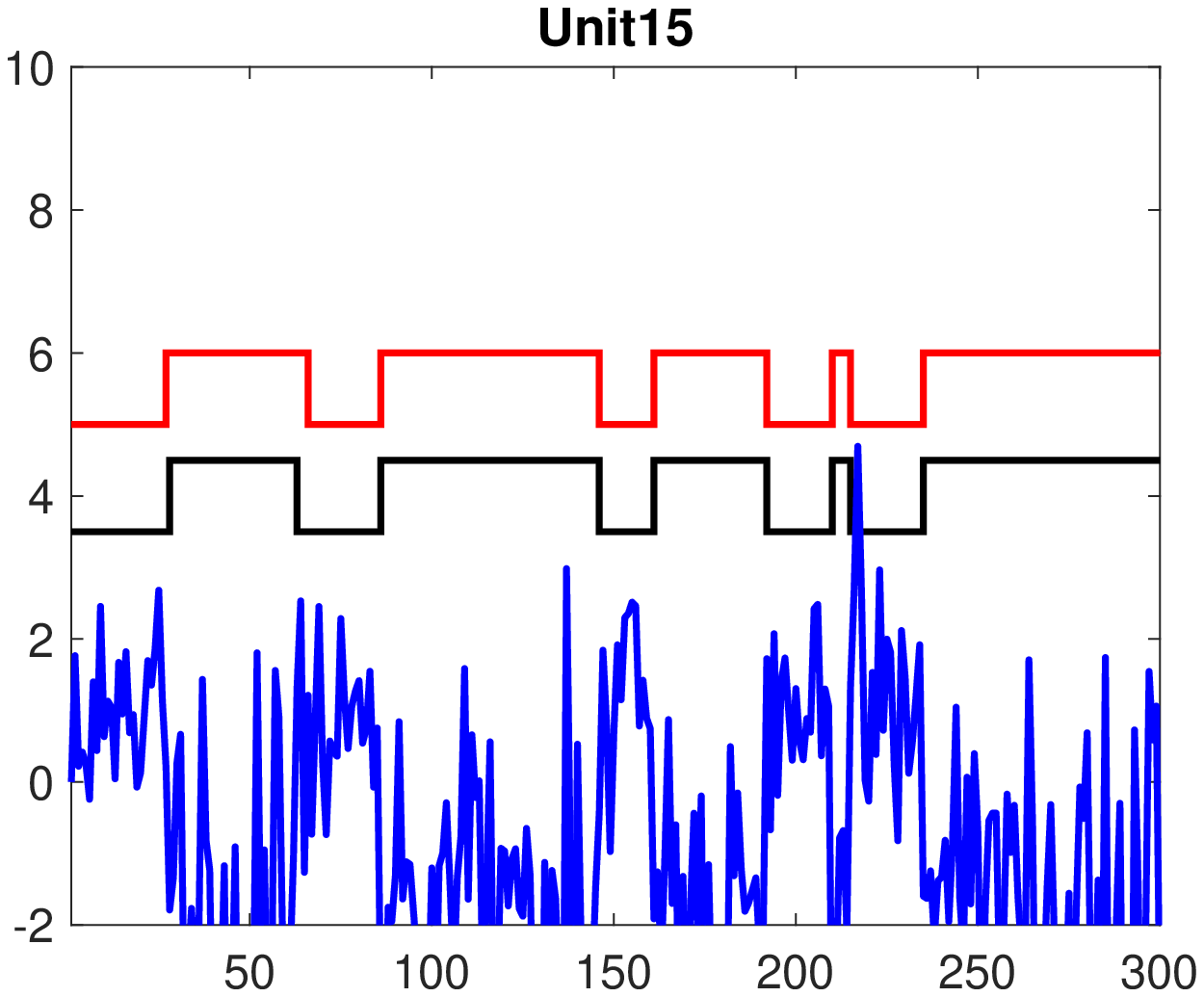}
\\\\\
\includegraphics[scale=0.18]{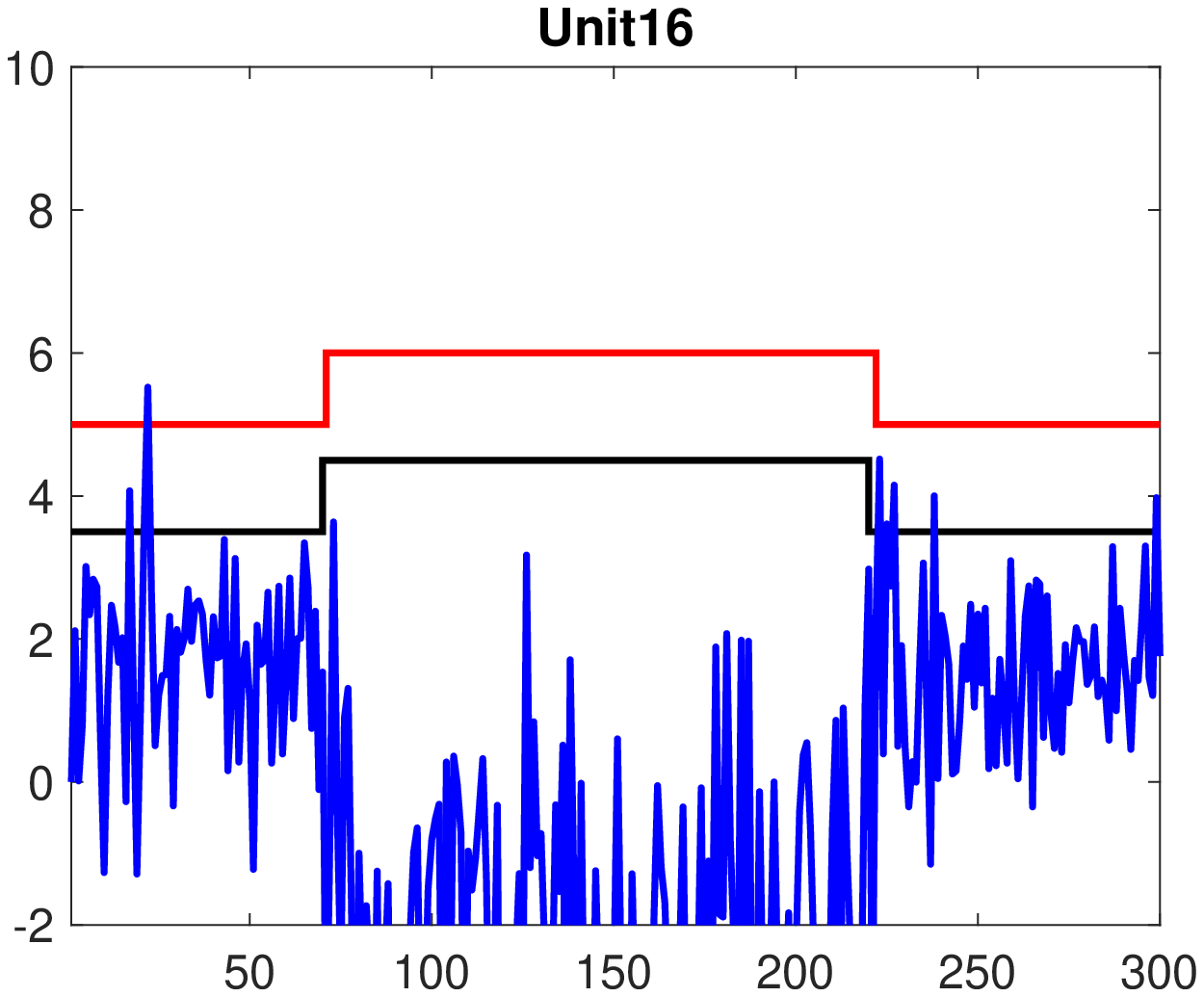}&
\includegraphics[scale=0.18]{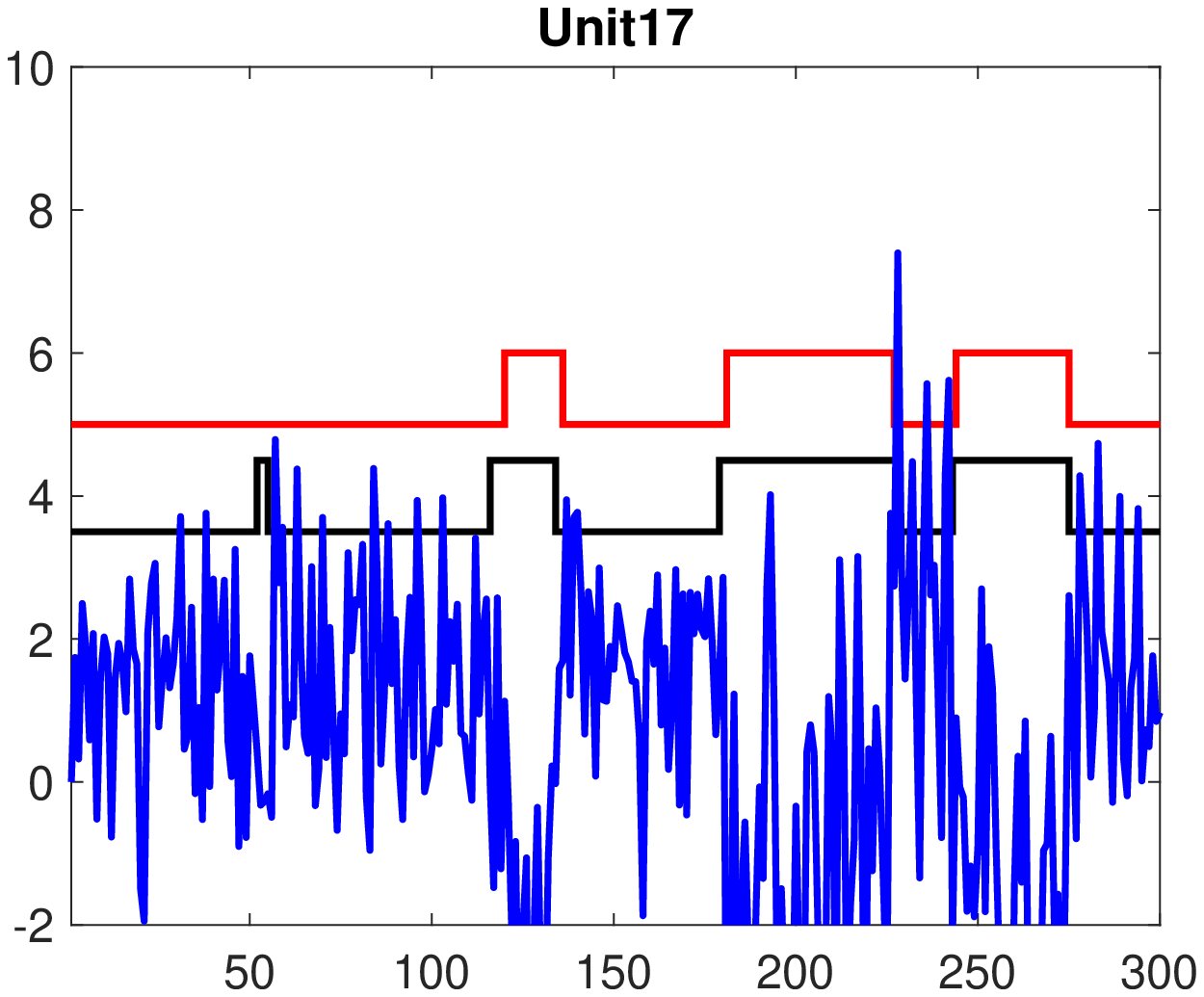}&
\includegraphics[scale=0.18]{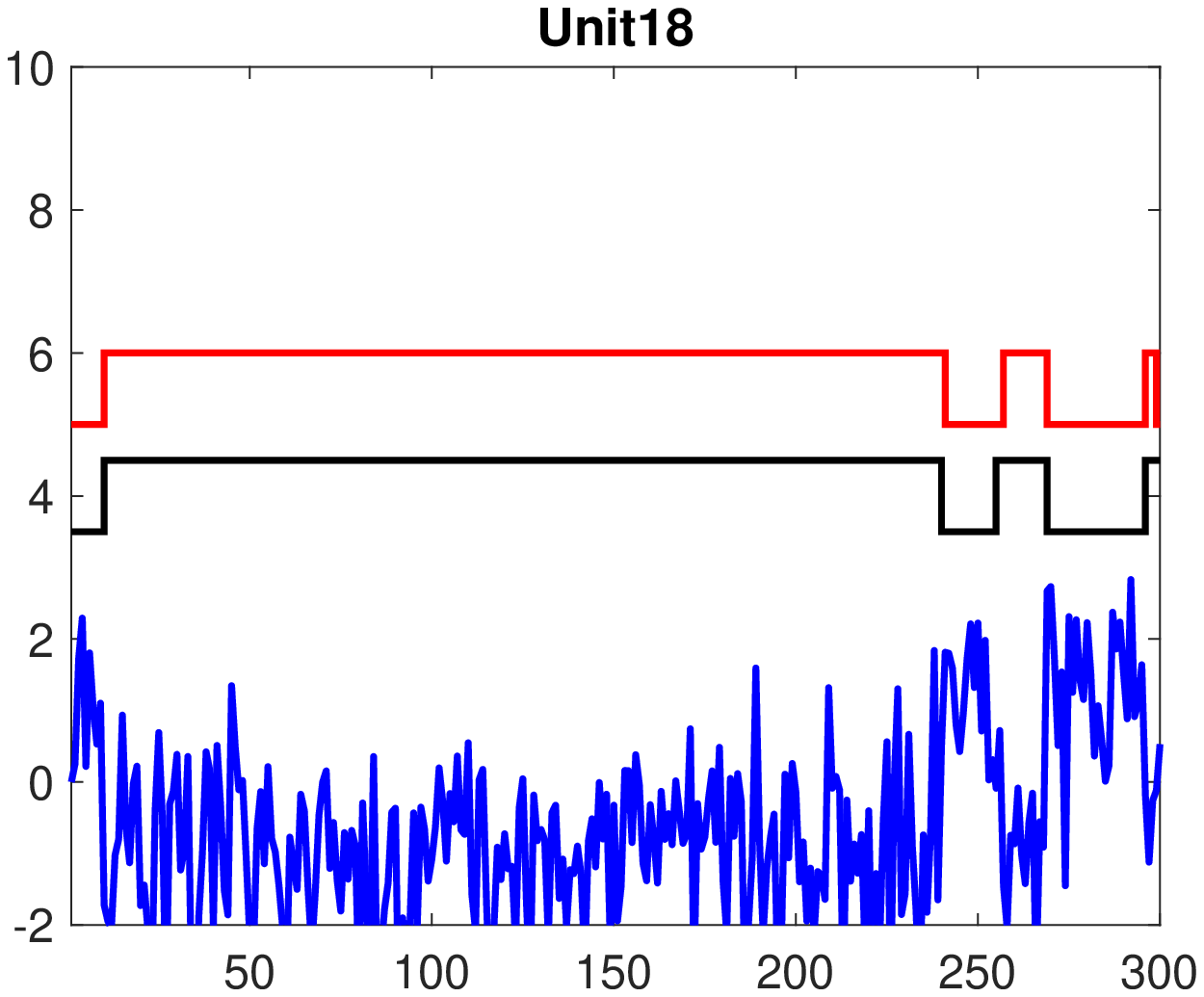}&
\includegraphics[scale=0.18]{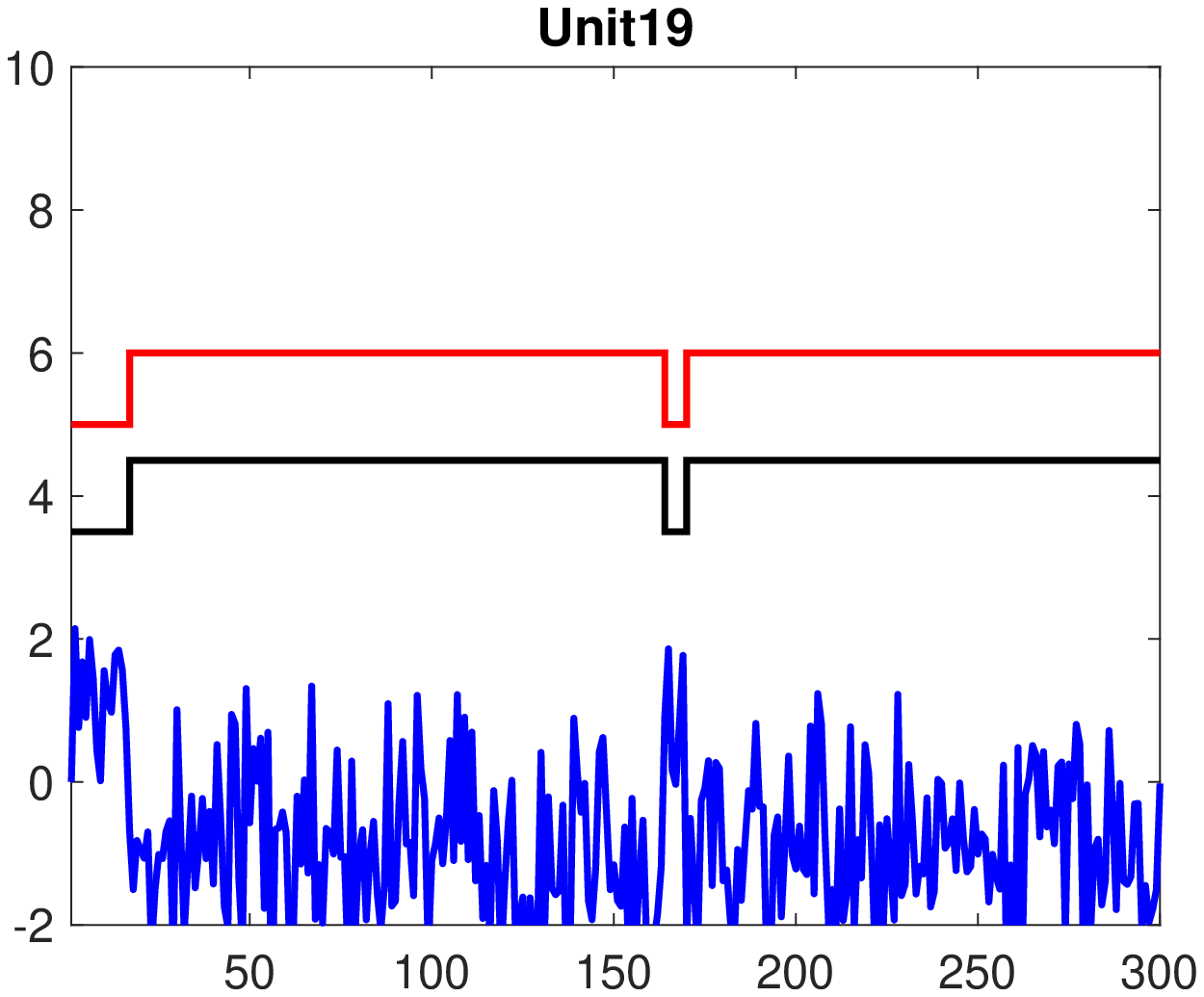}&
\includegraphics[scale=0.18]{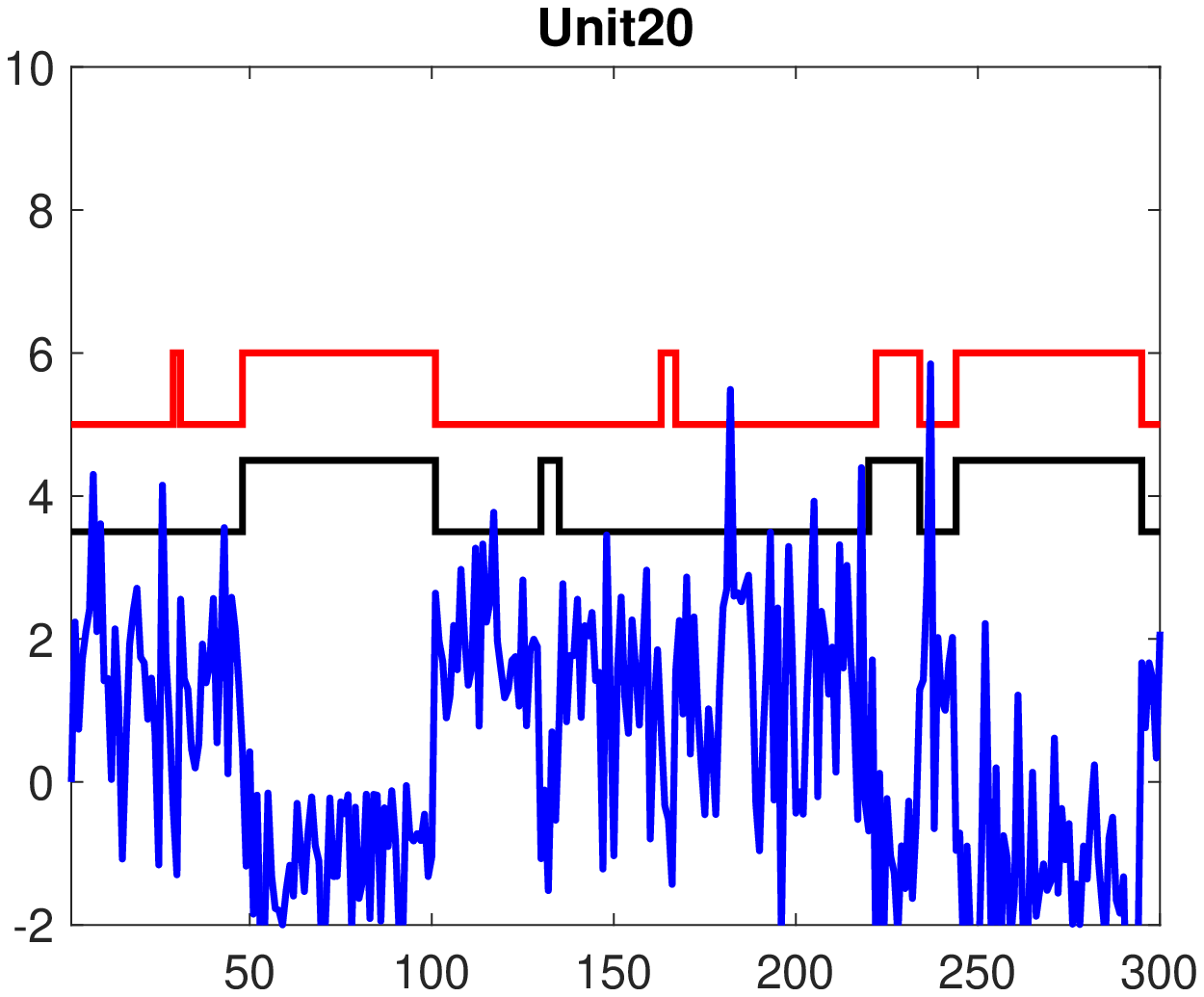}
\\\\\
\includegraphics[scale=0.18]{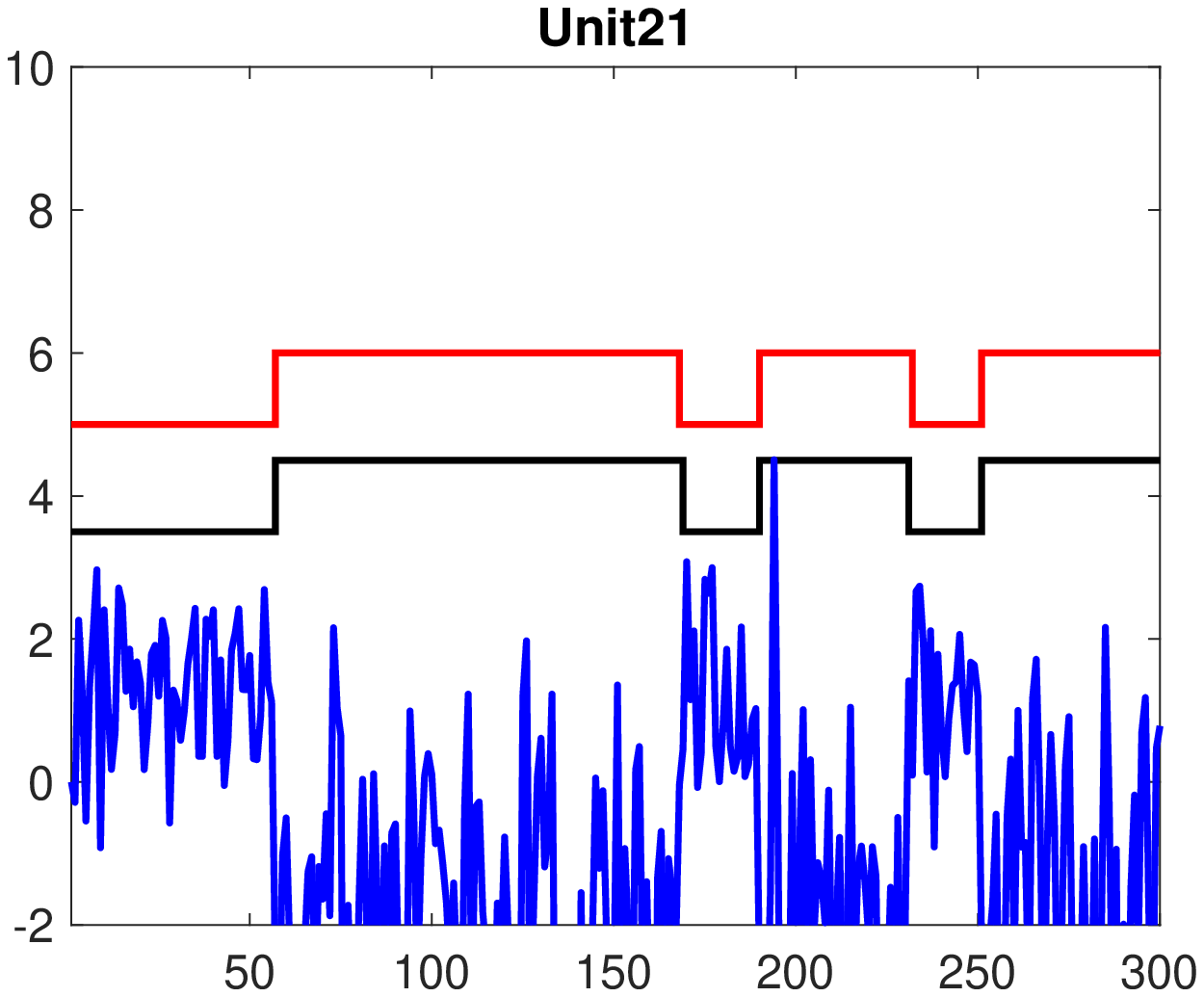}&
\includegraphics[scale=0.18]{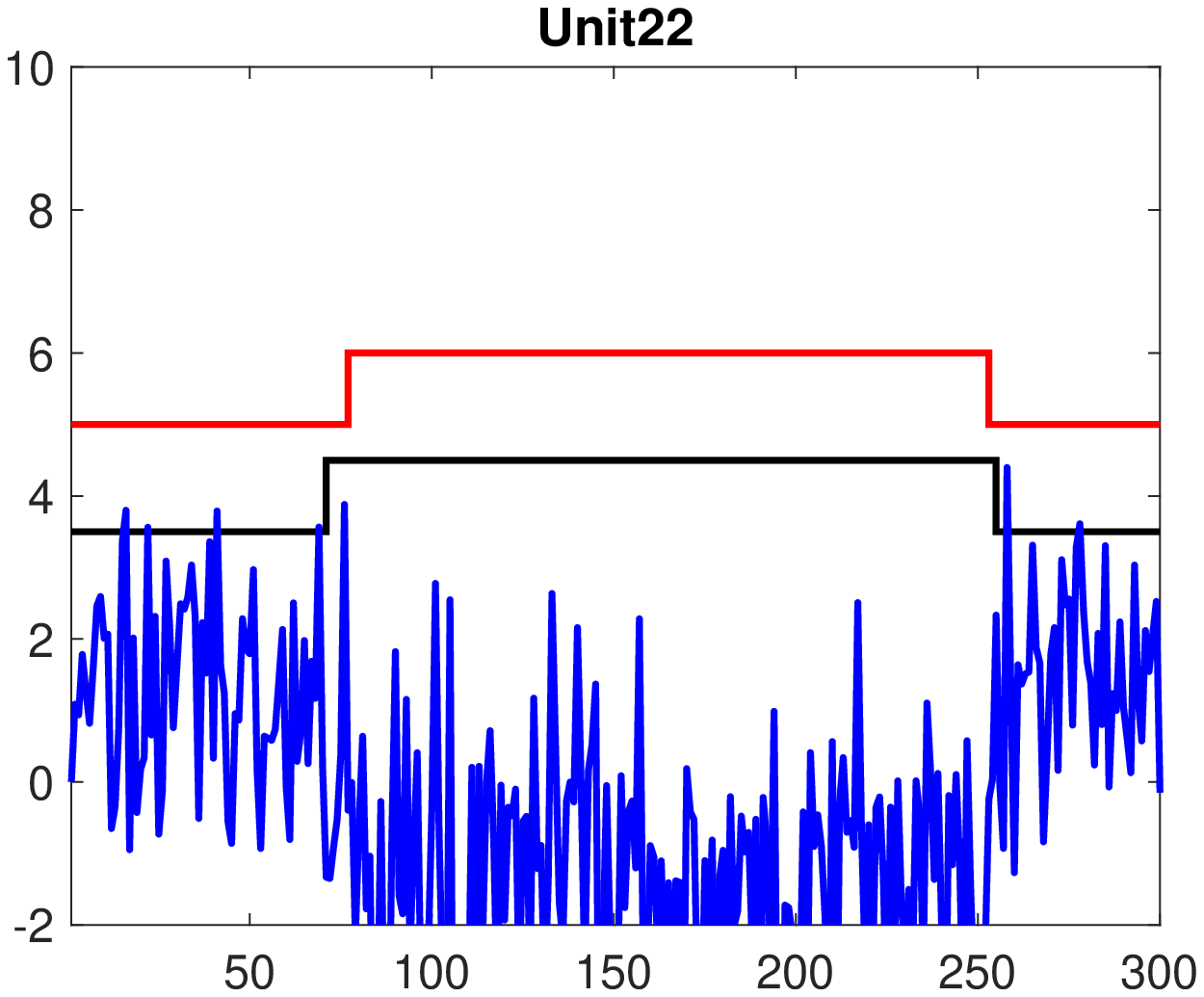}&
\includegraphics[scale=0.18]{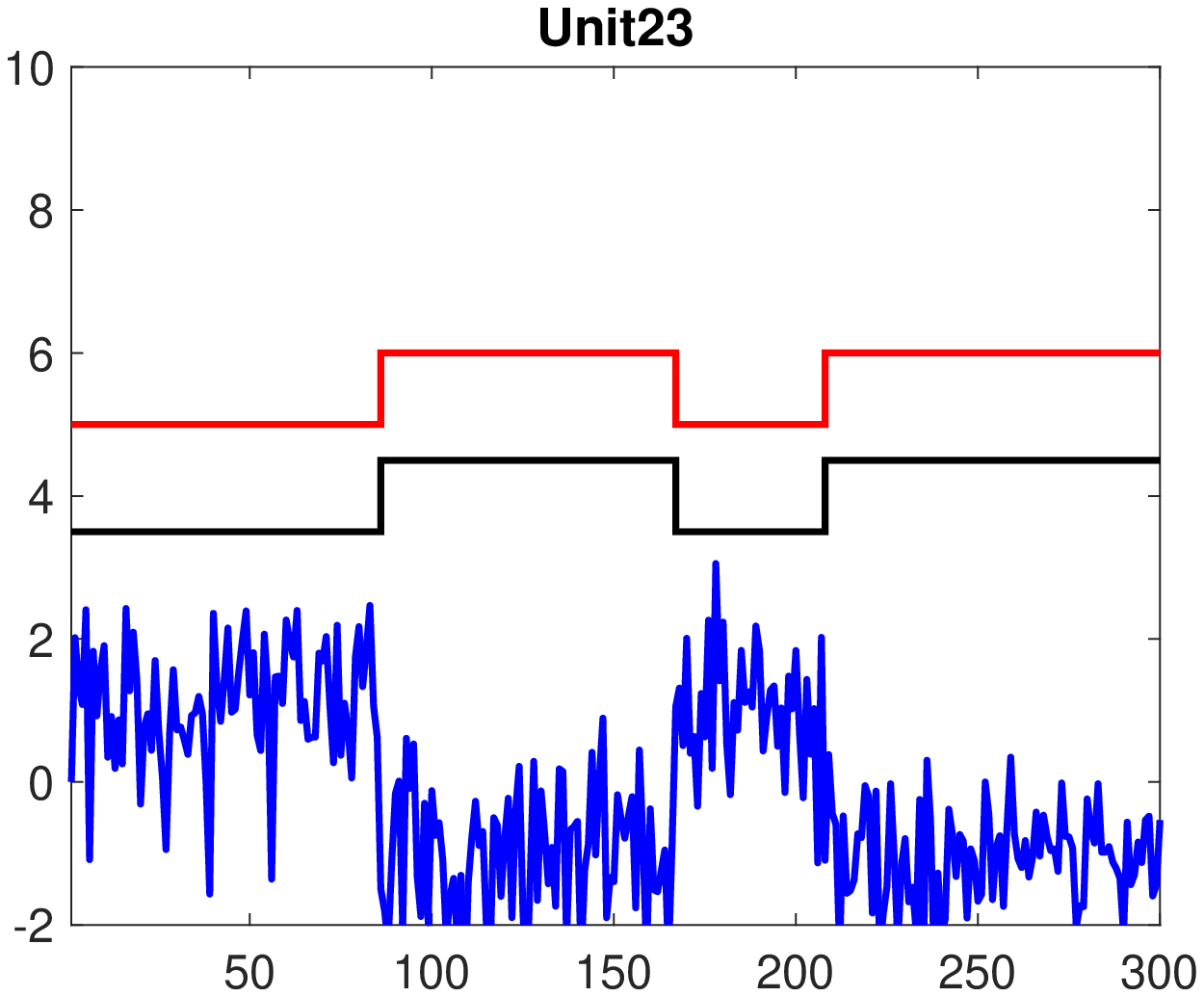}&
\includegraphics[scale=0.18]{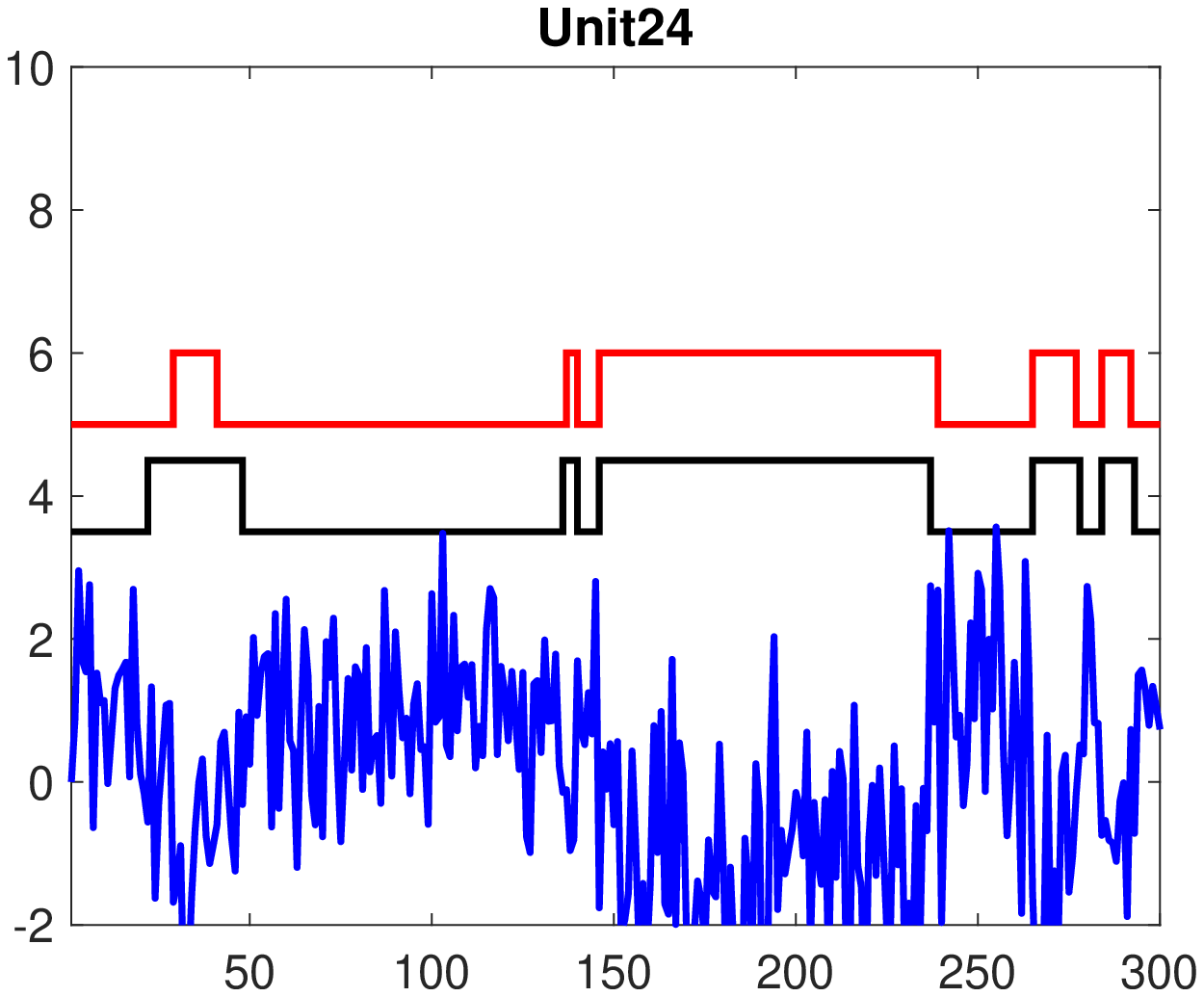}&
\includegraphics[scale=0.18]{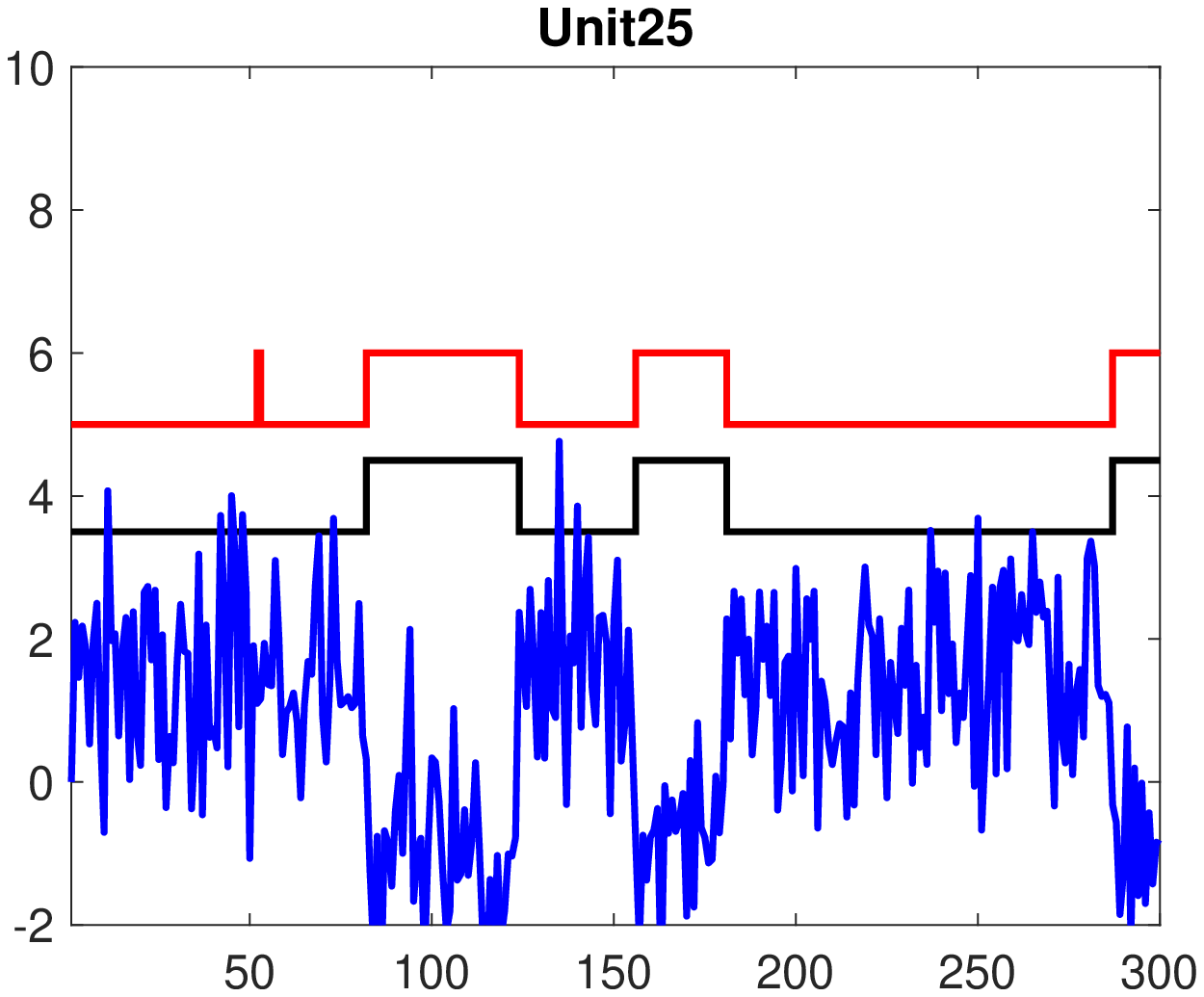}
\\\\\
\includegraphics[scale=0.18]{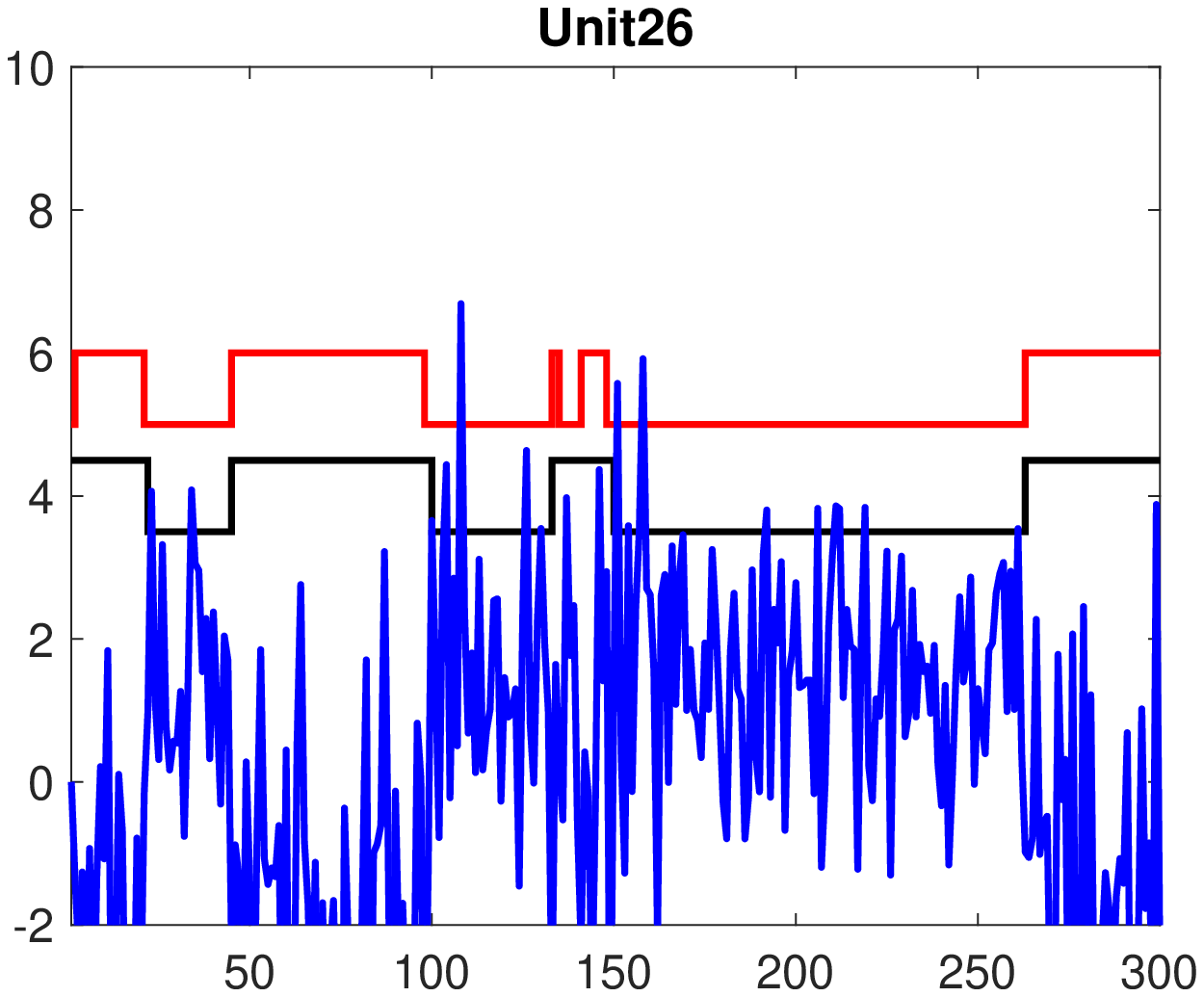}&
\includegraphics[scale=0.18]{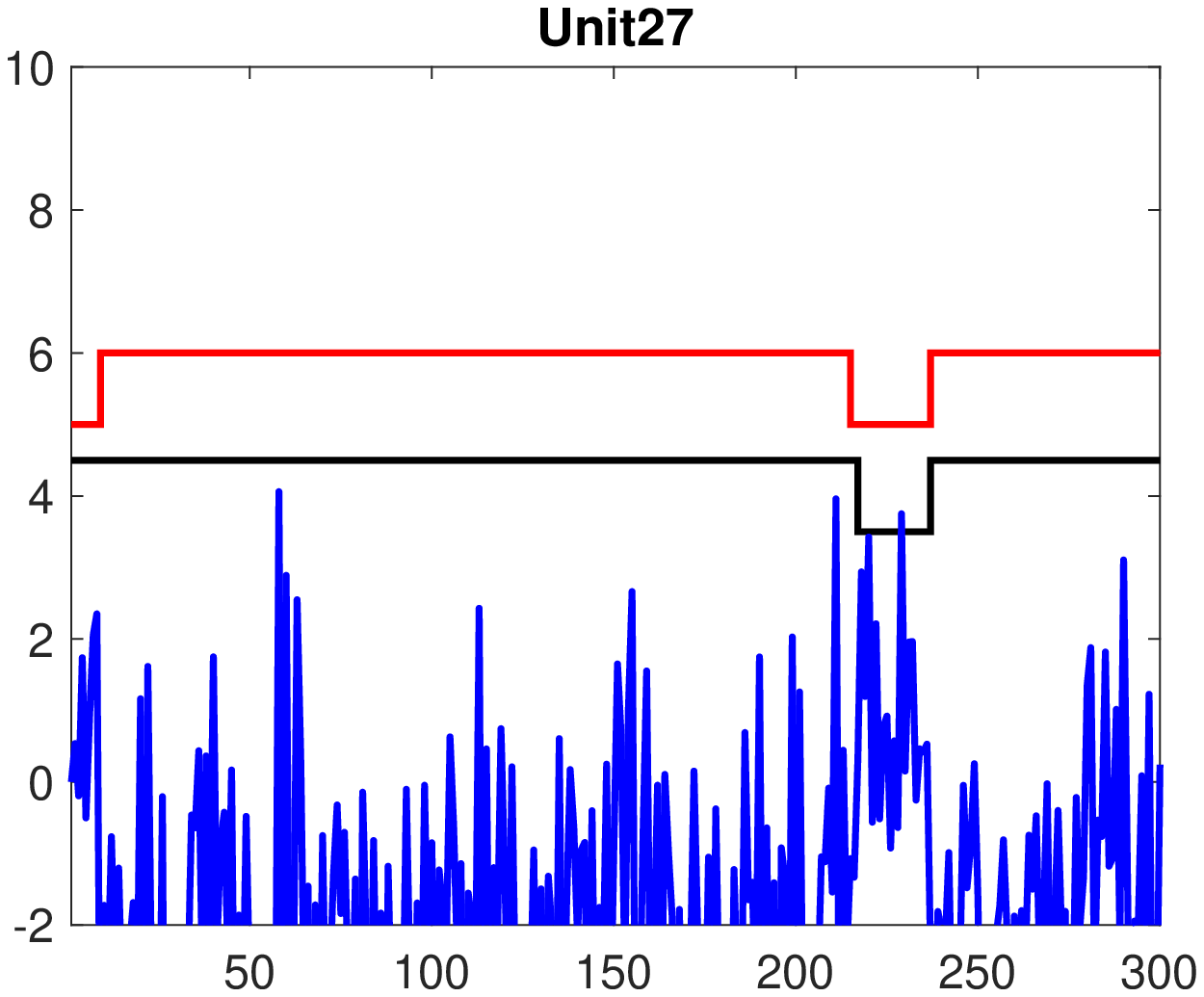}&
\includegraphics[scale=0.18]{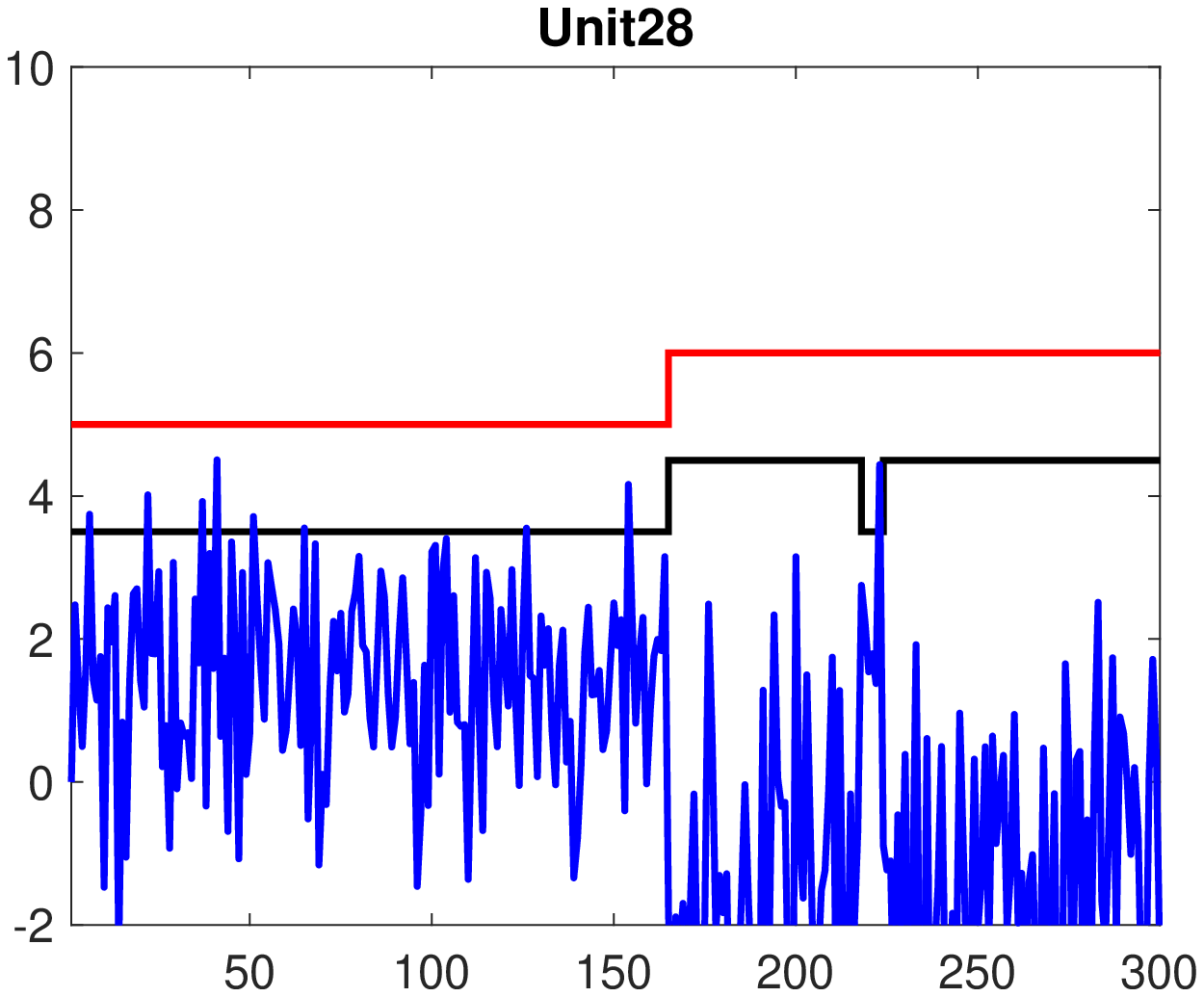}&
\includegraphics[scale=0.18]{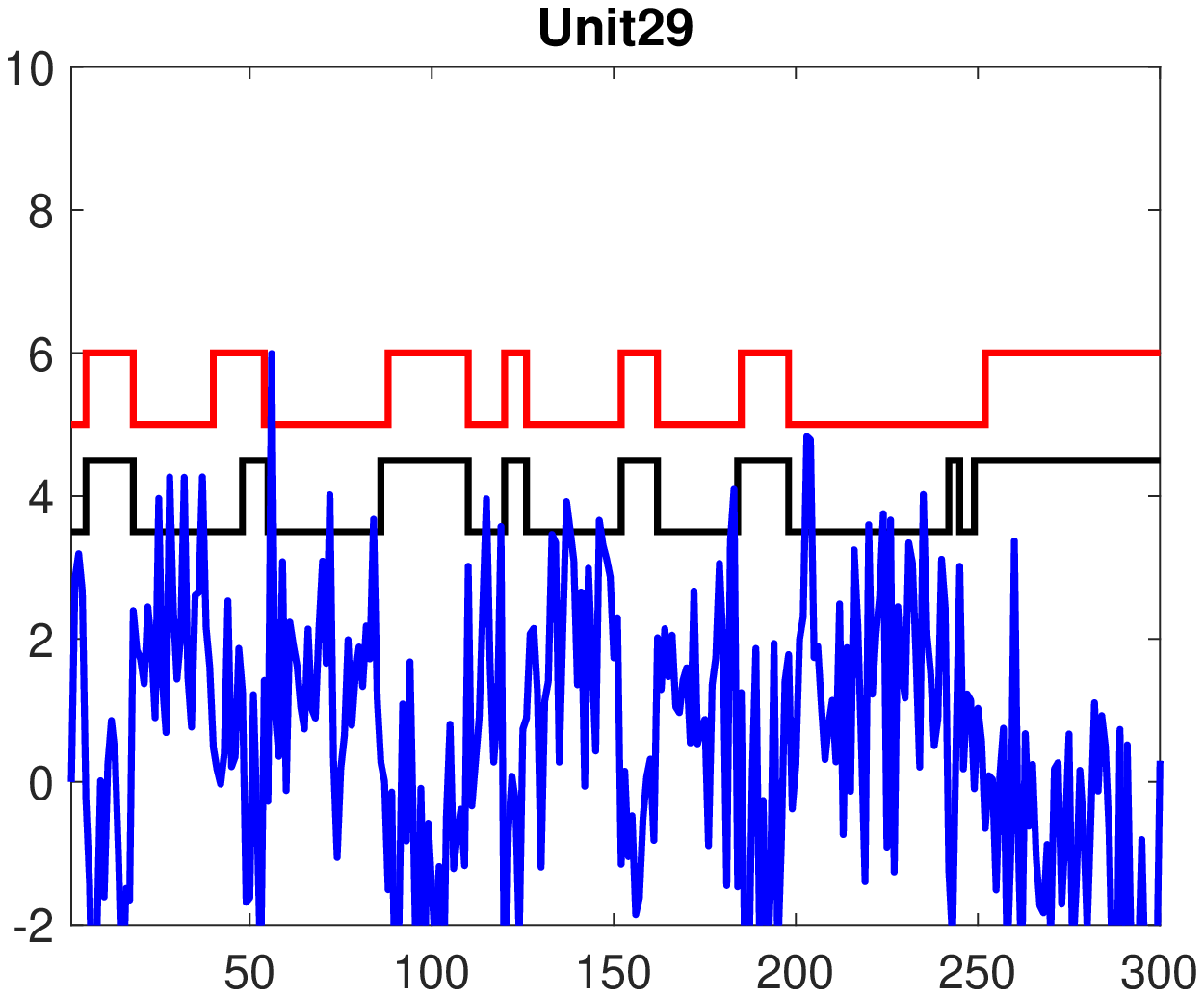}&
\includegraphics[scale=0.18]{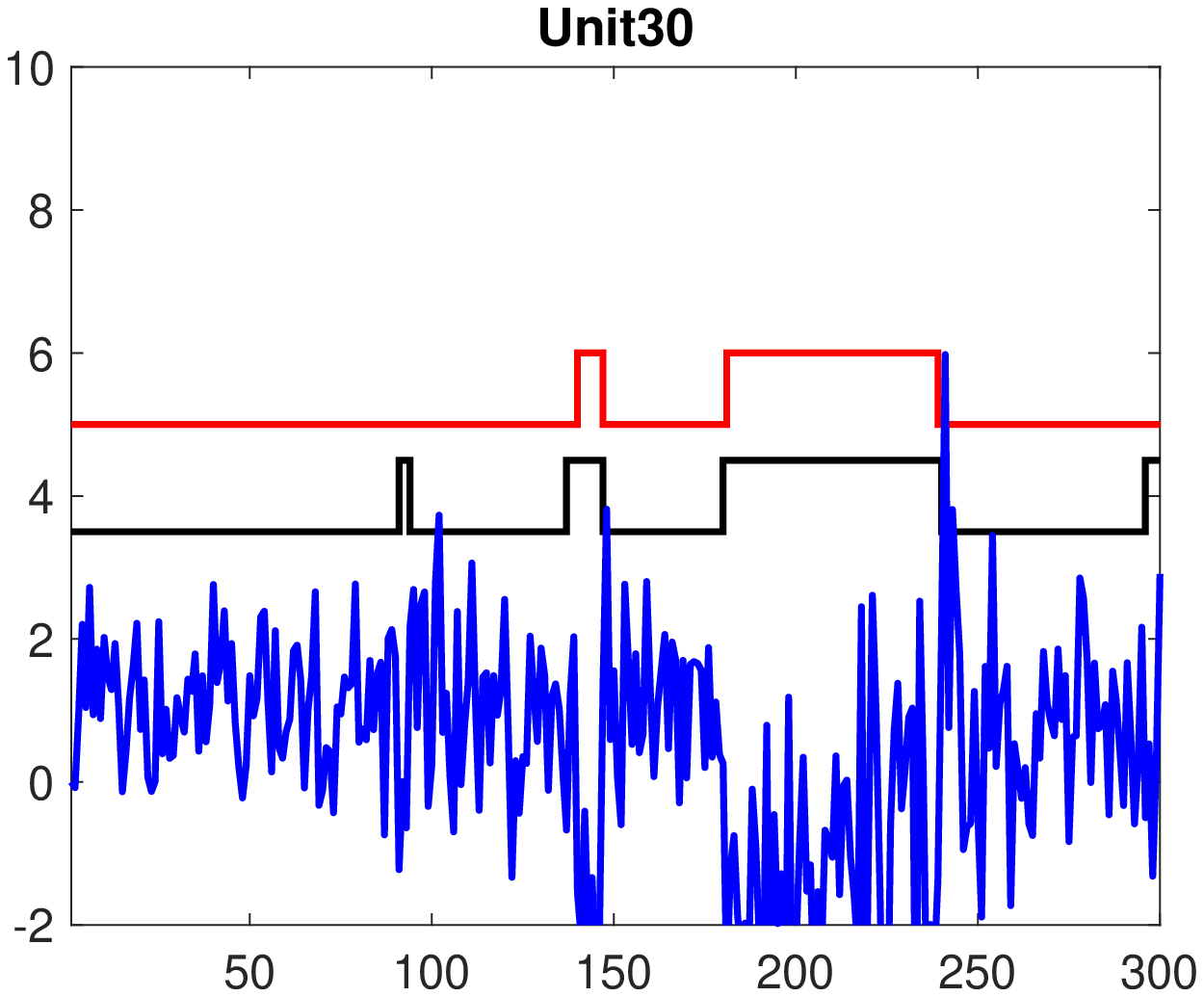}
\end{tabular}
\end{center}
\caption{In each plot, the true (red) and estimated (black) trajectories of the hidden Markov chain process and the observed process $y_{it}$ (blue line).}
\end{figure}

\pagebreak

$\,$

\vfill

\newpage

\section{Further details on the empirical application}\label{emp}
%Figure \label{ret} report the percentage log-returns from the $3^{rd}$ August 2000 to the $3^{rd}$ 

\begin{table}[h!]
\caption{\small Constituents of the S\&P100 at the $1^{st}$ October 2020. In the columns, the company label (Symbol), name (Name) and sector (S). Sector codes are: Mining and Quarrying (B), Financial and Insurance Activities (K), Information and Communication (J), Manufacturing (C), Real Estates Activities (L), Wholesale and Retail Trade; Repair of Motors (G), Accommodation and Food Service (I), Electricity Gas Steam and Air Cond. (D), Transp. and Storage (H), Professional Scientific and Technical Activities (M). The columns C indicates if a company is included in (1) in the analysis.}
\begin{center}
\vspace{-15pt}
\resizebox{\textwidth}{!}{
\begin{scriptsize}
\begin{tabular}{llll|llll}
\hline
Symbol & Name & S &C&Symbol & Name & S &C\\
\hline
    OXY   & Occidental Petroleum Corp & B     & 1     & COST  & Costco Wholesale Corp & G     & 1 \\
    COP   & ConocoPhillips & B     & 1     & TGT   & Target Corp & G     & 1 \\
    SLB   & Schlumberger NV & B     & 1     & LOW   & Lowe's Cos Inc & G     & 1 \\
    MDLZ  & Mondelez Int Inc & C     & 0     & CVS   & CVS Health Corp & G     & 1 \\
    BA    & Boeing Co/The & C     & 1     & UNP   & Union Pacific Corp & H     & 1 \\
    CAT   & Caterpillar Inc & C     & 1     & KMI   & Kinder Morgan Inc & H     & 0 \\
    CVX   & Chevron Corp & C     & 1     & FDX   & FedEx Corp & H     & 1 \\
    KO    & Coca-Cola Co/The & C     & 1     & UPS   & United Parcel Service Inc & H     & 1 \\
    XOM   & Exxon Mobil Corp & C     & 1     & MCD   & McDonald's Corp & I     & 1 \\
    GE    & General Electric Co & C     & 1     & SBUX  & Starbucks Corp & I     & 1 \\
    JNJ   & Johnson \& Johnson & C     & 1     & VZ    & Verizon Communications Inc & J     & 1 \\
    MRK   & Merck \& Co Inc & C     & 1     & DIS   & Walt Disney Co/The & J     & 1 \\
    MMM   & 3M Co & C     & 1     & IBM   & Int Business Machines Corp & J     & 1 \\
    PFE   & Pfizer Inc & C     & 1     & ACN   & Accenture PLC & J     & 0 \\
    PG    & Procter \& Gamble Co/The & C     & 1     & GOOG  & Alphabet Inc & J     & 0 \\
    RTX   & Raytheon Technologies Corp & C     & 1     & T     & AT\&T Inc & J     & 1 \\
    CSCO  & Cisco Systems Inc & C     & 1     & CHTR  & Charter Communications Inc & J     & 0 \\
    INTC  & Intel Corp & C     & 1     & MSFT  & Microsoft Corp & J     & 1 \\
    NVDA  & NVIDIA Corp & C     & 1     & BKNG  & Booking Holdings Inc & J     & 1 \\
    HON   & Honeywell Int Inc & C     & 1     & GOOGL  & Alphabet Inc & J     & 0 \\
    MO    & Altria Group Inc & C     & 1     & NFLX  & Netflix Inc & J     & 0 \\
    ABT   & Abbott Laboratories & C     & 1     & CRM   & salesforce.com Inc & J     & 0 \\
    TXN   & Texas Instruments Inc & C     & 0     & ADBE  & Adobe Inc & J     & 1 \\
    KHC   & Kraft Heinz Co/The & C     & 0     & CMCSA  & Comcast Corp & J     & 1 \\
    TMO   & Thermo Fisher Scientific Inc & C     & 1     & ORCL  & Oracle Corp & J     & 0 \\
    PM U  & Philip Morris International Inc & C     & 0     & FB    & Facebook Inc & J     & 0 \\
    BMY   & Bristol Myers Squibb Co & C     & 1     & AXP   & American Express Co & K     & 1 \\
    AAPL  & Apple Inc & C     & 1     & JPM   & JPMorgan Chase \& Co & K     & 1 \\
    CL    & Colgate-Palmolive Co & C     & 1     & BAC   & Bank of America Corp & K     & 1 \\
    ABBV  & AbbVie Inc & C     & 0     & C     & Citigroup Inc & K     & 1 \\
    DHR   & Danaher Corp & C     & 1     & AIG   & American International Group Inc & K     & 1 \\
    DOW   & Dow Inc & C     & 0     & GS    & Goldman Sachs Group Inc/The & K     & 1 \\
    GM    & General Motors Co & C     & 0     & UNH   & UnitedHealth Group Inc & K     & 1 \\
    EMR   & Emerson Electric Co & C     & 1     & BLK   & BlackRock Inc & K     & 1 \\
    F     & Ford Motor Co & C     & 1     & BK    & Bank of NY Mellon Corp/The & K     & 1 \\
    GD    & General Dynamics Corp & C     & 1     & MET   & MetLife Inc & K     & 0 \\
    QCOM  & QUALCOMM Inc & C     & 1     & BRK/B  & Berkshire Hathaway Inc & K     & 1 \\
    PEP   & PepsiCo Inc & C     & 0     & MA    & Mastercard Inc & K     & 0 \\
    LLY   & Eli Lilly and Co & C     & 1     & V     & Visa Inc & K     & 0 \\
    MDT   & Medtronic PLC & C     & 1     & PYPL  & PayPal Holdings Inc & K     & 0 \\
    LMT   & Lockheed Martin Corp & C     & 1     & USB   & US Bancorp & K     & 1 \\
    NKE   & NIKE Inc & C     & 1     & MS    & Morgan Stanley & K     & 1 \\
    DD    & DuPont de Nemours Inc & C     & 1     & ALL   & Allstate Corp/The & K     & 1 \\
    SO    & Southern Co/The & D     & 1     & COF   & Capital One Financial Corp & K     & 1 \\
    DUK   & Duke Energy Corp & D     & 1     & WFC   & Wells Fargo \& Co & K     & 1 \\
    EXC   & Exelon Corp & D     & 0     & AMT   & American Tower Corp & L     & 1 \\
    NEE   & NextEra Energy Inc & D     & 1     & SPG   & Simon Property Group Inc & L     & 1 \\
    AMZN  & Amazon.com Inc & G     & 1     & AMGN  & Amgen Inc & M     & 1 \\
    HD    & Home Depot Inc/The & G     & 1     & GILD U & Gilead Sciences Inc & M     & 1 \\
    WMT   & Walmart Inc & G     & 1     & BIIB  & Biogen Inc & M     & 1 \\
    WBA   & Walgreens Boots Alliance Inc & G     & 0     &       &       &       &  \\
\hline    
\end{tabular}
\end{scriptsize}
}
\vspace{-5pt}\end{center}
\label{tab}
\end{table}

%October 2020 for the constituents of the S\&P100. The log-volatility and log-kurtosis statistics on the whole sample.
\begin{figure}[h!]
\begin{center}
\setlength{\tabcolsep}{10pt}
\begin{tabular}{c}
\includegraphics[scale=0.6]{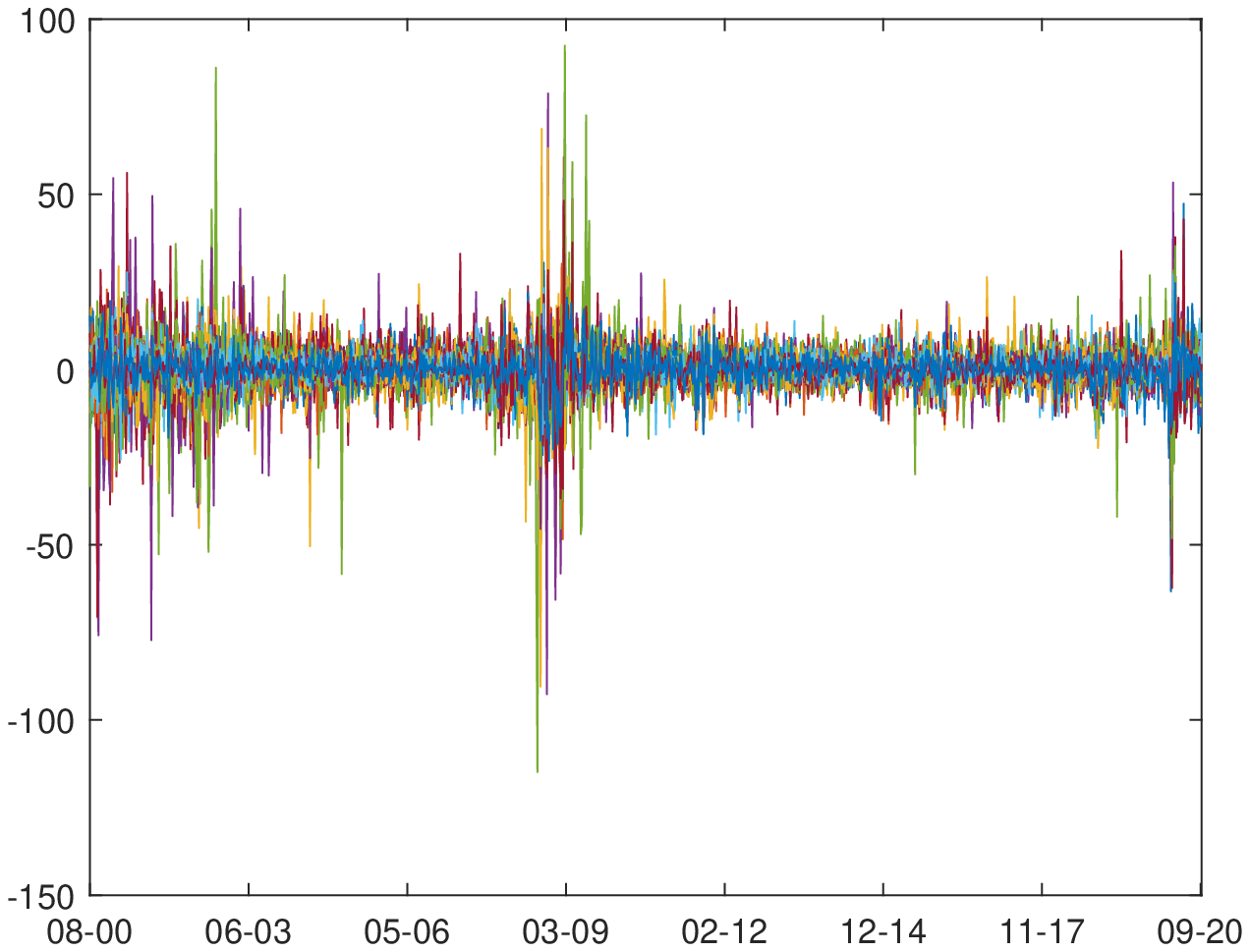}\vspace{10pt}\\
\includegraphics[scale=0.6]{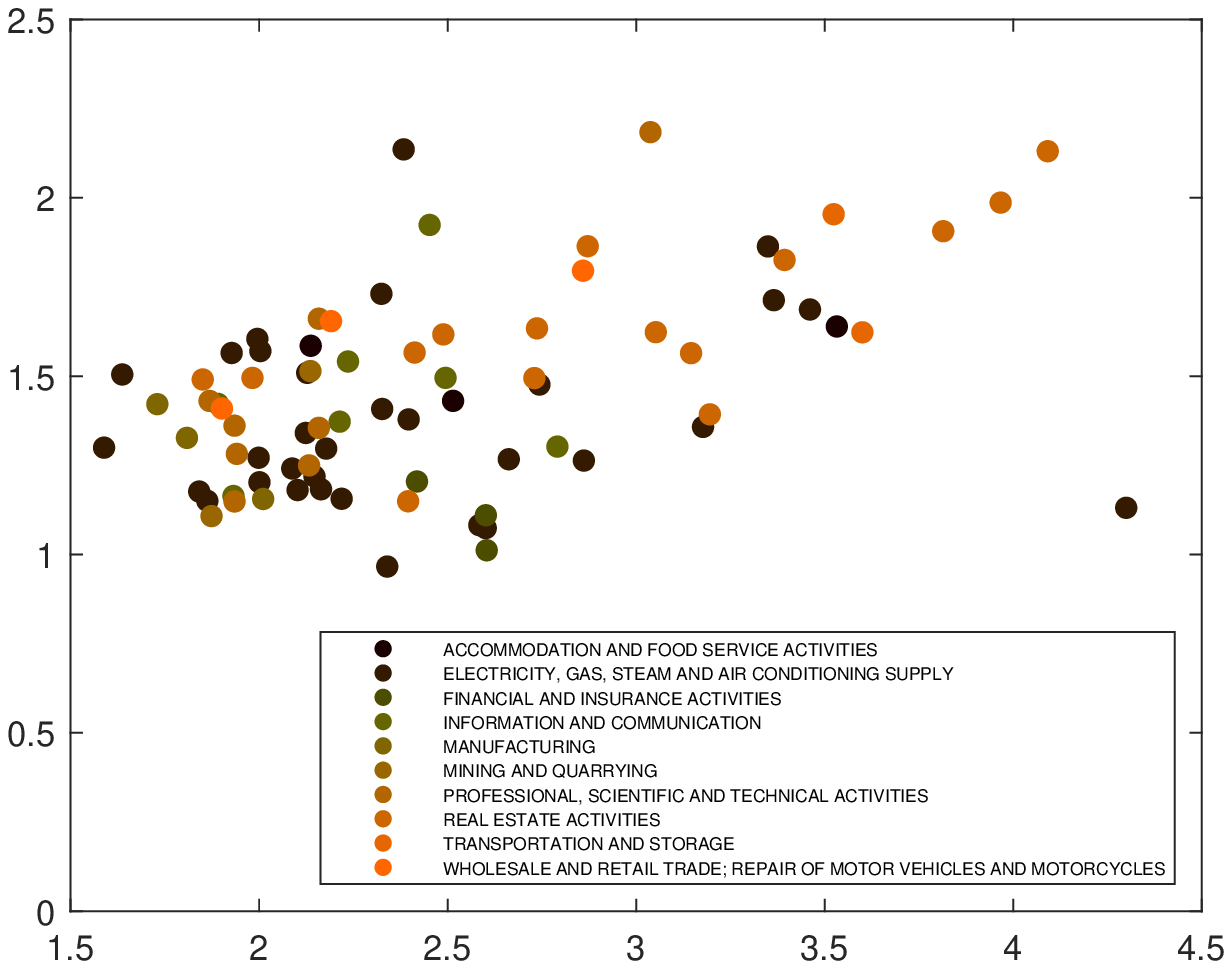}
\end{tabular}
\end{center}
\caption{Top: percentage log-returns of the SP\&100's constituents for the period $3^{rd}$ August 2000 to $3^{rd}$ October 2020. Bottom: scatter plot of the log-variance and log-kurtosis of the financial returns. Colours indicate a different sector.}\label{ret}
\end{figure}
%
%%\section{Empirical with $r=30$}
%\begin{figure}[h]
%\begin{center}
%\setlength{\tabcolsep}{10pt}
%\begin{tabular}{cc}
%%\hspace{-20pt}\includegraphics[scale=0.45]{Code/Figures/FigEmpiricalBNP_ANTr50medstar/Nclust1.eps}&
%%\hspace{-20pt}\includegraphics[scale=0.45]{Code/Figures/FigEmpiricalBNP_ANTr50medstar/Nclust2.eps}\\
%\includegraphics[scale=0.4]{Code/Figures/FigEmpiricalROBr50Full/AllocationRegime1.eps}&
%\includegraphics[scale=0.4]{Code/Figures/FigEmpiricalROBr50Full/AllocationRegime2.eps}
%\end{tabular}
%\end{center}
%\caption{Sector co-clustering matrix in regime 1 (left) and 2 (right).} %\label{NclustCoclust}
%\end{figure}

\begin{figure}[h!]
\begin{center}
\setlength{\tabcolsep}{10pt}
\begin{tabular}{cc}
\includegraphics[scale=0.4]{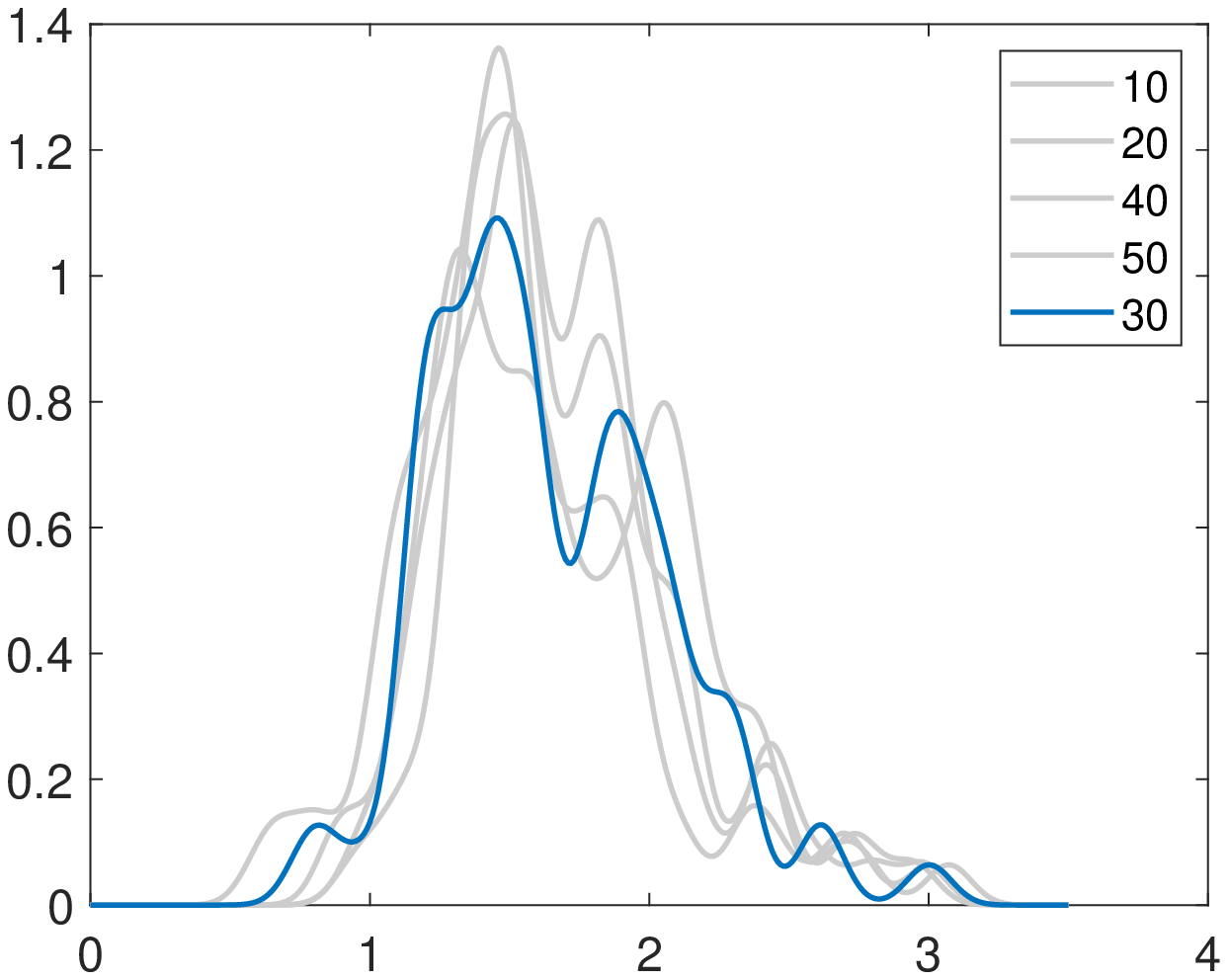}&
\includegraphics[scale=0.4]{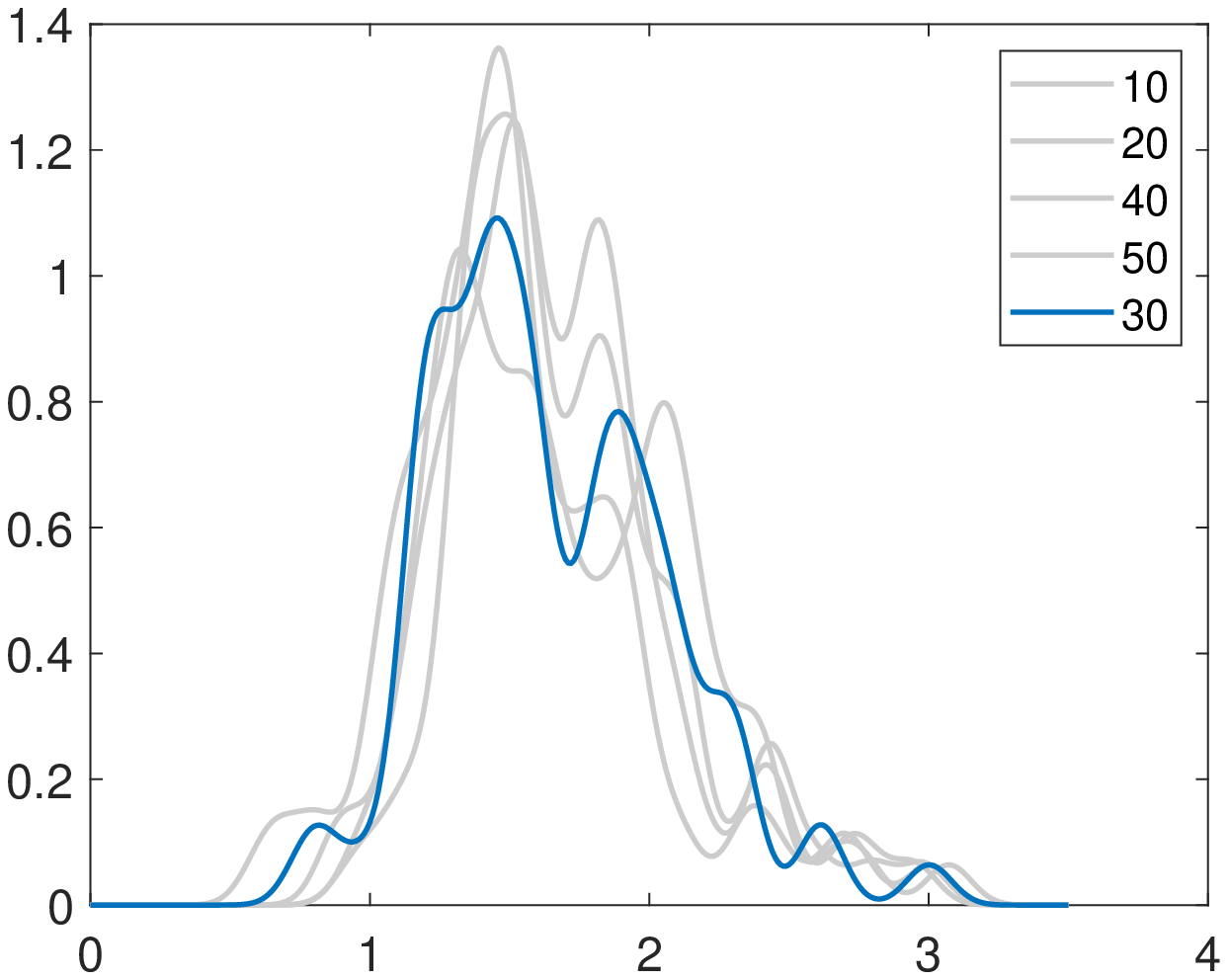}
\\\\\
\includegraphics[scale=0.4]{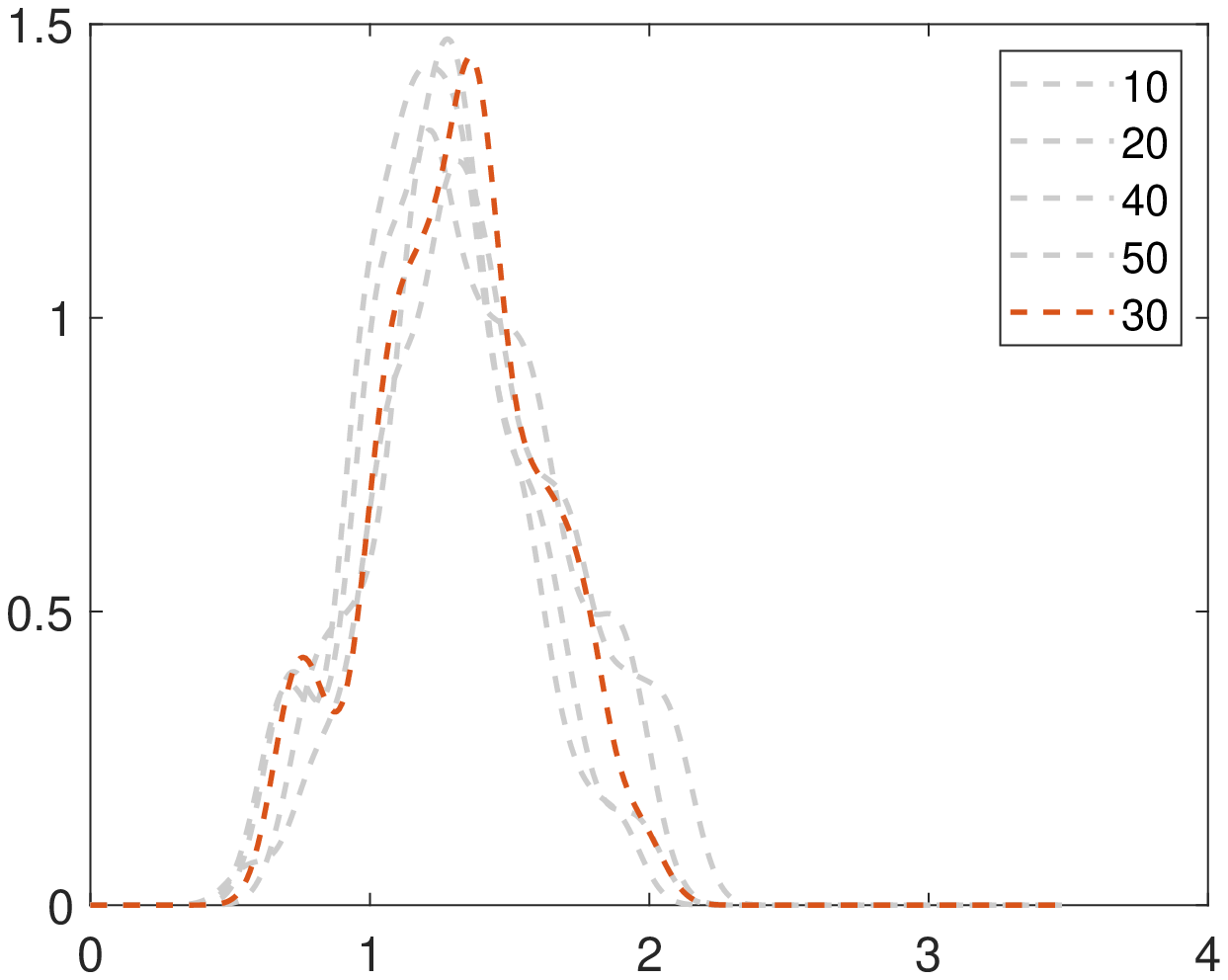}&
\includegraphics[scale=0.4]{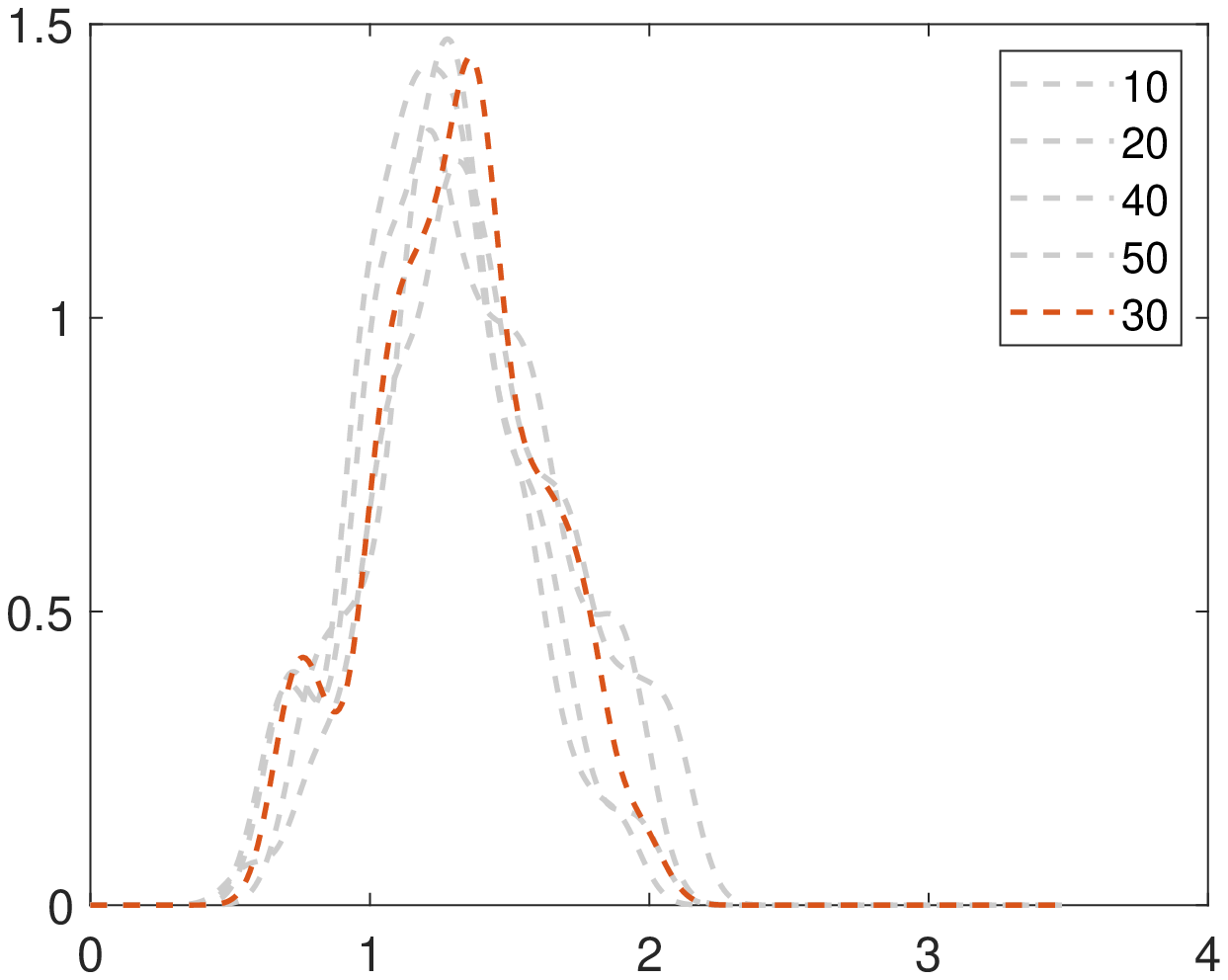}
\\\\\
\includegraphics[scale=0.4]{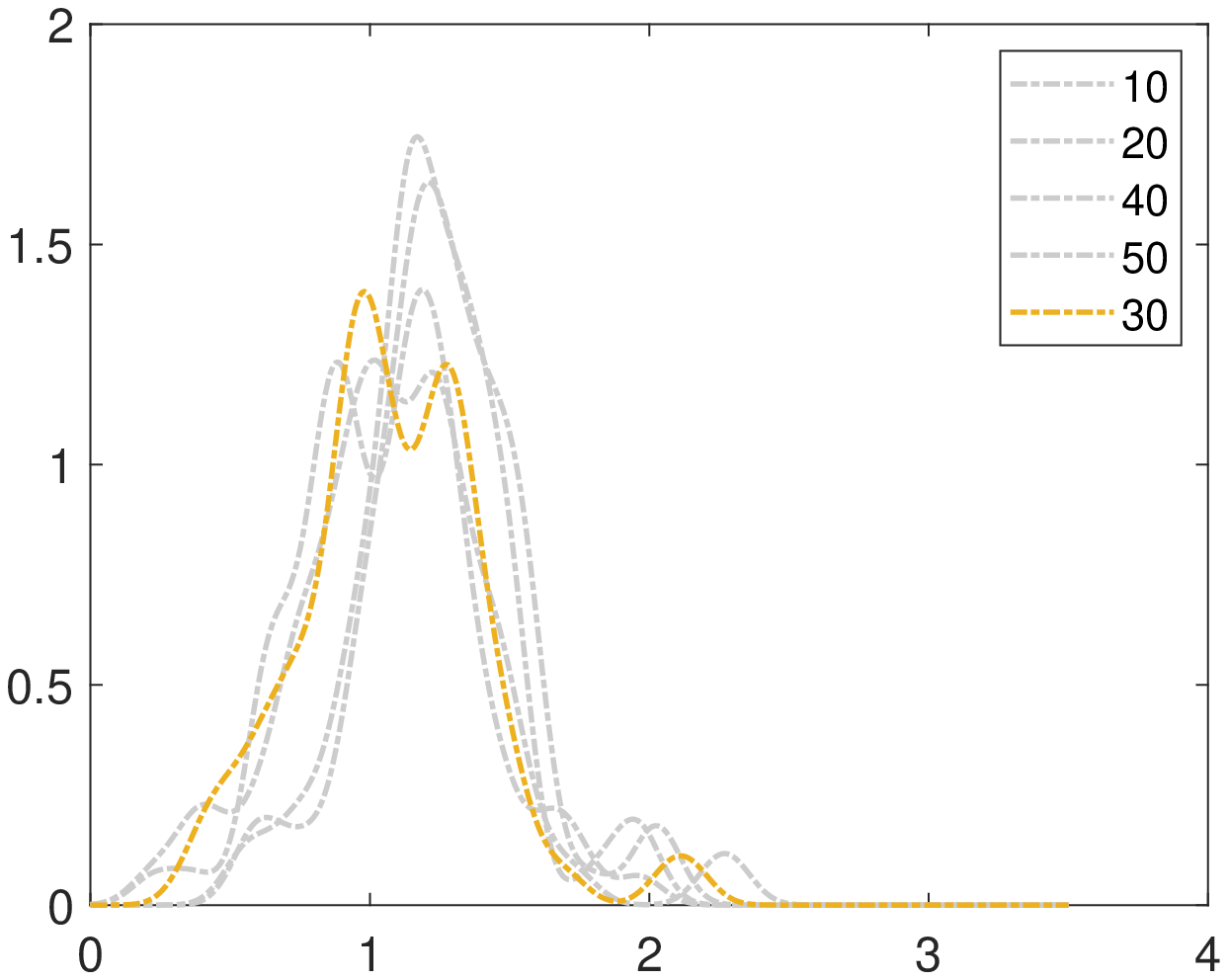}&
\includegraphics[scale=0.4]{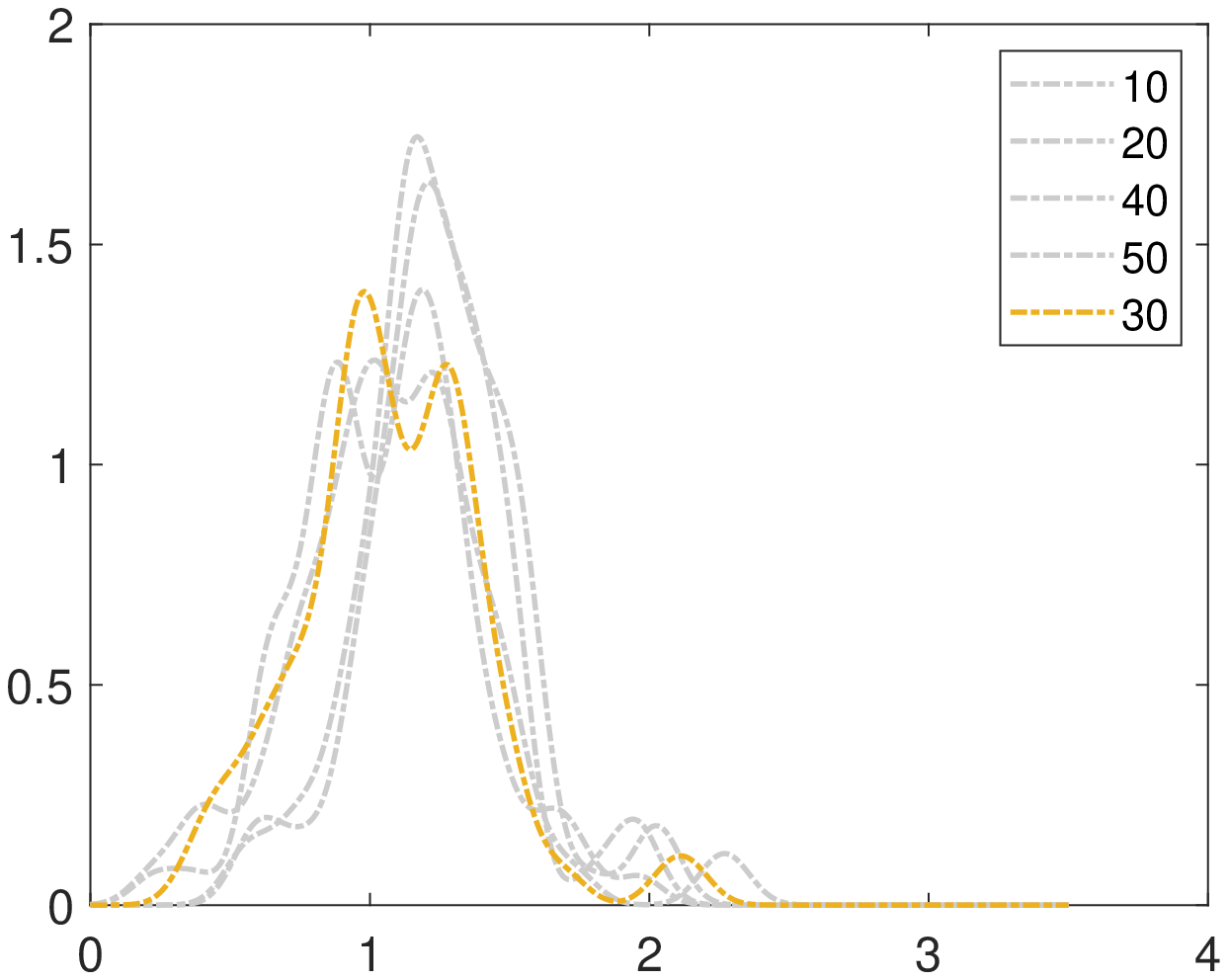}
\end{tabular}
\end{center}
\caption{Cross-sectional distribution of the log-volatility (left) and log-kurtosis (right) of the SP\&100's constituents log-returns in the three dates: $6^{th}$ July 2002, $23^{rd}$ August 2008 and $22^{nd}$ February 2020 (different rows). In each plot, the cross-section of statics is derived with different sizes of the rolling window (different lines).}\label{rob}
\end{figure}

\begin{table}[h!]
\caption{\small Cluster composition in Regime 1}\label{tab:addlabel}%
\vspace{-10pt}
\begin{center}
\begin{scriptsize}
    \begin{tabular}{lll|lll}
\hline       
\multicolumn{6}{c}{\textbf{Cluster 1}}  \\
\hline     
    Symbol &  Name  & S     & Symbol &  Name  & S \\
\hline     
  GD  & General Dynamics Corp  & C & BK   & Bank of New York Mellon Corp/The & K \\
  MDT & Medtronic PLC          & C & HD   & American Express Co              & K \\
  BA  & Boeing Co/The          & C & C    & Citigroup Inc                    & K \\
  KO  & Coca-Cola Co/The       & C & MS   & Morgan Stanley                   & K \\
  F   & Ford Motor Co          & C & ALL  & Allstate Corp/The                & K \\
  XOM & Exxon Mobil Corp       & C & WFC  & Wells Fargo \& Co                & K \\
  LLY & Eli Lilly and Co       & C & JPM  & JPMorgan Chase \& Co             & K \\
  GE  & General Electric Co    & C & AIG  & American Int. Group Inc & K \\
  CL  & Colgate-Palmolive Co   & C &  GS  & Goldman Sachs Group Inc/The      & K \\
 LMT  & Lockheed Martin Corp   & C & LOW  & Lowe's Cos Inc                   & G \\
 MRK  & Merck \& Co Inc        & C & TGT  & Target Corp                      & G \\
 PFE  & Pfizer Inc             & C & AXP  & Home Depot Inc/The               & G \\
 PG & Procter \& Gamble Co/The & C & CVS  & CVS Health Corp                  & G \\
RTX&Raytheon Technologies Corp & C & COST & Costco Wholesale Corp            & G \\
HON&Honeywell Int. Inc& C & WMT  & Walmart Inc                      & G \\
CSCO  & Cisco Systems Inc      & C & DUK  & Duke Energy Corp                 & D \\
 CAT  & Caterpillar Inc        & C & SO   & Southern Co/The                  & D \\
TMO & Thermo Fisher Scientific Inc &C&OXY & Occidental Petroleum Corp        & B \\
  CVX & Chevron Corp           & C & COP  & ConocoPhillips                   & B \\
VZ&  Verizon Communications Inc& J & AMT  & American Tower Corp              & L \\
MSFT  & Microsoft Corp         & J & SPG  & Simon Property Group Inc         & L \\
  DIS & Walt Disney Co/The     & J & GILD & Gilead Sciences Inc              & M \\
BKNG  & Booking Holdings Inc& J& BIIB  & Biogen Inc                          & M \\
IBM & Int. Business Machines Corp & J &FDX  & FedEx Corp            & H \\
          &       &       &       &       &  \\
\hline           
\multicolumn{6}{c}{\textbf{Cluster 2}}  \\
\hline     
    BMY   & Bristol Myers Squibb Co & C & UNH   & UnitedHealth Group Inc     & K \\
    JNJ   & Johnson \& Johnson      & C & BLK   & BlackRock Inc              & K \\
    AAPL  & Apple Inc               & C & COF   & Capital One Financial Corp & K \\
    MMM   & 3M Co                   & C & BRK/B & Berkshire Hathaway Inc     & K \\
    EMR   & Emerson Electric Co     & C & BAC   & Bank of America Corp       & K \\
    DHR   & Danaher Corp            & C & USB   & US Bancorp                 & K \\
    INTC  & Intel Corp              & C & CMCSA & Comcast Corp               & J \\
    QCOM  & QUALCOMM Inc            & C & ADBE  & Adobe Inc                  & J \\
    NVDA  & NVIDIA Corp             & C & T     & AT\&T Inc                  & J \\
    MO    & Altria Group Inc        & C & SBUX  & Starbucks Corp             & I \\
    ABT   & Abbott Laboratories     & C & MCD   & McDonald's Corp            & I \\
    NKE   & NIKE Inc                & C & NEE   & NextEra Energy Inc         & D \\
    DD    & DuPont de Nemours Inc   & C & AMGN  & Amgen Inc                  & M \\
  UPS   & United Parcel Service Inc & H & SLB   & Schlumberger NV            & B \\
   UNP   & Union Pacific Corp       & H & AMZN  & Amazon.com Inc             & G \\
\hline   
\end{tabular}%
\end{scriptsize}  
\vspace{-5pt}\end{center}  
\end{table}%

\begin{table}[h!]
    \caption{\small Cluster composition in Regime 2.}\label{tab:addlabe2}%
\vspace{-10pt}    
\begin{center}
\begin{scriptsize}
\begin{tabular}{lll|lll}
\hline
\multicolumn{6}{c}{\textbf{Cluster 1}}  \\
\hline 
Symbol &  Name  & S     & Symbol &  Name  & S \\
\hline 
    XOM   & Exxon Mobil Corp      & C  & GS    & Goldman Sachs Group Inc/The & K \\
    MMM   & 3M Co                 & C  & BK    & Bank of New York Mellon Corp/The & K \\
    BA    & Boeing Co/The         & C  & AXP   & American Express Co   & K \\
    CVX   & Chevron Corp          & C  & BRK/B & Berkshire Hathaway Inc & K \\
    KO    & Coca-Cola Co/The      & C  & OXY& American Int. Group Inc & K\\
    CL    & Colgate-Palmolive Co  & C  & WFC   & Wells Fargo \& Co & K   \\
    F     & Ford Motor Co         & C  & USB   & US Bancorp & K \\
    BMY  & Bristol Myers Squibb Co& C  & MS    & Morgan Stanley & K \\
TMO& Thermo Fisher Scientific Inc & C  & ALL   & Allstate Corp/The & K \\
    GD   & General Dynamics Corp  & C  &  HD    & Home Depot Inc/The & G \\
    LLY  & Eli Lilly and Co       & C  &  COST  & Costco Wholesale Corp & G\\
    MDT   & Medtronic PLC         & C  & CVS   & CVS Health Corp & G \\
    PFE   & Pfizer Inc            & C  & COP   & ConocoPhillips        & B \\
 PG    & Procter \& Gamble Co/The & C  & AIG & Occidental Petroleum Corp & B \\
 RTX & Raytheon Technologies Corp & C  & SPG   & Simon Property Group Inc & L \\
    MSFT  & Microsoft Corp        & J  & AMT   & American Tower Corp & L     \\
IBM   & Int. Business Machines Corp & J & DUK   & Duke Energy Corp & D \\
DIS   & Walt Disney Co/The       & J   &     SO    & Southern Co/The & D \\
          &       &       &       &       &  \\
\hline           
\multicolumn{6}{c}{\textbf{Cluster 2}}  \\
\hline     
    DHR   & Danaher Corp         & C     & JPM   & JPMorgan Chase \& Co & K  \\
    CAT   & Caterpillar Inc      & C     & BAC   & Bank of America Corp & K \\
    AAPL  & Apple Inc            & C     & T     & AT\&T Inc            & J \\
   DD    & DuPont de Nemours Inc & C     & ADBE  & Adobe Inc            & J\\
    JNJ   & Johnson \& Johnson   & C     & CMCSA & Comcast Corp         & J \\
    QCOM  & QUALCOMM Inc         & C     & VZ    & Verizon Communications Inc & J\\
    EMR   & Emerson Electric Co  & C     & UNP   & Union Pacific Corp   & H\\
    NKE   & NIKE Inc             & C     & UPS   & United Parcel Service Inc & H \\
    INTC  & Intel Corp           & C     & MCD   & McDonald's Corp      & I\\
    NVDA  & NVIDIA Corp          & C     & SBUX  & Starbucks Corp       & I \\
    ABT   & Abbott Laboratories  & C     & NEE   & NextEra Energy Inc   & D \\
COF & Capital One Financial Corp & K     & SLB   & Schlumberger NV      & B \\
    BLK   & BlackRock Inc        & K     & AMGN  & Amgen Inc            & M\\
  UNH   & UnitedHealth Group Inc & K   & AMZN  & Amazon.com Inc         & G \\    
          &       &       &       &       &  \\
\hline           
\multicolumn{6}{c}{\textbf{Cluster 3}}  \\
\hline         
    GE    & General Electric Co    & C     & LOW   & Lowe's Cos Inc & G \\
    MRK   & Merck \& Co Inc        & C     & WMT   & Walmart Inc & G   \\    
    LMT   & Lockheed Martin Corp   & C     & TGT   & Target Corp & G \\
    CSCO  & Cisco Systems Inc      & C     & GILD  & Gilead Sciences Inc & M \\
    MO    & Altria Group Inc       & C     & BIIB  & Biogen Inc & M \\
    HON   & Honeywell International Inc & C& BKNG  & Booking Holdings Inc & J \\
    C UN  & Citigroup Inc          & K     & FDX   & FedEx Corp & H \\
\hline    
\end{tabular}%
\end{scriptsize}  
    \vspace{-5pt}\end{center}  
\end{table}%

\begin{table}[h!]
    \caption{\small Cluster composition by market capitalization, small (bottom 30\%), medium (middle 40\%) and big (top 30\%) companies, in the two regimes (panel (a) and (b).}\label{tab:addlabe3}%
%\vspace{-10pt}    
  \centering
\resizebox{\textwidth}{!}{
    \begin{tabular}{ccccccccc}
    \multicolumn{9}{c}{(a) Regime 1}\\
    \hline
    \multicolumn{3}{c}{Cluster 1}     & \multicolumn{3}{c}{Cluster 2}      &       &       &  \\
    \hline
    Small   & Medium & Big& Small & Medium & Big&       &       &  \\
    \hline
    AIG   & AMT   & CSCO  & CAT   & ABT   & AAPL  &       &       &  \\
    ALL   & AXP   & HD    & COF   & ADBE  & AMZN  &       &       &  \\
    BIIB  & BA    & KO    & DD    & AMGN  & BAC   &       &       &  \\
    BK    & C     & MRK   & EMR   & BLK   & BRK/B  &       &       &  \\
    BKNG  & COST  & MSFT  & SLB   & BMY   & CMCSA &       &       &  \\
    CL    & CVS   & PFE   & USB   & DHR   & CVX   &       &       &  \\
    COP   & GE    & PG    &       & MCD   & DIS   &       &       &  \\
    DUK   & GILD  & WMT   &       & MMM   & INTC  &       &       &  \\
    F     & HON   & XOM   &       & MO    & JNJ   &       &       &  \\
    FDX   & IBM   &       &       & NEE   & JPM   &       &       &  \\
    GD    & LLY   &       &       & NKE   & NVDA  &       &       &  \\
    GS    & LMT   &       &       & QCOM  & T     &       &       &  \\
    MS    & LOW   &       &       & SBUX  & UNH   &       &       &  \\
    OXY   & MDT   &       &       & UNP   & VZ    &       &       &  \\
    SO    & RTX   &       &       & UPS   &       &       &       &  \\
    SPG   & TMO   &       &       &       &       &       &       &  \\
    TGT   & WFC   &       &       &       &       &       &       &  \\
          &       &       &       &       &       &       &       &  \\
    \multicolumn{9}{c}{(b) Regime 2}\\          
    \multicolumn{3}{c}{Cluster 1}       &\multicolumn{3}{c}{Cluster 2}     & \multicolumn{3}{c}{Cluster 3}  \\
\hline    
    Small & Medium & Big & Small & Medium & Big & Small & Medium & Big\\
\hline    
    AIG   & AMT   & BRK/B & CAT   & ABT   & AAPL  & BIIB  & C     & CSCO  \\
    ALL   & AXP   & CVX   & COF   & ADBE  & AMZN  & BKNG  & GE    & MRK  \\
    BK    & BA    & DIS   & DD    & AMGN  & BAC   & FDX   & GILD  & WMT \\
    CL    & BMY   & HD    & EMR   & BLK   & CMCSA & TGT   & HON   &  \\
    COP   & COST  & KO    & SLB   & DHR   & INTC  &       & LMT   &  \\
    DUK   & CVS   & MSFT  &       & MCD   & JNJ   &       & LOW   &  \\
    F     & IBM   & PFE   &       & NEE   & JPM   &       & MO    &  \\
    GD    & LLY   & PG    &       & NKE   & NVDA  &       &       &  \\
    GS    & MDT   & XOM   &       & QCOM  & T     &       &       &  \\
    MS    & MMM   &       &       & SBUX  & UNH   &       &       &  \\
    OXY   & RTX   &       &       & UNP   & VZ    &       &       &  \\
    SO    & TMO   &       &       & UPS   &       &       &       &  \\
    SPG   & WFC   &       &       &       &       &       &       &  \\
    USB   &       &       &       &       &       &       &       &  \\
    \end{tabular}%
}
\end{table}%

\begin{figure}[h!]
\begin{center}
\setlength{\tabcolsep}{10pt}
\renewcommand{\arraystretch}{1.1}
\begin{tabular}{cc}
$p_{i,11}$ vs $p_{i,22}$ & $\mu_{ik}$ vs $\gamma_{ik}$ \vspace{8pt}\\
\includegraphics[scale=0.4]{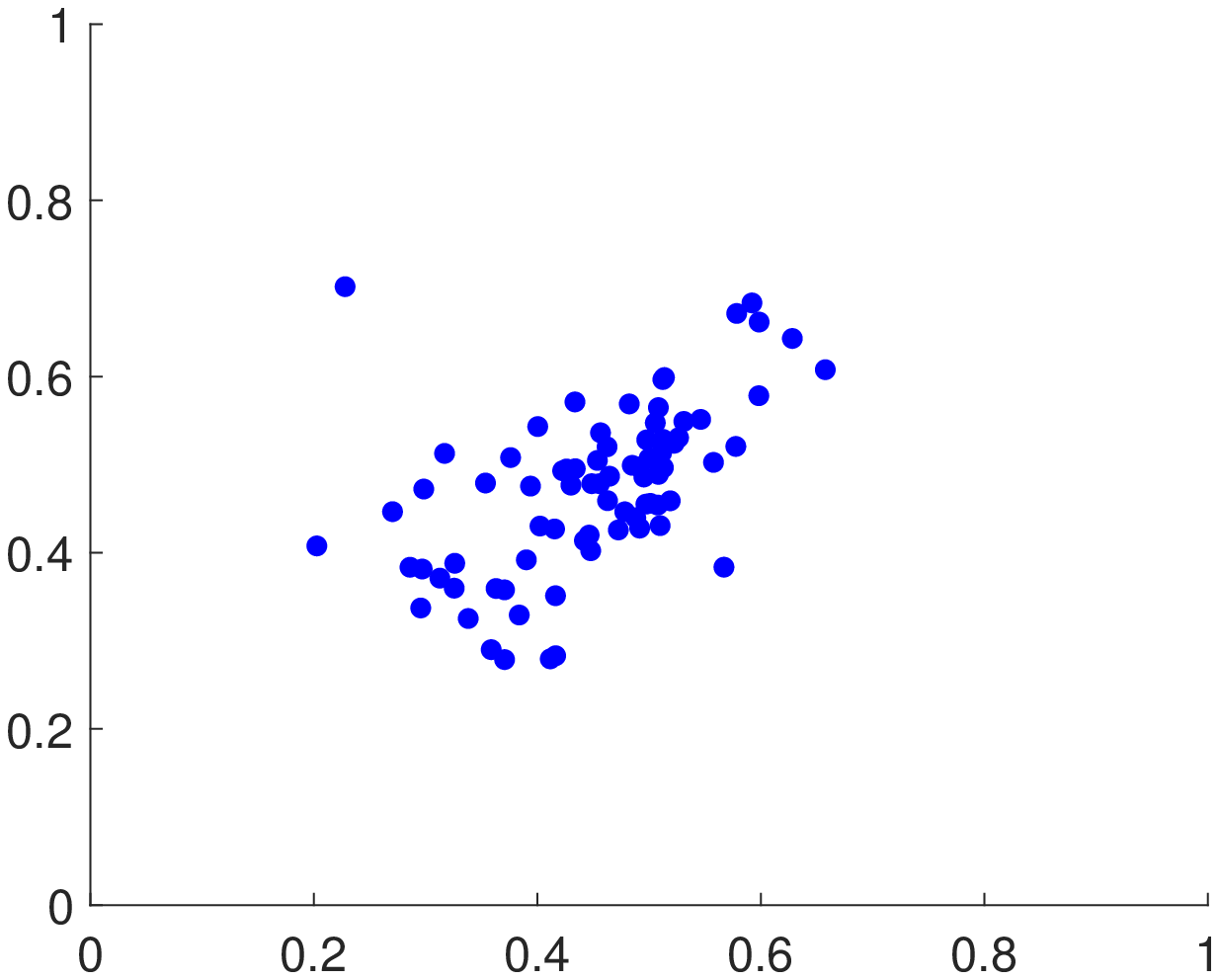}&
\includegraphics[scale=0.4]{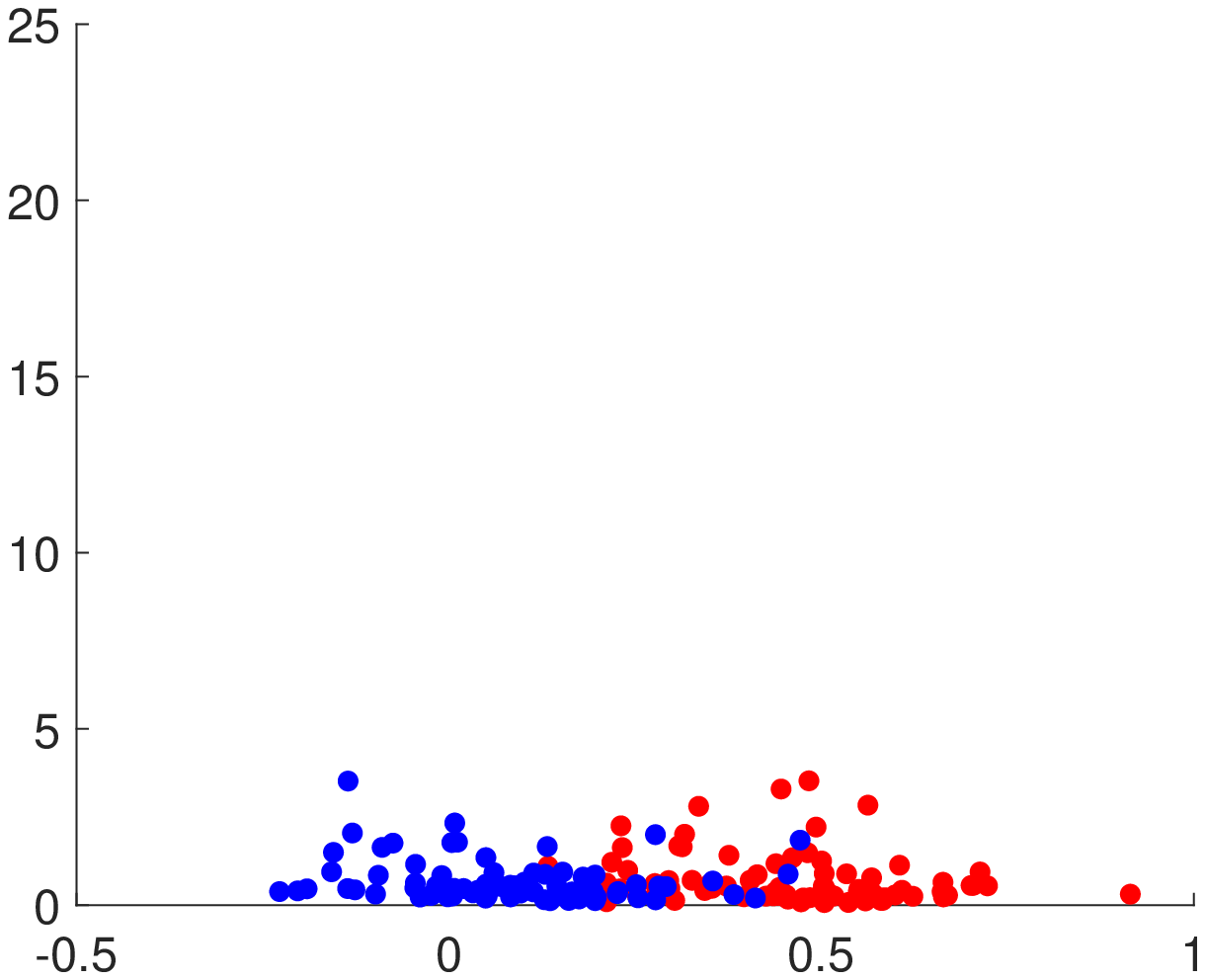}\\
\multicolumn{2}{c}{$\alpha_{ik}$ vs $\beta_{ik}$}\vspace{8pt}\\
\multicolumn{2}{c}{\includegraphics[scale=0.4]{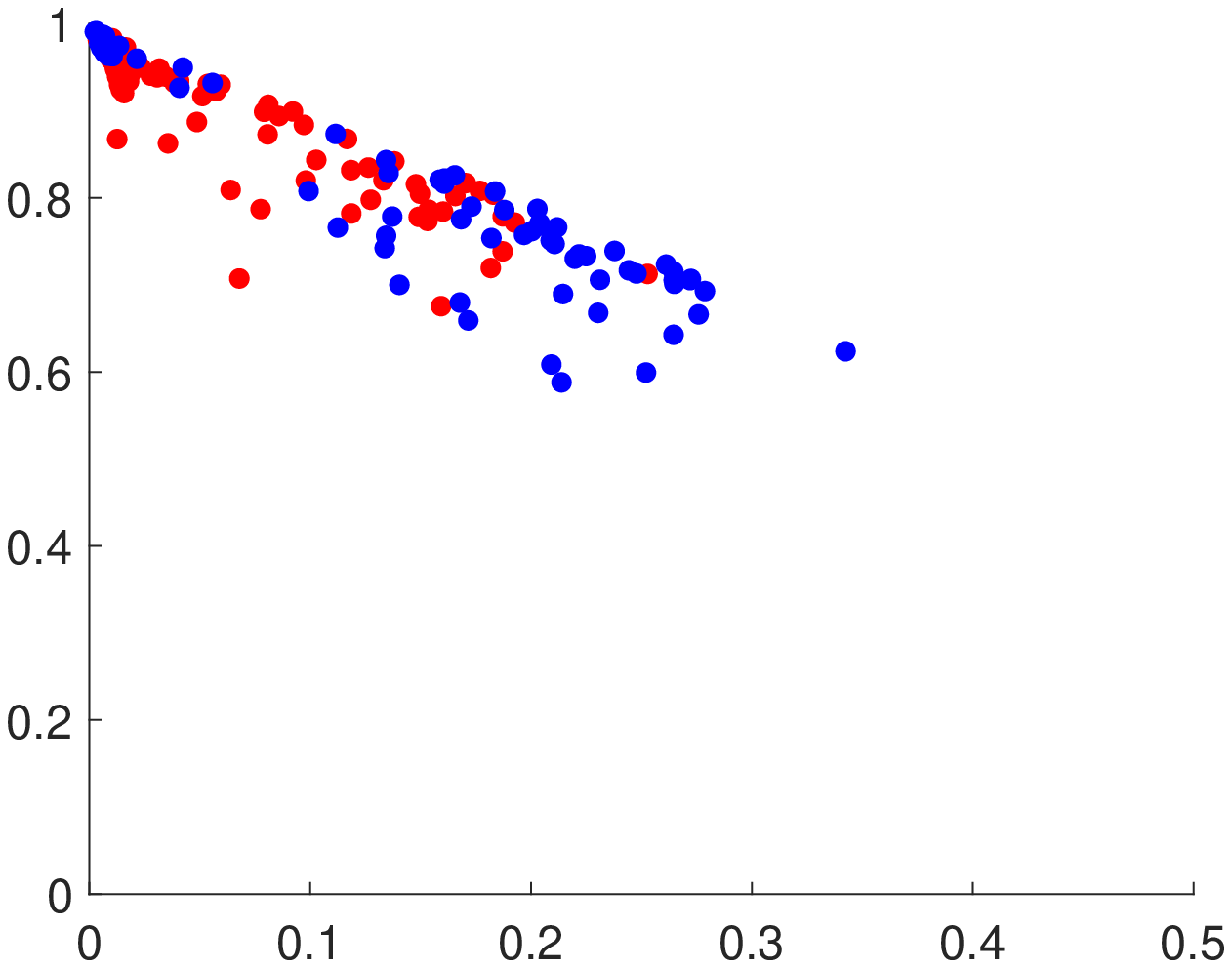}}
\end{tabular}
\end{center}
\caption{Parameter estimates. Colors indicate regime-specific parameter values with under-performance regime in blue (regime $k=1$) and over-performance regime in red (regime $k=2$).} \label{ParamBNPFina}
\end{figure}

\end{document}